\documentclass[twocolumn,showpacs,preprintnumbers,amsmath,amssymb]{revtex4}
\usepackage[table]{xcolor}
\usepackage{graphicx}
\usepackage{epsfig}
\usepackage{epstopdf}
\usepackage{dcolumn}
\usepackage{bm}
\usepackage{esvect}

\usepackage{hyperref}
\hypersetup{colorlinks=true,linkcolor=magenta,anchorcolor=green,citecolor=cyan,filecolor=black,menucolor=black,urlcolor=brown}
\usepackage{slashed}

\newcommand{\be}{\begin{equation}}
\newcommand{\ee}{\end{equation}}
\newcommand{\ba}{\begin{eqnarray}}
\newcommand{\ea}{\end{eqnarray}}

\begin{document}

\title{Lagrangian-space Gaussian ansatz for the matter redshift-space power spectrum and correlation function}

\author{Patrick Valageas}
\affiliation{Institut de Physique Th\'eorique, Universit\'e  Paris-Saclay, CEA, CNRS, F-91191 Gif-sur-Yvette Cedex, France}

\author{Takahiro Nishimichi}
\affiliation{Center for Gravitational Physics, Yukawa Institute for Theoretical Physics, Kyoto University, Kyoto 606-8502, Japan}
\affiliation{Kavli Institute for the Physics and Mathematics of the Universe (WPI), The University of Tokyo Institutes for Advanced Study (UTIAS), The University of Tokyo, Chiba 277-8583, Japan}

\begin{abstract}

We study the predictions for the matter redshift-space power spectrum and correlation function
of a Lagrangian-space Gaussian ansatz introduced in a previous work. This model is a natural
extension of the Zeldovich approximation, where the displacement and velocity power spectra
are determined by the equations of motion, instead of being set equal to the linear power
spectrum. It does not contain any free parameter. As for the real-space statistics, we
find that this Lagrangian-space approach is much more efficient for the correlation functions
than for the power spectra. The damping of the BAO oscillations is well recovered but there
is a large smooth drift from the simulations in the power spectra. The multipoles of the correlation functions
are well recovered on BAO scales, with an accuracy of $2\%$ for $\xi^s_0$ down to
$10 h^{-1}$ Mpc, and of $3\%$ for $\xi^s_2$ down to $26 h^{-1}$ Mpc, at $z \geq 0.35$.

\end{abstract}

\date{\today}
\preprint{YITP-20-50}

\maketitle


\section{Introduction}
\label{sec:introduction}

The large-scale structure of the Universe is a key probe of cosmological scenarios and
gravitational physics. For instance, the baryon acoustic oscillations (BAO) of the matter and
galaxy power spectra, which appear as a peak at about $100 h^{-1} {\rm Mpc}$
in their correlation functions, provide a standard ruler \cite{Eisenstein:2006nj}.
In combination with other probes such as the cosmic microwave background (CMB) and
the Hubble expansion rate measured from distant supernovae, this constrains the
parameters of the standard $\Lambda$-CDM model \cite{Alam:2016hwk}
and alternative dark-energy or modified-gravity scenarios.
The growth rate of large-scale structures,
measured for instance from the shape and the evolution with redshift of the galaxy
power spectrum, also constrains the underlying cosmology and gravity on large scales.
This has led to various observational programs
\cite{Ross:2016gvb,Blake:2011en,Martini:2018kdj,Laureijs:2011gra,Abell:2009aa},
which require theoretical modelling of these large-scale structures in order to compare
the predictions of various scenarios with the data.

These predictions are often done through numerical simulations, which can handle
complicated nonlinear effects, such as the nonlinear mode coupling of the gravitational
dynamics or baryonic feedback associated with cooling, star formation and active galactic
nuclei.
However, analytical approaches remain useful on large weakly nonlinear scales,
where they are reliable and facilitate the scanning of a large parameter space,
e.g. if one wishes to investigate alternative scenarios. They also clarify the
main features of the gravitational dynamics that govern the growth of structures.

In Eulerian space, the main analytical methods are the standard perturbation theory
(SPT) \cite{Goroff:1986ep,Bernardeau:2001qr}, and its various partial resummations
\cite{Crocce:2005xy,Valageas:2006bi,Bernardeau:2008fa,Taruya:2012ut}.
However, going to high orders does not ensure a systematically greater accuracy
\cite{Carlson:2009it,Valageas:2010rx,Blas:2013aba,Valageas:2013hxa}
and the Euler equation itself is only an approximation that breaks down after shell crossing.
This can be handled by explicit coarse-graining \cite{Pietroni:2011iz} or by
effective field theory (EFT) methods
\cite{Baumann:2010tm,Carrasco:2012cv,Lewandowski:2015ziq,delaBella:2017qjy}.
In practice, usual EFT schemes only take into account part of the impact of small-scale
nonlinearities as they neglect vorticity, but this could be added to the formalism.

An alternative is to work in Lagrangian space, where we follow the trajectories of particles
\cite{Zeldovich:1969sb,Buchert:1992ya,Bouchet:1992uh,Matsubara:2007wj,Vlah:2014nta,McDonald:2017ths} and shell crossing is not necessarily a problem.
In a recent work \cite{Valageas:2020}, we have presented a new approach to follow the gravitational
dynamics. The idea is to follow the evolution of the probability distribution
${\cal P}({\bf x},{\bf u};t)$ of the displacement and velocity fields by considering
a simplified ansatz for ${\cal P}$, characterized for instance by its low-order correlations,
and to use the equations of motion to derive as many constraints as needed to
fully determine these parameters. At the lowest order, we considered a Lagrangian-space
curl-free Gaussian ansatz, where ${\cal P}({\bf x},{\bf u};t)$ is Gaussian and we only
need to follow the evolution of the displacement and velocity power spectra.
Because we do not expand on the displacement and velocity fields, this provides
a nonperturbative scheme with a damping of their power spectra on nonlinear scales
that arises from the equations of motion.

For real-space statistics of the density field, the predictions of this method coincide
with a truncated Zeldovich approximation \cite{Coles:1992vr}.
However, in contrast with the truncated Zeldovich approximation, the displacement and
velocity power spectra are different. This implies that redshift-space statistics
no longer coincide with the predictions of any truncated Zeldovich approximation.
We investigate in this paper the predictions of this Lagrangian-space Gaussian ansatz
for the matter density redshift-space power spectrum and correlation function,
which we compare with numerical simulations and the standard Zeldovich approximation.
Redshift-space anisotropies, due to the velocity of the tracers along the line of sight,
actually provide an additional probe of the growth of large-scale structures and
cosmological scenarios
\cite{Cole:1993kh,Hamilton:1997zq,Reid:2012sw,Samushia:2012iq,Reid:2014iaa,Alam:2016hwk,Bose:2017dtl,Zarrouk:2018vwy}.
For biased tracers, such as galaxies, one also needs
to model the bias to compare with data \cite{Kaiser:1984sw,Desjacques:2016bnm}.
We leave this second step for future works
and we focus on the matter clustering in this paper, as our aim is to investigate
the properties of this new Lagrangian-space Gaussian ansatz.

This paper is organized as follows. We recall in Sec.~\ref{sec:Gaussian-ansatz}
the Lagrangian-space Gaussian ansatz developed in \cite{Valageas:2020} and we give its prediction
for the redshift-space matter power spectrum in Sec.~\ref{sec:power-spectrum}.
Then, we compare our results with the Zeldovich approximation and numerical simulations.
We study the redshift-space correlation function in Sec.~\ref{sec:correlation}.
We compare our method with other approaches in Sec.~\ref{sec:other-approaches}
and we conclude in Sec.~\ref{sec:conclusion}.
We describe our numerical procedure for the computation of the power spectrum
in the appendix~\ref{app:power-spectrum}.

\section{Lagrangian-space curl-free Gaussian ansatz}
\label{sec:Gaussian-ansatz}

We recall in this section the Lagrangian-space curl-free Gaussian ansatz introduced
in \cite{Valageas:2020}. It is based on a Lagrangian framework, where we follow the comoving trajectories of
dark matter particles as
\be
{\bf x}({\bf q},t) = {\bf q} + {\bf \Psi}({\bf q},t) ,
\ee
where ${\bf q}$ is the initial (Lagrangian) coordinate of the particle and ${\bf\Psi}({\bf q},t)$
the displacement field.
We simultaneously keep track of the particle velocities, ${\bf u}({\bf q},t)$, defined by
\be
{\bf u}({\bf q},\eta) \equiv \frac{\partial{\bf\Psi}}{\partial\eta} ,
\label{eq:u-def}
\ee
where we use $\eta=\ln D_+(t)$ as the time coordinate.
Here $D_+(t)$ is the linear growing mode and the linear growth rate $f(t)$ is given by
\be
f(t) = \frac{d\ln D_+}{d\ln a} = \frac{\dot{D}_+}{H D_+} ,
\ee
where $H(t)$ is the Hubble expansion rate and the dot denotes the derivative with respect
to cosmic time $t$.
As in \cite{Valageas:2020}, we consider a curl-free ansatz, where the displacement and velocity fields
are fully defined by their divergences $\chi$ and $\theta$,
\be
\chi({\bf q},\eta) = - \nabla_{\bf q} \cdot{\bf\Psi} , \;\;\;
\theta({\bf q},\eta) = - \nabla_{\bf q} \cdot{\bf u} ,
\label{eq:chi-theta-def}
\ee
which also read in Fourier space as
\be
{\bf \Psi}({\bf k}) = \frac{i {\bf k}}{k^2} \chi({\bf k}) , \;\;\;
{\bf u}({\bf k}) = \frac{i {\bf k}}{k^2} \theta({\bf k}) .
\label{eq:chi-theta-Fourier}
\ee

Discarding non-gravitational interactions, the equation of motion of the particles is
\be
\ddot{\bf \Psi} + 2 H \dot{\bf \Psi} = - \frac{\nabla_{\bf x} \Phi}{a^2} ,
\label{eq:Psi-eom}
\ee
where $\Phi$ is the gravitational potential.
The gravitational force on the particle ${\bf q}$ can be written as
\cite{McDonald:2017ths,Valageas:2020}
\be
{\bf F}({\bf q},\eta) =  \int \frac{d{\bf q}'d{\bf k}}{(2\pi)^3} \,
 e^{i {\bf k} \cdot [ {\bf x}({\bf q}) - {\bf x}({\bf q}') ]} \frac{i {\bf k}}{k^2+\mu^2} ,
\label{eq:Fq}
\ee
where $\mu\to 0$ provides a convenient regularization of infrared divergences
associated with the homogeneous background. This corresponds to the well-known
Jeans ``swindle'' \cite{Kiessling:1999eq,2009PhRvE..80d1108G,Gabrielli:2010uk}.
The expression (\ref{eq:Fq}) of the force is well suited to Lagrangian space,
as it sums the gravitational attraction from all particles ${\bf q}'$, at distance
${\bf x}({\bf q}') - {\bf x}({\bf q})$ from the particle ${\bf q}$.
Using $\eta$ as the time coordinate the equation of motion (\ref{eq:Psi-eom}) becomes
\be
\frac{\partial^2 {\bf \Psi}}{\partial\eta^2} + \left( \frac{3\Omega_{\rm m}}{2f^2} - 1 \right)
\frac{\partial {\bf \Psi}}{\partial\eta} = \frac{3\Omega_{\rm m}}{2f^2} {\bf F} .
\label{eq:eom-k-F}
\ee
This implies for the Fourier-space power spectra of the displacement and velocity fields
the exact equations
\ba
&& \frac{\partial P_{\chi\chi}}{\partial\eta} = 2 P_{\chi\theta} ,
\label{eq:dPchichi-deta} \\
&& \frac{\partial P_{\chi\theta}}{\partial\eta} = P_{\theta\theta}
+ \left( 1 - \frac{3\Omega_{\rm m}}{2f^2} \right) P_{\chi\theta}
+ \frac{3\Omega_{\rm m}}{2f^2} P_{\chi\zeta} ,
\label{eq:dPchitheta-deta} \\
&& \frac{\partial P_{\theta\theta}}{\partial\eta} = \left( 2 - \frac{3\Omega_{\rm m}}{f^2} \right)
P_{\theta\theta} + \frac{3\Omega_{\rm m}}{f^2} P_{\theta\zeta} .
\label{eq:dPthetatheta-deta}
\ea
This system is not closed, as it involves cross correlations with the Lagrangian-space
divergence of the gravitational force $\zeta$,
\be
\zeta({\bf q},\eta) = - \nabla_{\bf q} \cdot{\bf F}  .
\label{eq:zeta-def}
\ee

The method used in \cite{Valageas:2020} to close this system is to take a curl-free Gaussian 
ansatz for the displacement and velocity fields. Thus, taking the displacement and velocity fields
to have the curl-free form (\ref{eq:chi-theta-Fourier}) with $\chi$ and $\theta$ being Gaussian
fields, we can exactly compute the cross power spectra $P_{\chi\zeta}$ and
$P_{\theta\zeta}$ at each time. The latter are nonlinear functionals of the displacement
and velocity fields, using the expression (\ref{eq:Fq}) of the gravitational force.
Then, the system (\ref{eq:dPchichi-deta})-(\ref{eq:dPthetatheta-deta}) determines the
evolution with time of the displacement and velocity power spectra.
This scheme is nonperturbative, as we do not expand the equations of motion
(\ref{eq:dPchichi-deta})-(\ref{eq:dPthetatheta-deta}) nor the nonlinear expression
(\ref{eq:Fq}) of the gravitational force.

The approximation enters at the level of the curl-free Gaussian ansatz for the probability
distribution ${\cal P}({\bf \Psi},{\bf u};\eta)$.
In particular, the exact probability distribution ${\cal P}$ obeys an infinite number of constraints,
e.g. the evolution equations of displacement and velocity polyspectra at all orders
(bispectra, trispectra, and so on).
By imposing a Gaussian ansatz, fully defined by the three power spectra
$\{P_{\chi\chi},P_{\chi\theta},P_{\theta\theta}\}$, we can only keep track of three of
these constraints. Then, it is natural to consider
Eqs.(\ref{eq:dPchichi-deta})-(\ref{eq:dPthetatheta-deta}) that directly follow the evolution with
time of these three power spectra.

This improves over the Zeldovich approximation \cite{Zeldovich:1969sb}
 in the sense that we derive the ``best''
Gaussian ansatz for the displacement and velocity fields, as defined by the requirement to
fulfil the exact constraints (\ref{eq:dPchichi-deta})-(\ref{eq:dPthetatheta-deta}),
instead of simply setting $\{P_{\chi\chi},P_{\chi\theta},P_{\theta\theta}\}$ equal to the
linear-theory power spectrum.
As seen in \cite{Valageas:2020}, this automatically yields a self-truncation of these power spectra
at high $k$.
For the real-space matter density power spectrum, this is equivalent to a truncated
Zeldovich approximation \cite{Coles:1992vr}. However, in contrast with the
standard truncated Zeldovich approximation, the truncation is not put by hand,
with some free parameters fitted to numerical simulations. It automatically arises from the equations of
motion (\ref{eq:dPchichi-deta})-(\ref{eq:dPthetatheta-deta}).
Moreover, the displacement and velocity power spectra become different on nonlinear scales.
This implies that our model is different from a truncated Zeldovich approximation
for the redshift-space matter density power spectrum.

We refer the reader to \cite{Valageas:2020} for details on the numerical computation of the
displacement and velocity power spectra $\{P_{\chi\chi},P_{\chi\theta},P_{\theta\theta}\}$
from the equations of motion (\ref{eq:dPchichi-deta})-(\ref{eq:dPthetatheta-deta}).

\section{Redshift-space matter density power spectrum}
\label{sec:power-spectrum}

\subsection{Analytical expressions}
\label{sec:analytical-Pk}

The redshift-space coordinate ${\bf s}$ differs from the real-space coordinate
${\bf x}$ by the Doppler effect associated with the peculiar velocity $v_z$ along
the line of sight \cite{Kaiser:1987qv,Hamilton:1997zq}
\be
{\bf s} = {\bf x} + \frac{{\bf v} \cdot {\bf e}_{z}}{a H} {\bf e}_z ,
\ee
where ${\bf e}_z$ is the outward unit vector along the line of sight and the peculiar velocity
${\bf v}$ is defined as
\be
{\bf v} = a \dot{\bf \Psi} = a f H {\bf u} .
\ee
This gives in terms of the displacement ${\bf\Psi}$ and of the velocity ${\bf u}$ introduced
in Eq.(\ref{eq:u-def})
\be
{\bf s}({\bf q}) = {\bf q} +{\bf\Psi} + f u_z {\bf e}_z .
\ee
The conservation of matter gives for the redshift-space matter density field
$\rho^s({\bf s}) d{\bf s} = \bar\rho d{\bf q}$ in the single-stream regime. After shell crossing
we need to sum over all streams, but in both cases the redshift-space matter density
power spectrum can be written as \cite{Taylor:1996ne}
\be
P^s({\bf k}) = \int \frac{d{\bf q}}{(2\pi)^3} \langle e^{i {\bf k}\cdot [ {\bf s}(q) - {\bf s}(0)]} \rangle ,
\label{eq:Ps-sq}
\ee
where we used the flat-sky limit.
In this regime this expression is exact, but in general the average of the exponential term
is difficult to compute. However, as for the Zeldovich approximation, for the Gaussian ansatz
described in Sec.~\ref{sec:Gaussian-ansatz} this is a simple Gaussian average.
This gives  \cite{Taylor:1996ne}
\be
P^s({\bf k}) \! = \! \int \!\!\! \frac{d{\bf q}}{(2\pi)^3} e^{i {\bf k}\cdot{\bf q} - \frac{1}{2}
\langle[ {\bf k} \cdot ( {\bf \Psi}(q) - {\bf \Psi}(0) ) + f k_z ( v_z({\bf q}) - v_z(0) ) ]^2 \rangle } .
\label{eq:Psk-Psi-q-0}
\ee
For the curl-free displacement and velocity fields (\ref{eq:chi-theta-Fourier}), this reads
\be
P^s({\bf k}) = \int \frac{d{\bf q}}{(2\pi)^3}  e^{i {\bf k}\cdot{\bf q}}
e^{-A_{\chi\chi} - 2f A_{\chi\theta} - f^2 A_{\theta\theta}} ,
\label{eq:Psk-A-def}
\ee
with
\ba
&& A_{\chi\chi} = \int d{\bf k}' [1-\cos({\bf k}'\cdot{\bf q})] \frac{({\bf k}\cdot{\bf k}')^2}{k'^4}
P_{\chi\chi}(k') ,
\label{eq:A-chichi-def} \\
&& A_{\chi\theta} = \int d{\bf k}' [1-\cos({\bf k}'\cdot{\bf q})] \frac{({\bf k}\cdot{\bf k}') k_z k'_z}{k'^4}
P_{\chi\theta}(k') , \hspace{1cm}
\label{eq:A-chitheta-def} \\
&& A_{\theta\theta} = \int d{\bf k}' [1-\cos({\bf k}'\cdot{\bf q})] \frac{(k_z k'_z)^2}{k'^4}
P_{\theta\theta}(k') .
\label{eq:A-thetatheta-def}
\ea
It is convenient to define the relative displacement and velocity variances
\ba
&& \alpha_{**}(q) = \frac{4\pi}{3} \int_0^\infty dk \, P_{**}(k)
[ 1 - j_0(k q) - j_2(k q) ] , \hspace{0.7cm}
\label{eq:alpha-def} \\
&& \beta_{**}(q) = 4\pi \int_0^\infty dk \, P_{**}(k) j_2(k q) ,
\label{eq:beta-def}
\ea
where $**$ stands for $\{\chi\chi\}$, $\{\chi\theta\}$ or $\{\theta\theta\}$.
Then, the quantities $A_{**}$ introduced in
Eqs.(\ref{eq:A-chichi-def})-(\ref{eq:A-thetatheta-def}) read
\ba
&& A_{\chi\chi} = \alpha_{\chi\chi} k^2 + \beta_{\chi\chi} \frac{({\bf k}\cdot{\bf q})^2}{q^2} ,
\label{eq:A-chichi} \\
&& A_{\chi\theta} = \alpha_{\chi\theta} k_z^2 + \beta_{\chi\theta}
\frac{({\bf k}\cdot{\bf q}) k_z q_z}{q^2} ,
\label{eq:A-chitheta} \\
&& A_{\theta\theta} = \alpha_{\theta\theta} k_z^2 + \beta_{\theta\theta}
\frac{(k_z q_z)^2}{q^2} .
\label{eq:A-thetatheta}
\ea
Substituting into Eq.(\ref{eq:Psk-A-def}) we obtain for the redshift-space power spectrum,
\ba
&& P^s({\bf k}) = \int \frac{d{\bf q}}{(2\pi)^3}  \; e^{i {\bf k}\cdot{\bf q}} \;
e^{-\alpha_{\chi\chi} k^2 - (2f \alpha_{\chi\theta}+f^2\alpha_{\theta\theta}) k_z^2}
\nonumber \\
&& \times e^{- [\beta_{\chi\chi} ({\bf k}\cdot{\bf q})^2
+ 2f \beta_{\chi\theta} ({\bf k}\cdot{\bf q}) k_z q_z + f^2 \beta_{\theta\theta} (k_z q_z)^2 ]/q^2} ,
\label{eq:Ps-k-alpha-beta}
\ea
which depends on both the norm $k$ of the wave 
vector and the cosine of its angle with the line of sight, $\mu=k_z/k$.
Because we work in a Lagrangian framework and do not perform any perturbative expansion,
the power spectrum (\ref{eq:Ps-k-alpha-beta}) does not suffer from the infrared divergences
or artificially large contributions that affect Eulerian approaches and require specific care
\cite{Senatore:2014via,Vlah:2015zda,Blas:2016sfa,Ivanov:2018gjr}.
Indeed, the argument of the exponential only depends on relative displacements
and velocities, as seen in Eq.(\ref{eq:Psk-Psi-q-0}).
Therefore, it is insensitive to uniform displacements and velocities and does not break
Galilean invariance (or the weak equivalence principle in the relativistic context).
We describe in the appendix~\ref{app:power-spectrum} our numerical procedure
to compute Eq.(\ref{eq:Ps-k-alpha-beta}).

It is usual to expand the redshift-space power spectrum over the Legendre polynomials
\cite{Hamilton:1997zq},
${\cal P}_{\ell}(\mu)$,
\be
P^s(k,\mu) = \sum_{\ell=0}^{\infty} P^s_{2\ell}(k) {\cal P}_{2\ell}(\mu) .
\ee
We obtain these multipoles from the integration over $\mu$,
\be
P^s_{2\ell}(k) = (4\ell+1) \int_0^1 d\mu P^s(k,\mu) {\cal P}_{2\ell}(\mu) .
\ee

\subsection{Zeldovich approximation}
\label{sec:Zeldovich}

In the Zeldovich approximation \cite{Zeldovich:1969sb},
we use the linear theory to obtain the displacement and
velocity fields. Therefore, the expression (\ref{eq:Ps-k-alpha-beta}) remains valid,
where we replace the variance $\alpha_{**}$ and $\beta_{**}$ by the linear
variances $\alpha_L$ and $\beta_L$. Thus, Eq.(\ref{eq:Ps-k-alpha-beta}) simplifies as
\cite{Taylor:1996ne,Valageas:2010rx,Vlah:2018ygt}
\ba
&& P^s_{\rm Zel}({\bf k}) = \int \frac{d{\bf q}}{(2\pi)^3}  \; e^{i {\bf k}\cdot{\bf q}} \;
e^{-\alpha_L [ k^2 + (2f+f^2) k_z^2 ]} \nonumber \\
&& \times e^{- \beta_L [ ({\bf k}\cdot{\bf q})^2 + 2f ({\bf k}\cdot{\bf q}) k_z q_z
+ f^2 (k_z q_z)^2 ]/q^2} .
\label{eq:Ps-Zel}
\ea
For the numerical computations, we again use the method described in the
appendix~\ref{app:power-spectrum}.

\subsection{Linear power spectrum}
\label{sec:linear-power}

At linear order over the initial power spectrum $P_L$, the redshift-space power
spectrum is given by the standard Kaiser expression \cite{Kaiser:1987qv}
\be
P^s_L(k,\mu) = (1+f\mu^2)^2 P_L(k) .
\label{eq:Ps-L}
\ee
This gives the multipoles
\ba
&& P^s_{L0}(k) = \left( 1 + \frac{2f}{3} + \frac{f^2}{5} \right) P_L(k) , \nonumber \\
&& P^s_{L2}(k) = \left( \frac{4f}{3} + \frac{4f^2}{7} \right) P_L(k), \;\;
P^s_{L4}(k) = \frac{8f^2}{35} P_L(k) . \nonumber \\
&&
\label{eq:PsL-multi}
\ea

\subsection{Numerical results}
\label{sec:power-spectrum-numerical}

\begin{figure*}
\begin{center}
\epsfxsize=6.4 cm \epsfysize=7 cm {\epsfbox{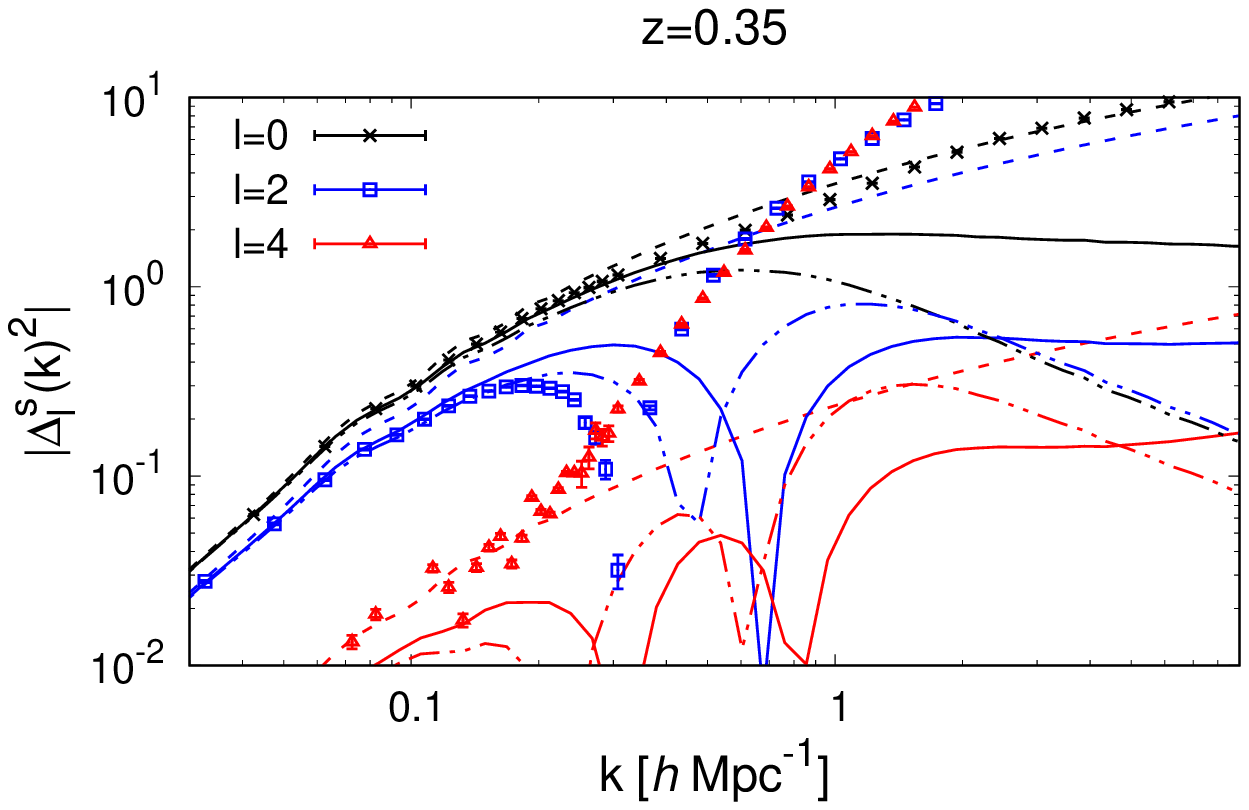}}
\epsfxsize=5.65 cm \epsfysize=7 cm {\epsfbox{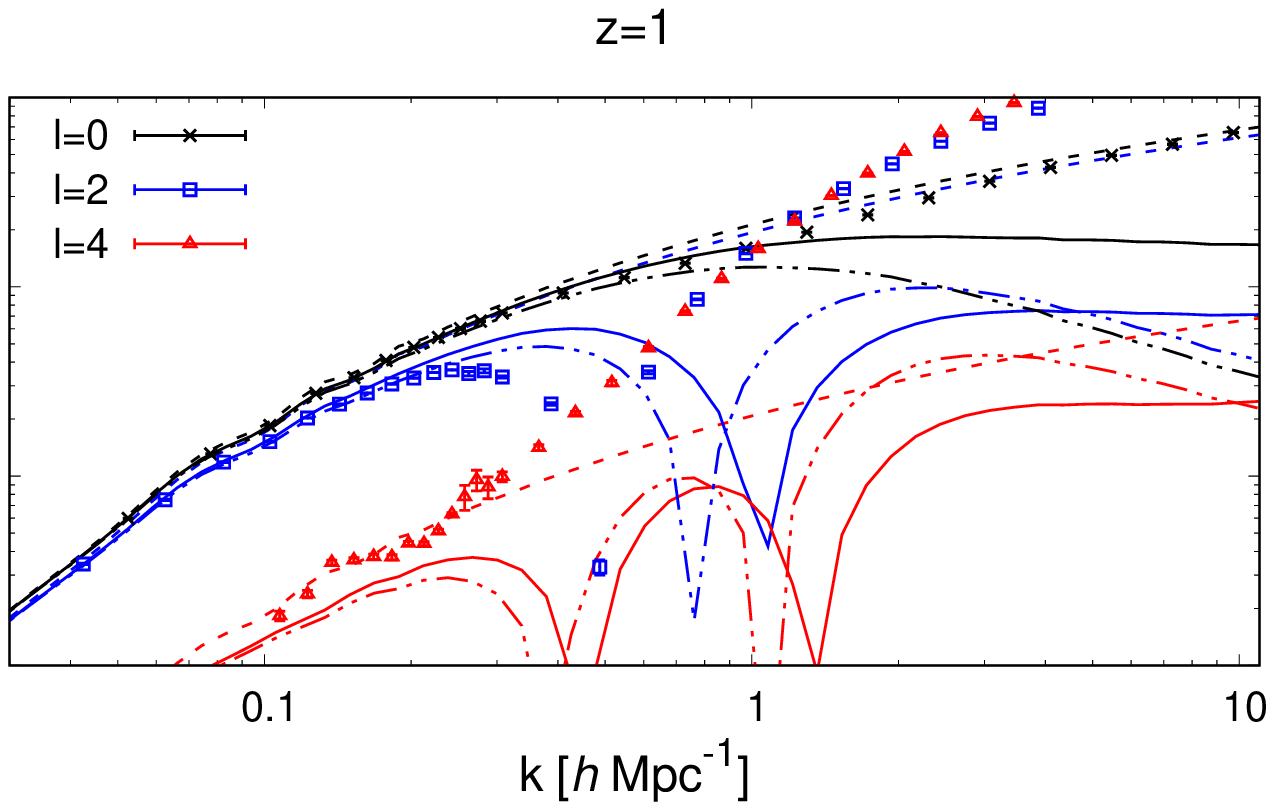}}
\epsfxsize=5.65 cm \epsfysize=7 cm {\epsfbox{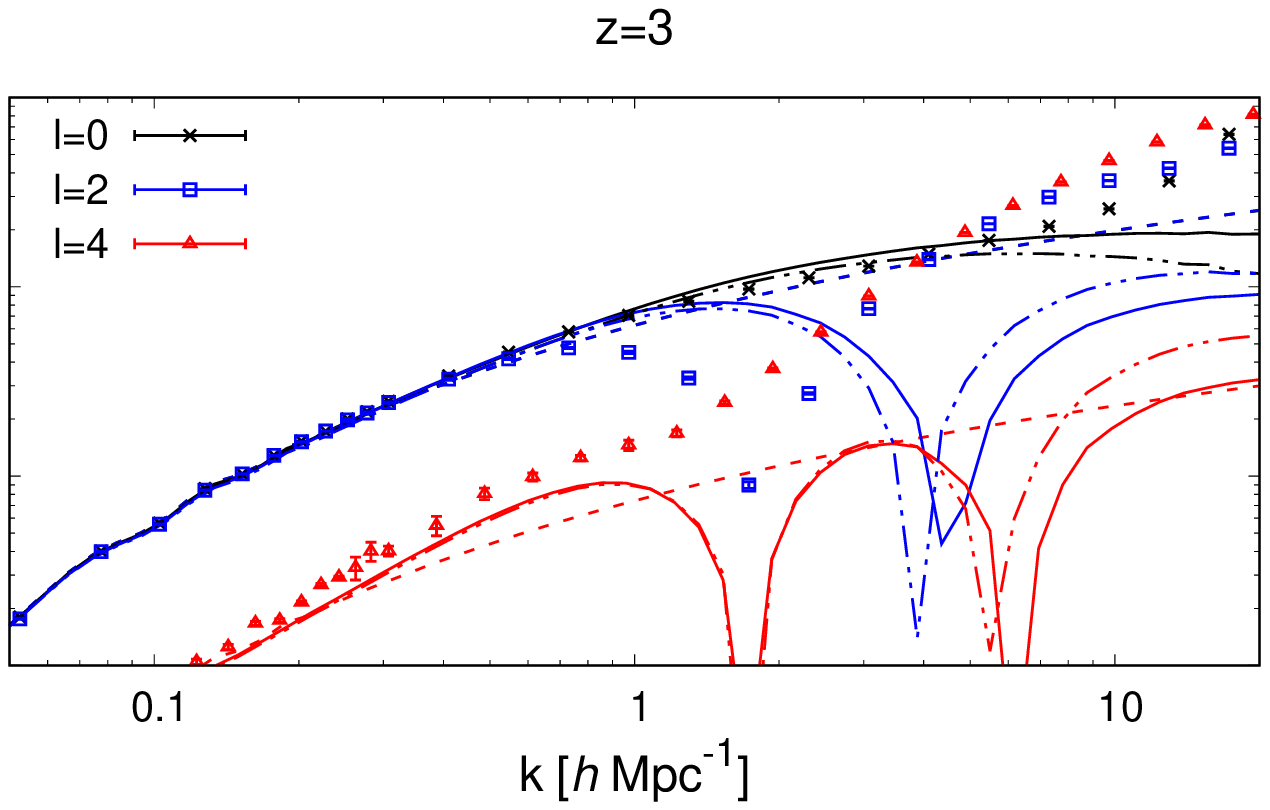}}
\end{center}
\caption{Logarithmic power spectra for multipoles $\ell=0$ (black crosses),
$\ell=2$ (blue squares) and $\ell=4$ (red triangles). We show the linear prediction ``L''
(dashed lines), our model ``Ga'' (solid lines) and the Zeldovich approximation ``Zel''
(dot-dashed lines). The symbols are the results from numerical simulations.
We show our results at redshifts $z=0.35$, $1$ and $3$.}
\label{fig_Deltak}
\end{figure*}

We show in Fig.~\ref{fig_Deltak} the logarithmic power spectrum multipoles,
$\Delta^s_{\ell}(k)^2 = 4\pi k^3 P^s_{\ell}(k)$.
We take the data points of the N-body simulations presented in \cite{Taruya:2012ut}
based on 60 random realizations of a flat $\Lambda$CDM universe consistent with the
five-year observation by the WMAP satellite (\cite{Komatsu:2009}; $\Omega_m=0.279$,
$\Omega_b/\Omega_m=0.165$, $h=0.701$, $n_s=0.96$, and $\sigma_8=0.8159$) with $1024^3$
particles performed in comoving periodic cubes with volume $(2048\,h^{-1}$Mpc$)^3$.
To study higher wavenumbers, where these simulations are not converged, we switch
to those done in \cite{Valageas:2010yw} with $2048^3$ particles in either
$(2048\,h^{-1}$Mpc$)^3$, $(1024\,h^{-1}$Mpc$)^3$ or $(512\,h^{-1}$Mpc$)^3$.
The first set of simulations cover wavenumbers up to $k=0.25\,h$Mpc$^{-1}$, where the BAO wiggles are prominent.
We compare to numerical simulations
the linear theory labelled ``L'', the Zeldovich approximation labelled ``Zel'',
and our model, labelled `Ga'' for Gaussian ansatz.
As for the real-space power spectrum, the logarithmic linear power spectra keep
increasing on nonlinear scales, the Zeldovich approximations decay and our model
predictions go to a constant.
As is well known, this is because in the Zeldovich approximation the large initial power
on small scales makes particles stream through overdensities and particles do not remain
trapped in gravitational potential wells. This erases structures on scales below the
nonlinear scale $x_{\rm NL}$, that is, at high wave numbers above $k_{\rm NL}$,
defined by $\Delta^2(k_{\rm NL}) = 1$.
In contrast, in our approach the equations of motion
(\ref{eq:dPchichi-deta})-(\ref{eq:dPthetatheta-deta}) generate a damping of the
displacement and velocity power spectra on nonlinear scales.
This arises from the fact that the force cross power spectra $P_{\chi\zeta}$ and
$P_{\theta\zeta}$, which are positive and equal to $P_L(k)$ on linear scales,
become negative on nonlinear scales for the curl-free Gaussian ansatz
(\ref{eq:chi-theta-Fourier}), as seen in \cite{Valageas:2020}. 
This effective anti-correlation is akin to a repulsive force
that stabilizes the nonlinear overdensities.
In practice, this coincides with a truncated Zeldovich approximation for the real-space
power spectrum, but with a truncation that is not set by hand and arises from the
equations of motion (\ref{eq:dPchichi-deta})-(\ref{eq:dPthetatheta-deta}).
For the redshift-space power spectrum, this goes beyond the truncated Zeldovich
approximation, as the displacement and velocity power spectra are different,
but the logarithmic power spectrum $4\pi k^3 P^s_{\ell}(k)$ again goes to a constant at high $k$.
Although this is a significant improvement over the standard Zeldovich approximation,
it cannot describe highly nonlinear scales associated with virialized halos,
where the true logarithmic power spectrum keeps growing.

Both the Zeldovich approximation and our Gaussian ansatz recover the change of sign of the
quadrupole $P^s_2(k)$ near the nonlinear transition, although they do not predict
its location with a good accuracy.
This is already a significant improvement over the linear theory, which does not
change sign, and it shows that this feature is associated with the mildly nonlinear stages
of the formation of large-scale structures.
In contrast, these two approximations predict two successive changes of sign of the
hexadecapole $P^s_4(k)$ while the numerical simulations do not show any change
of sign. This is another illustration of the well-known fact that the hexadecapole is
much more difficult to model and is sensitive to the details of the nonlinear dynamics.
In this case, the Zeldovich approximation and our Gaussian ansatz only give a significant
improvement over linear theory at high redshift, $z \gtrsim 3$.

\begin{figure*}
\begin{center}
\epsfxsize=6.4 cm \epsfysize=4 cm {\epsfbox{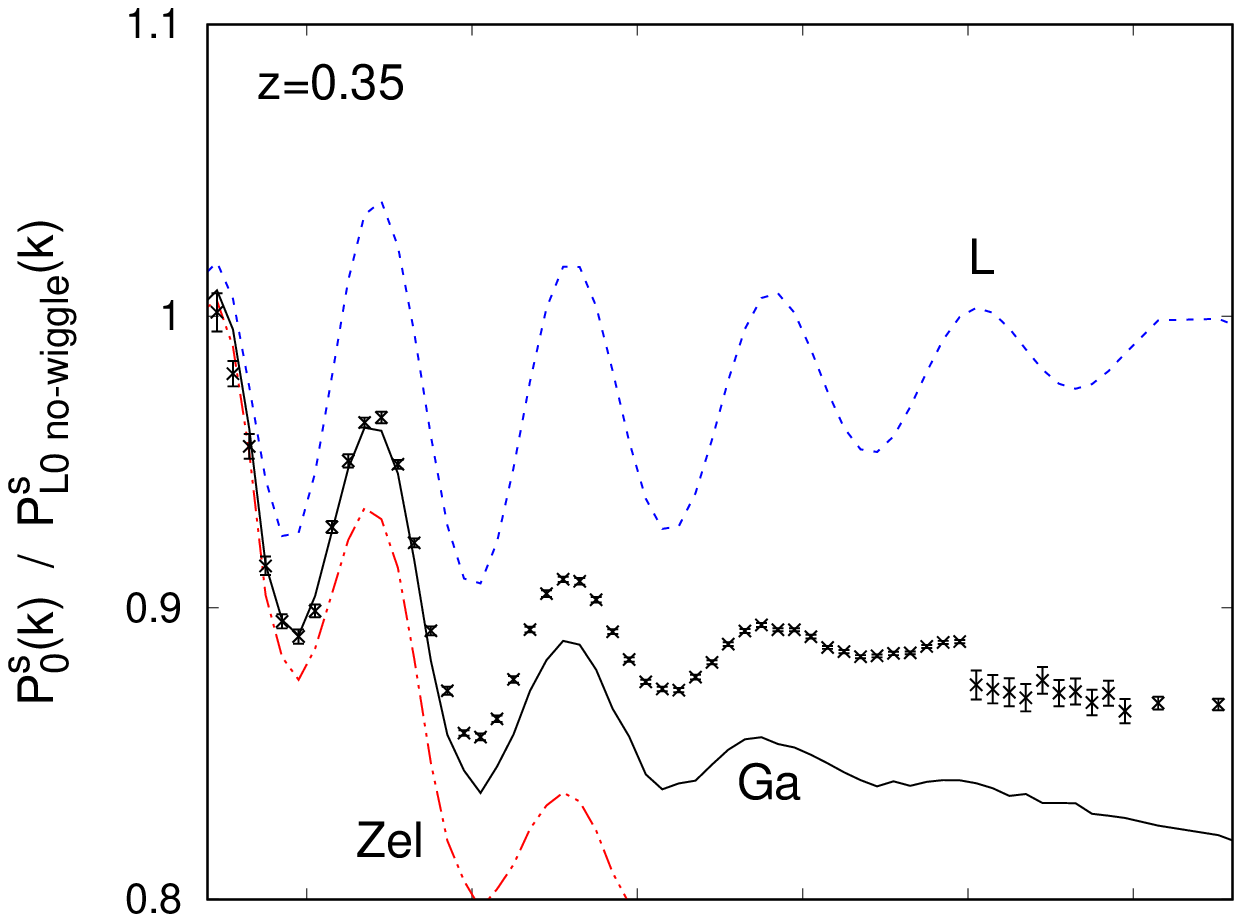}}
\epsfxsize=5.65 cm \epsfysize=4 cm {\epsfbox{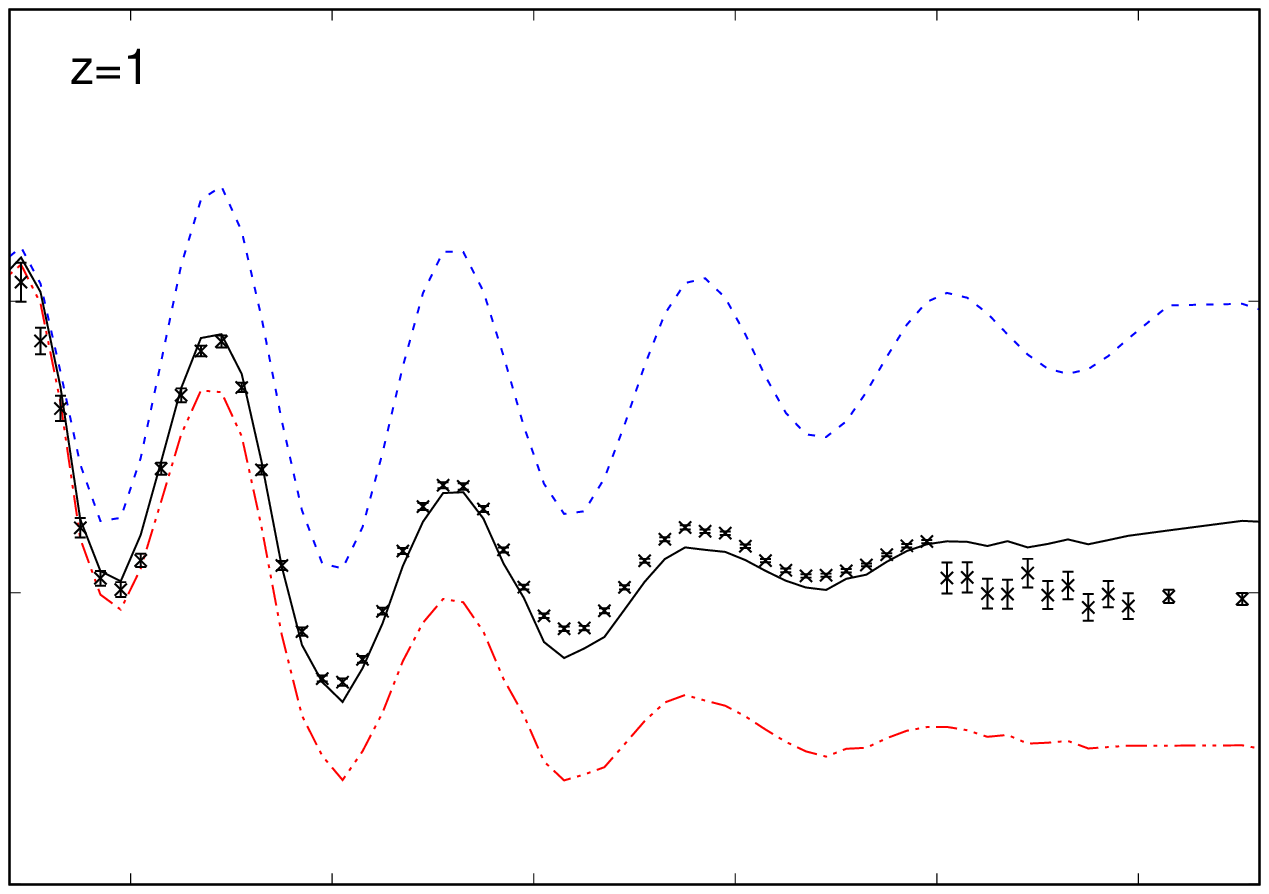}}
\epsfxsize=5.65 cm \epsfysize=4 cm {\epsfbox{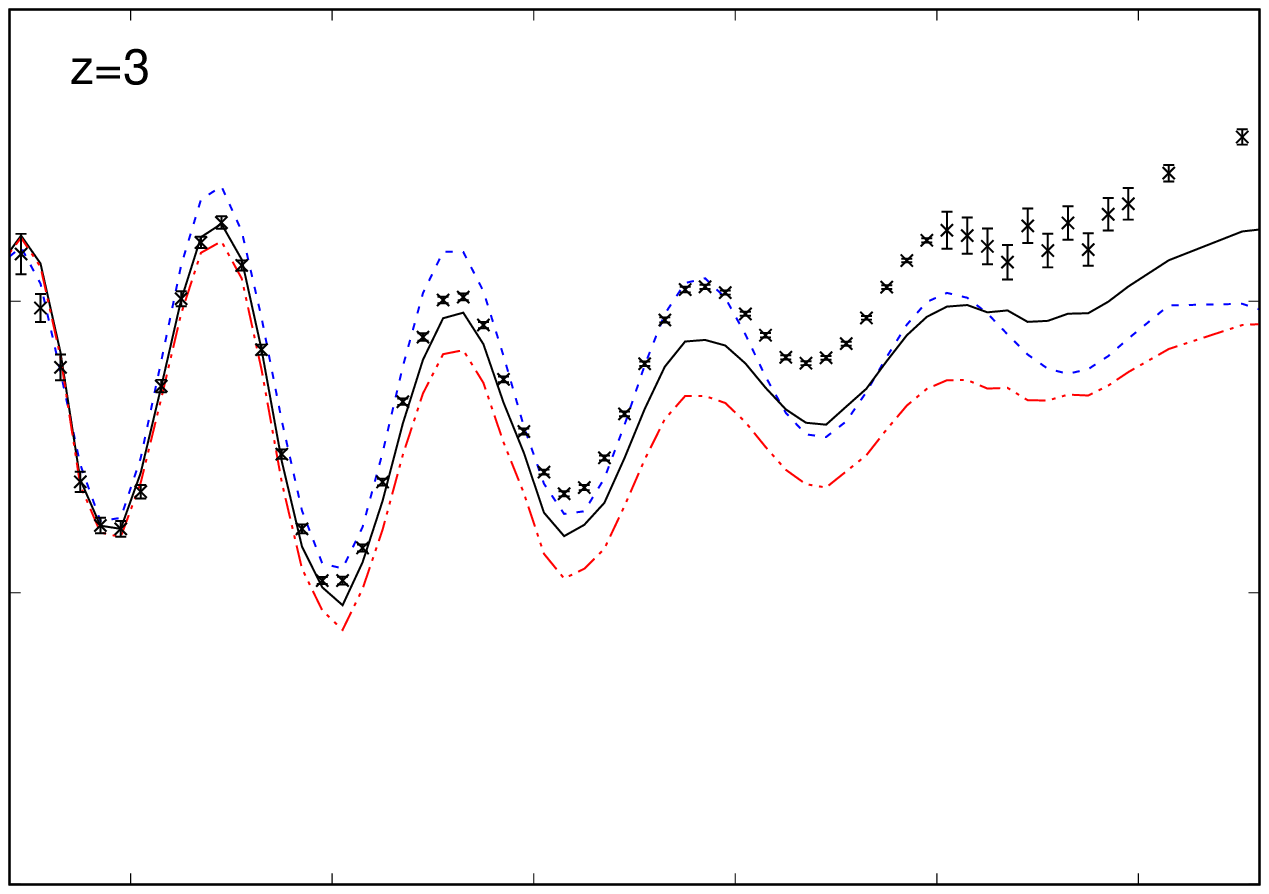}}\\
\epsfxsize=6.4 cm \epsfysize=5 cm {\epsfbox{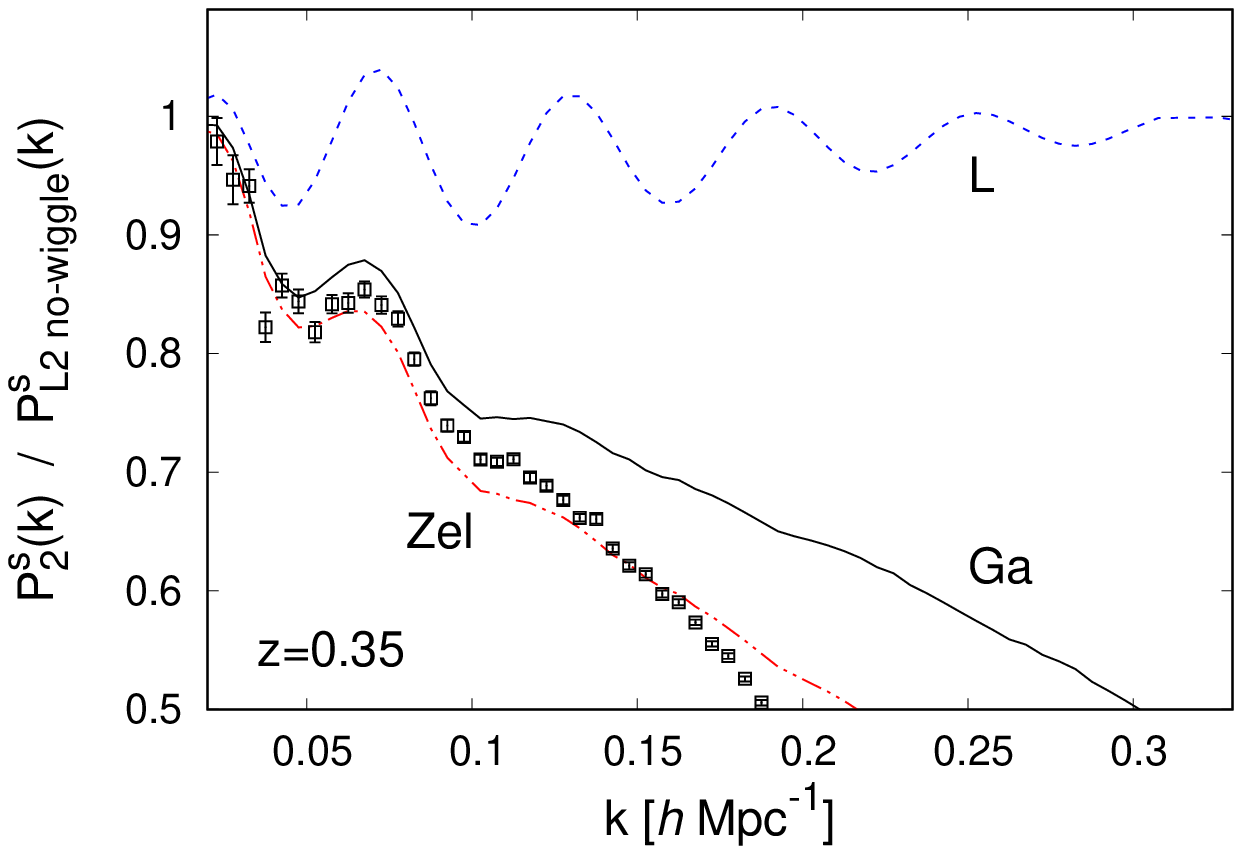}}
\epsfxsize=5.65 cm \epsfysize=5 cm {\epsfbox{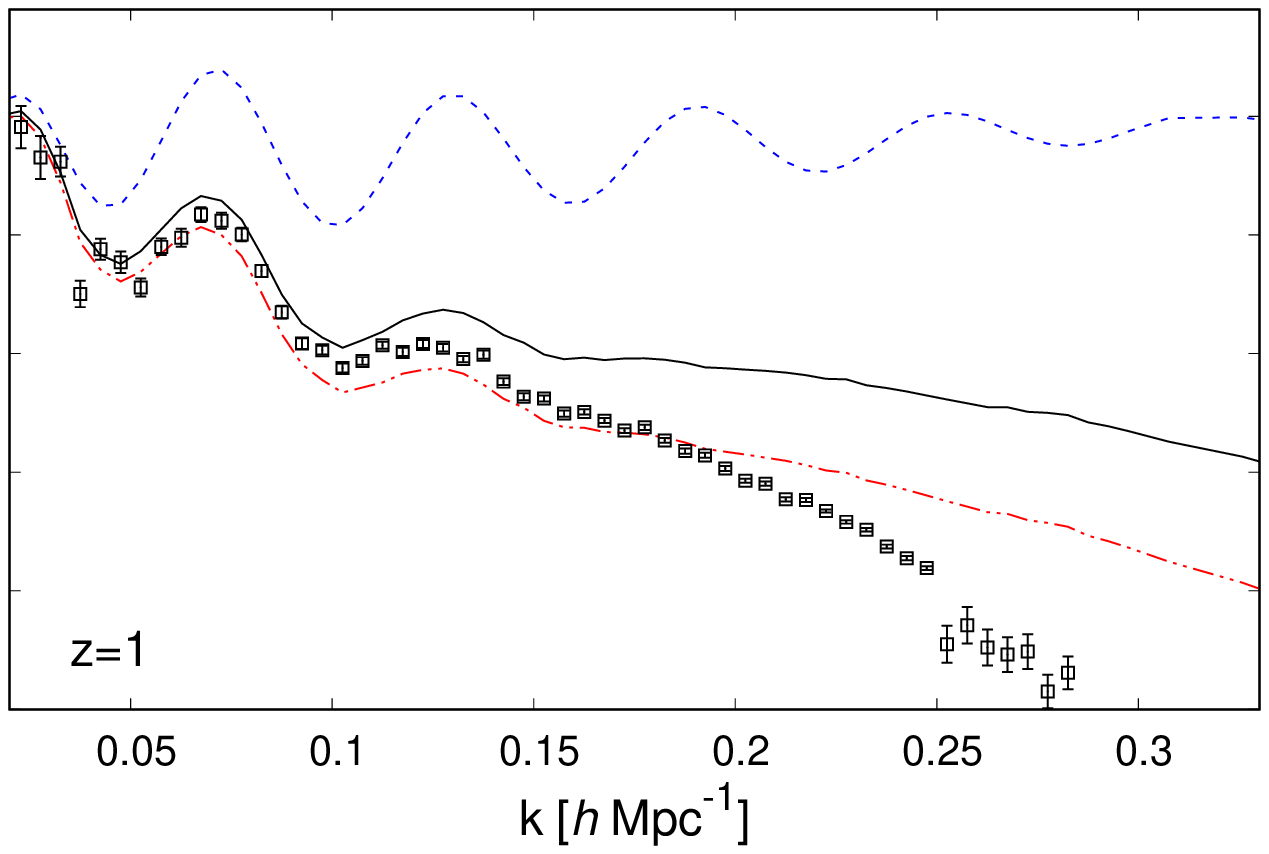}}
\epsfxsize=5.65 cm \epsfysize=5 cm {\epsfbox{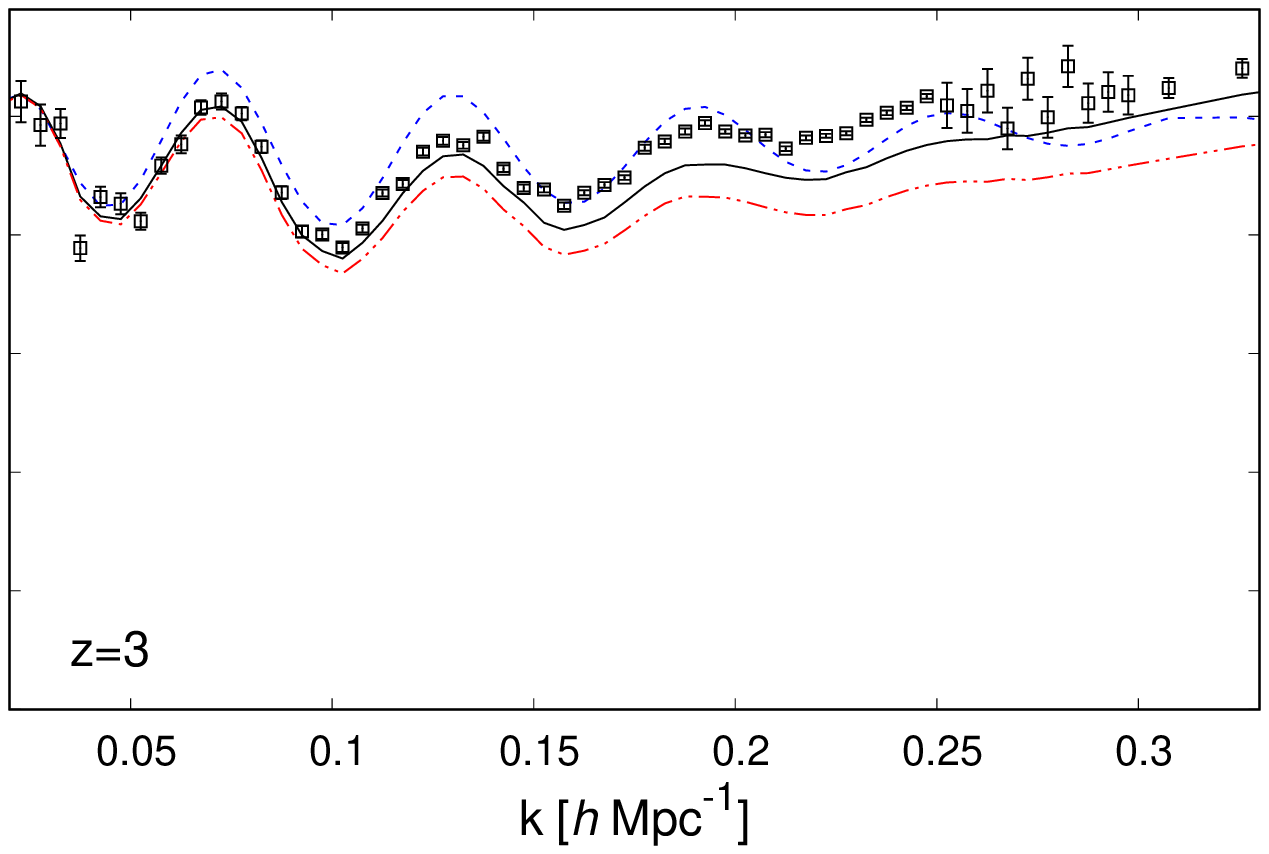}}
\end{center}
\caption{{\it Upper panels:} ratio of the power spectrum monopole $P^s_0(k)$ to
a reference linear power spectrum monopole without baryonic oscillations.
We show the linear prediction ``L'' (blue dashed lines), our model ``Ga'' (black solid lines)
and the Zeldovich approximation ``Zel'' (red dot-dashed lines), at redshifts $z=0.35$, $1$
and $3$.
{\it Lower panels:} ratio of the power spectrum quadrupole $P^s_2(k)$ to
a reference linear power spectrum quadrupole without baryonic oscillations.}
\label{fig_rPk}
\end{figure*}

\begin{figure*}
\begin{center}
\epsfxsize=6.4 cm \epsfysize=4 cm {\epsfbox{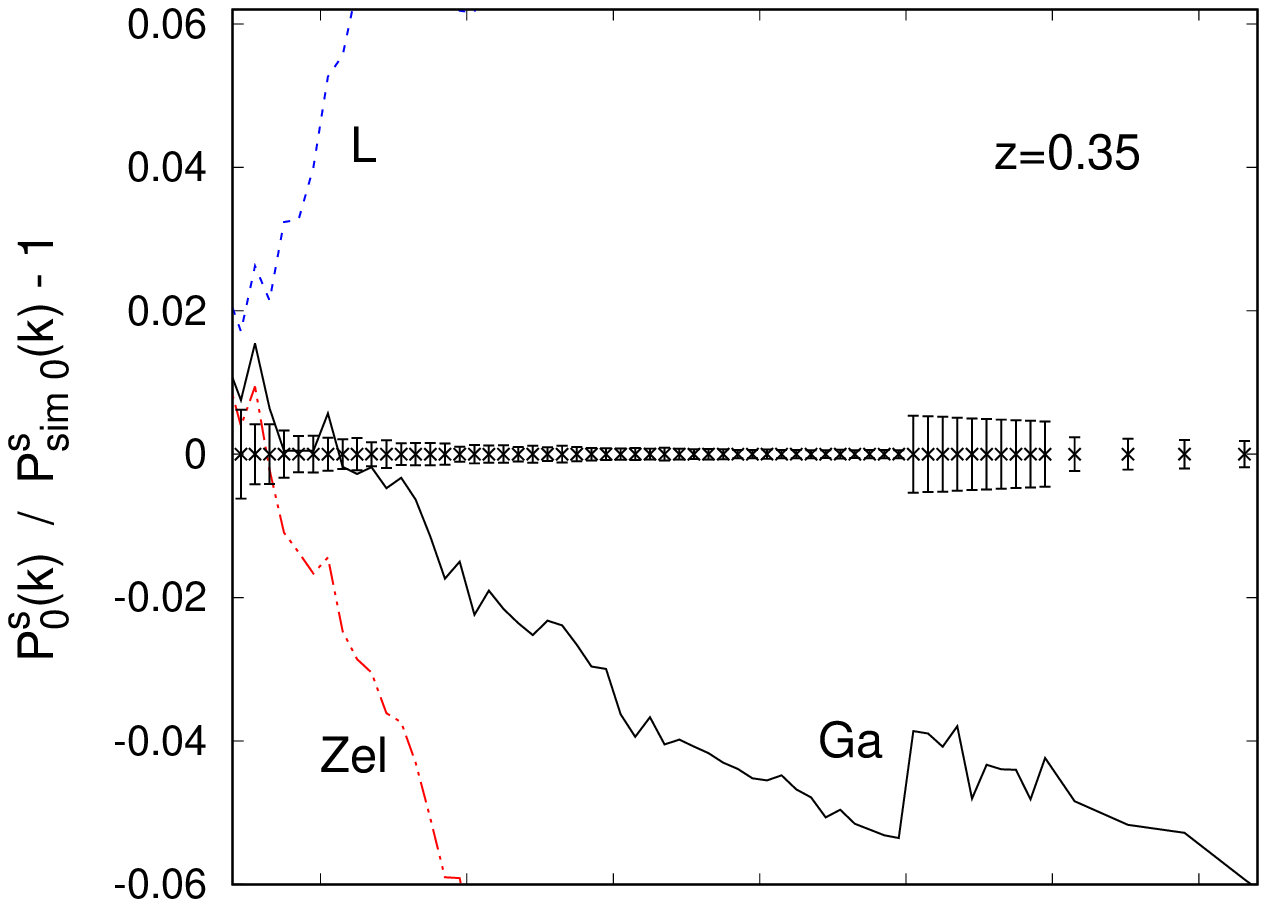}}
\epsfxsize=5.65 cm \epsfysize=4 cm {\epsfbox{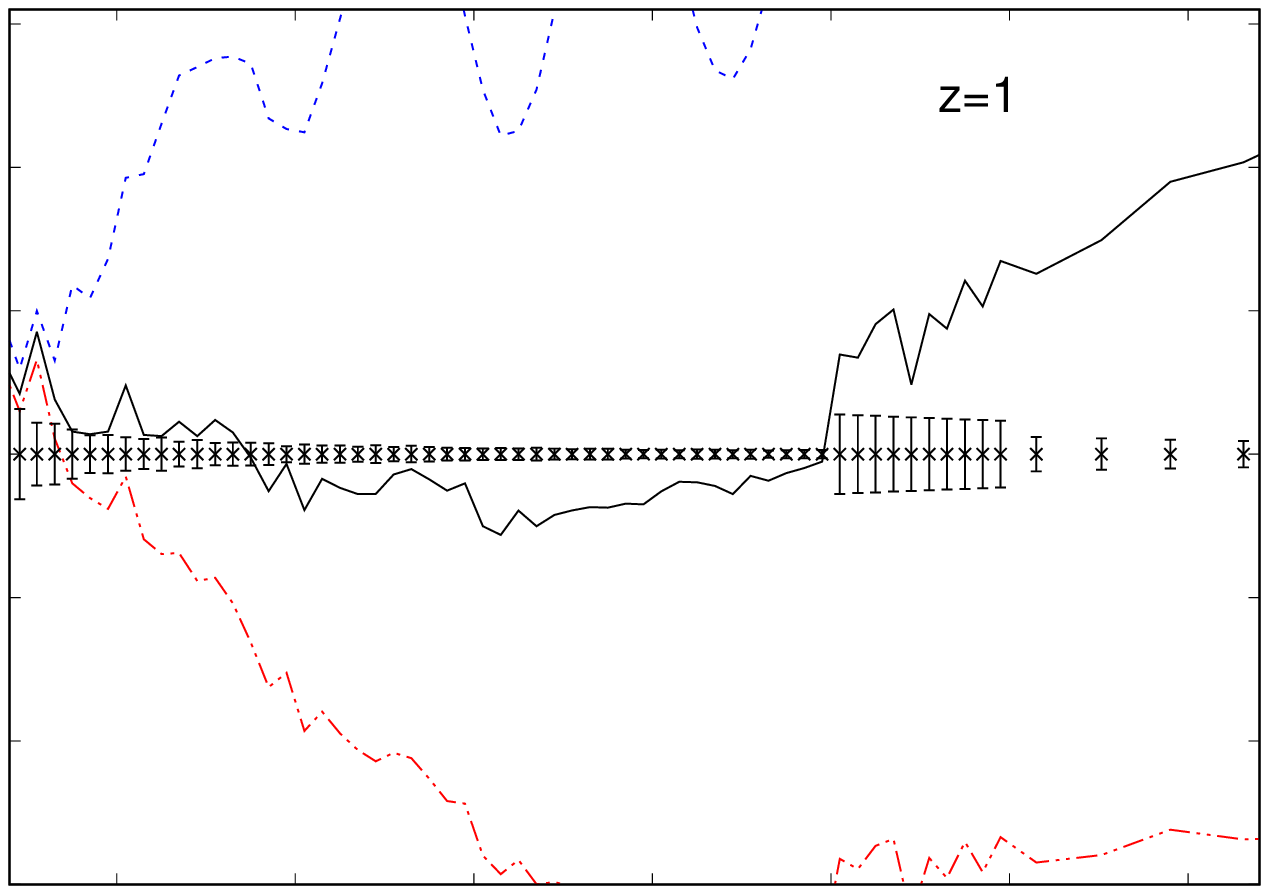}}
\epsfxsize=5.65 cm \epsfysize=4 cm {\epsfbox{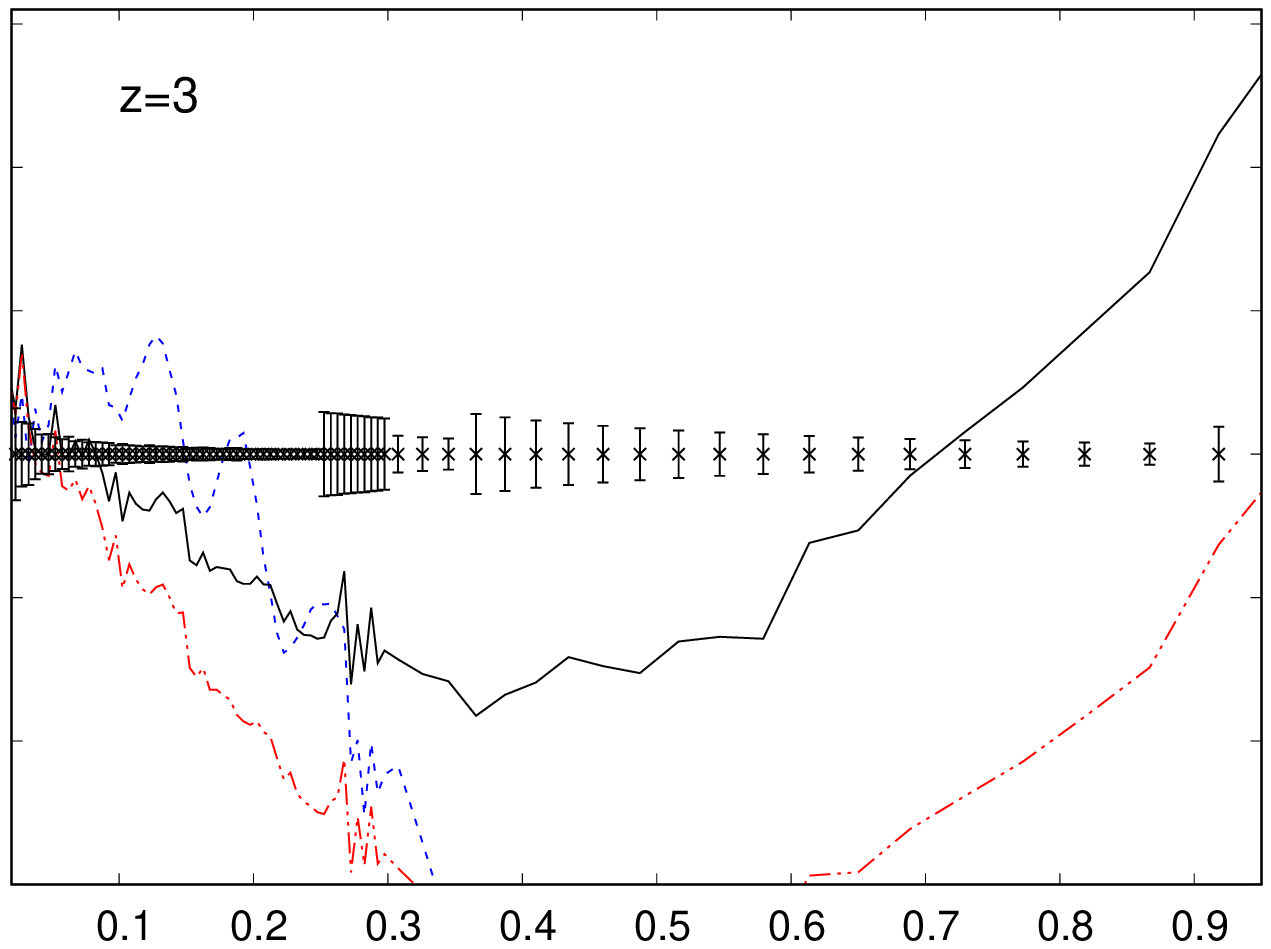}}\\
\epsfxsize=6.4 cm \epsfysize=5 cm {\epsfbox{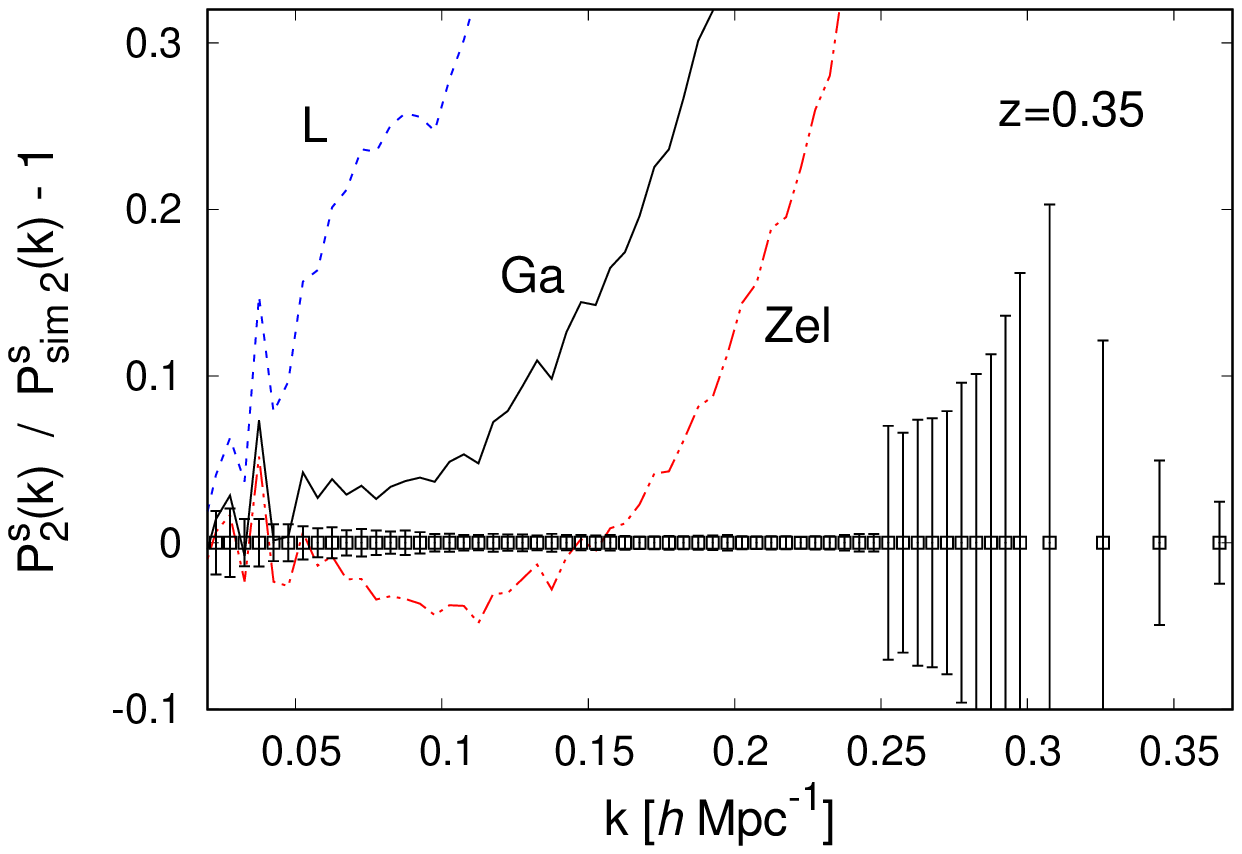}}
\epsfxsize=5.65 cm \epsfysize=5 cm {\epsfbox{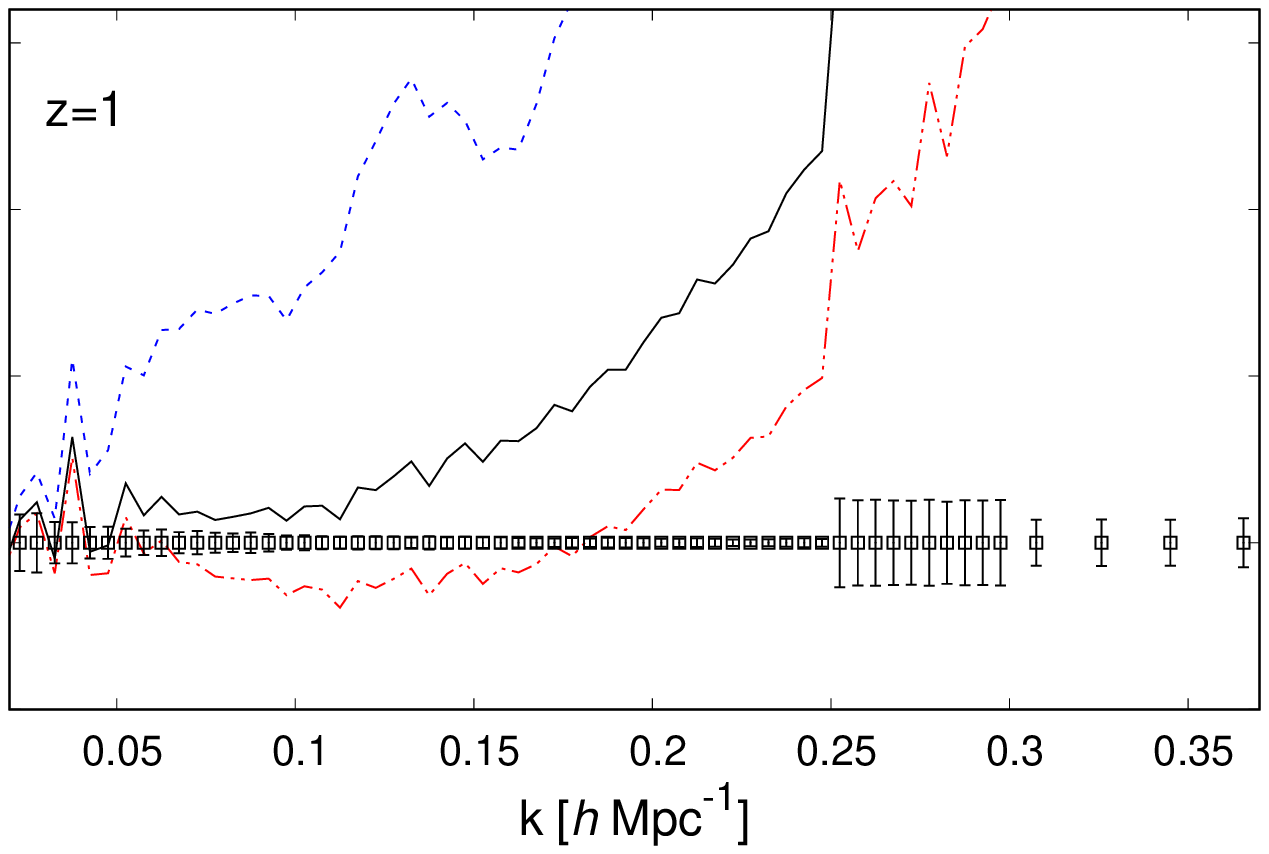}}
\epsfxsize=5.65 cm \epsfysize=5 cm {\epsfbox{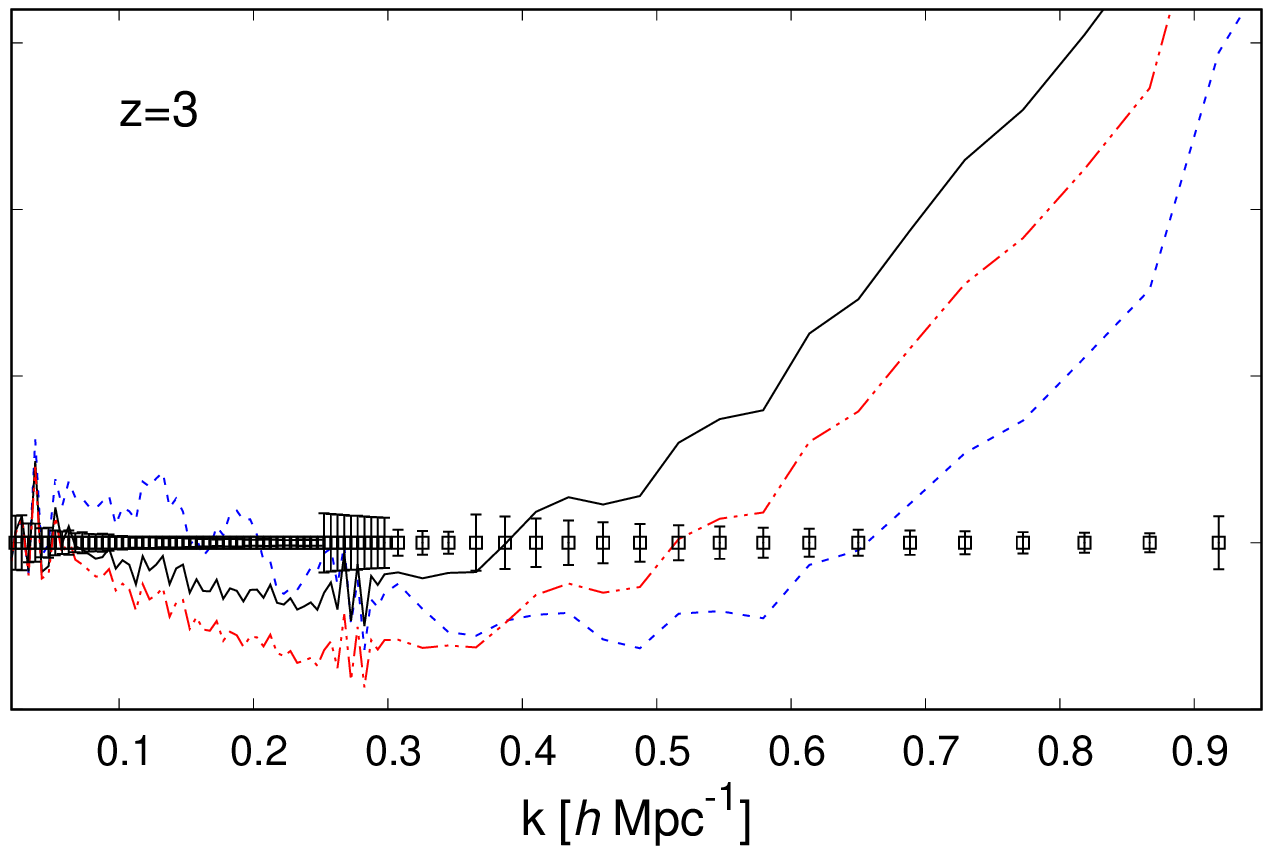}}
\end{center}
\caption{Relative deviation from the numerical simulations of the analytical predictions
for the redshift-space power spectrum. We show the multipoles $\ell=0$ (upper row)
and $\ell=2$ (lower row).}
\label{fig_dPk}
\end{figure*}

We zoom on the BAO scales in Fig.~\ref{fig_rPk}. We show the ratio of the numerical
simulations, the linear theory, the Zeldovich approximation and the Gaussian ansatz
with respect to the mulipoles (\ref{eq:PsL-multi}) of a wiggle-free linear power spectrum.
Because we saw in Fig.~\ref{fig_Deltak} that the models do not perform very well for
the hexadecapole, we focus on the monopole and quadrupole.

As for the real-space power spectrum, we find that the nonlinear damping of the baryon
acoustic oscillations is well recovered but there is a smooth drift with respect to the
numerical simulations. As we shall see in Sec.~\ref{sec:correlation}, the accuracy is much
greater for the configuration-space correlation function.
This is because the Lagrangian framework, common to both the Zeldovich approximation
and our Gaussian ansatz, is better suited to configuration-space statistics.
This can be seen from the fact that the fundamental objects are the
configuration-space displacement and velocity fields, as in (\ref{eq:Ps-sq}).
More generally, in contrast with the linearized dynamics, where Fourier modes are decoupled,
nonlinear processes that are local in configuration space, such as the trapping of particles
inside collapsed halos, should be easier to describe in configuration space, where
they should generate weak correlations across scales
\cite{Valageas:2013hxa,Tassev:2013rta}.
Then, even if the configuration-space correlation is well described except on small-scales,
the power spectrum it defines by a Fourier transform can show large deviations
from the exact results down to low $k$. For instance, adding a localized Dirac term
$\delta_D({\bf x})$ to the correlation $\xi^s({\bf x})$ gives a constant shot-noise contribution
to $P^s({\bf k})$ that will even dominate for $k \to 0$.

This behavior is common to both the monopole and quadrupole. It means that
the Zeldovich approximation and the Gaussian ansatz are not competitive with other
models for the power spectrum, which reach a better agreement with simulations
\cite{delaBella:2017qjy}.
However, if we are able to extract the oscillatory feature
of the power spectra, or if we add a few free parameters that describe the smooth drift
of the power spectra, they may fare as well as other approaches.
Moreover, because there are no free parameters to marginalize over (unless one adds
these background additional ingredients), the constraining power may compete with more
accurate methods that involve several free parameters.
We will investigate this point in future works.

We show the relative deviation of these power spectra from the numerical simulations
in Fig.~\ref{fig_dPk}. We clearly see for the monopole the improvement of the
Gaussian ansatz over both the linear prediction and the Zeldovich approximation.
This is not surprising. As compared with the Zeldovich approximation,
the Gaussian ansatz satisfies the additional constraints
(\ref{eq:dPchichi-deta})-(\ref{eq:dPthetatheta-deta}). It is then expected to give
a more realistic description of the dynamics.
For the quadrupole, the Zeldovich approximation fares better at $z=0.35$ and $z=1$.
However, because of the worse agreement at $z=3$ and for the monopole at all
redshifts, this is likely to be a coincidence.

At $z=0.35$, for the monopole we obtain an accuracy of about $5\%$ up to
$0.3 h {\rm Mpc}^{-1}$, and for the quadrupole of $25\%$ up to
$0.18 h {\rm Mpc}^{-1}$.
For comparison, we note that the Lagrangian approach of \cite{Matsubara:2007wj}
obtains at $z=0.3$ an accuracy of $5\%$ up to $0.11 h {\rm Mpc}^{-1}$ for $P^s_0$.
The TNS model of \cite{Taruya:2010mx}, which combines SPT with a damping prefactor
fitted to simulations, gives at $z=0$ an accuracy of $5\%$ up to
$0.3 h {\rm Mpc}^{-1}$ for $P^s_0$, and of $10\%$ up to $0.23 h {\rm Mpc}^{-1}$
for $P^s_2$ \cite{GilMarin:2012nb}.
Adding a partial resummation of Eulerian perturbation theory to this approach
\cite{Taruya:2013my} gives at $z=0$ an accuracy of $5\%$ up to
$0.24 h {\rm Mpc}^{-1}$ for $P^s_0$, and of $10\%$ up to $0.23 h {\rm Mpc}^{-1}$ for $P^s_2$.
Using an EFT approach, \cite{delaBella:2017qjy} obtains at $z=0$
an accuracy of $5\%$ up to $0.4 h {\rm Mpc}^{-1}$ for $P^s_0$, and of $25\%$
up to $0.4 h {\rm Mpc}^{-1}$ for $P^s_2$; while
\cite{Lewandowski:2015ziq} obtain at $z=0.56$
an accuracy of $5\%$ up to $0.24 h {\rm Mpc}^{-1}$ for $P^s_0$, and of $10\%$ up to
$0.20 h {\rm Mpc}^{-1}$ for $P^s_2$, with five parameters fitted to simulations.
The ``time-sliced perturbation theory'' approach of \cite{Ivanov:2018gjr} gives at $z=0$
an accuracy of $5\%$ up to $0.1 h {\rm Mpc}^{-1}$ for $P^s_0$, and of $25\%$ up to
$0.1 h {\rm Mpc}^{-1}$ for $P^s_2$.
Thus, the Lagrangian-space Gaussian ansatz studied in this paper gives an accuracy
that falls in between these various methods, but is significantly below that reached by the
most efficient schemes like the EFT study \cite{delaBella:2017qjy}.
This is not so surprising, as our model is only correct up to linear order over $P_L$.
To go to higher orders, one needs to go beyond the Gaussian and include higher-order
correlations, which will be governed by additional constraints similar to
Eqs.(\ref{eq:dPchichi-deta})-(\ref{eq:dPthetatheta-deta}), again derived from the equation of
motion (\ref{eq:Psi-eom}).

\section{Redshift-space matter density correlation function}
\label{sec:correlation}

\begin{figure*}
\begin{center}
\epsfxsize=6.4 cm \epsfysize=4.6 cm {\epsfbox{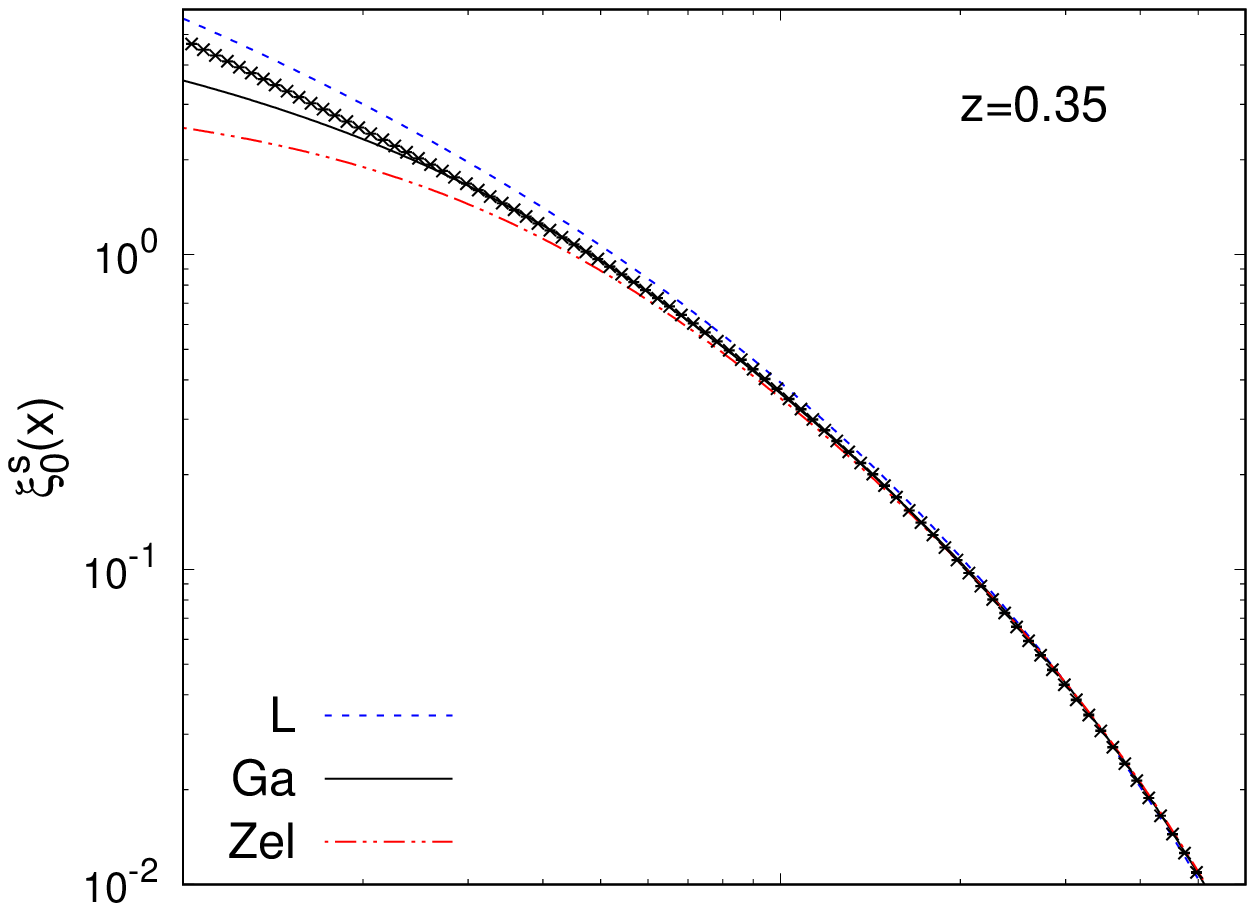}}
\epsfxsize=5.65 cm \epsfysize=4.6 cm {\epsfbox{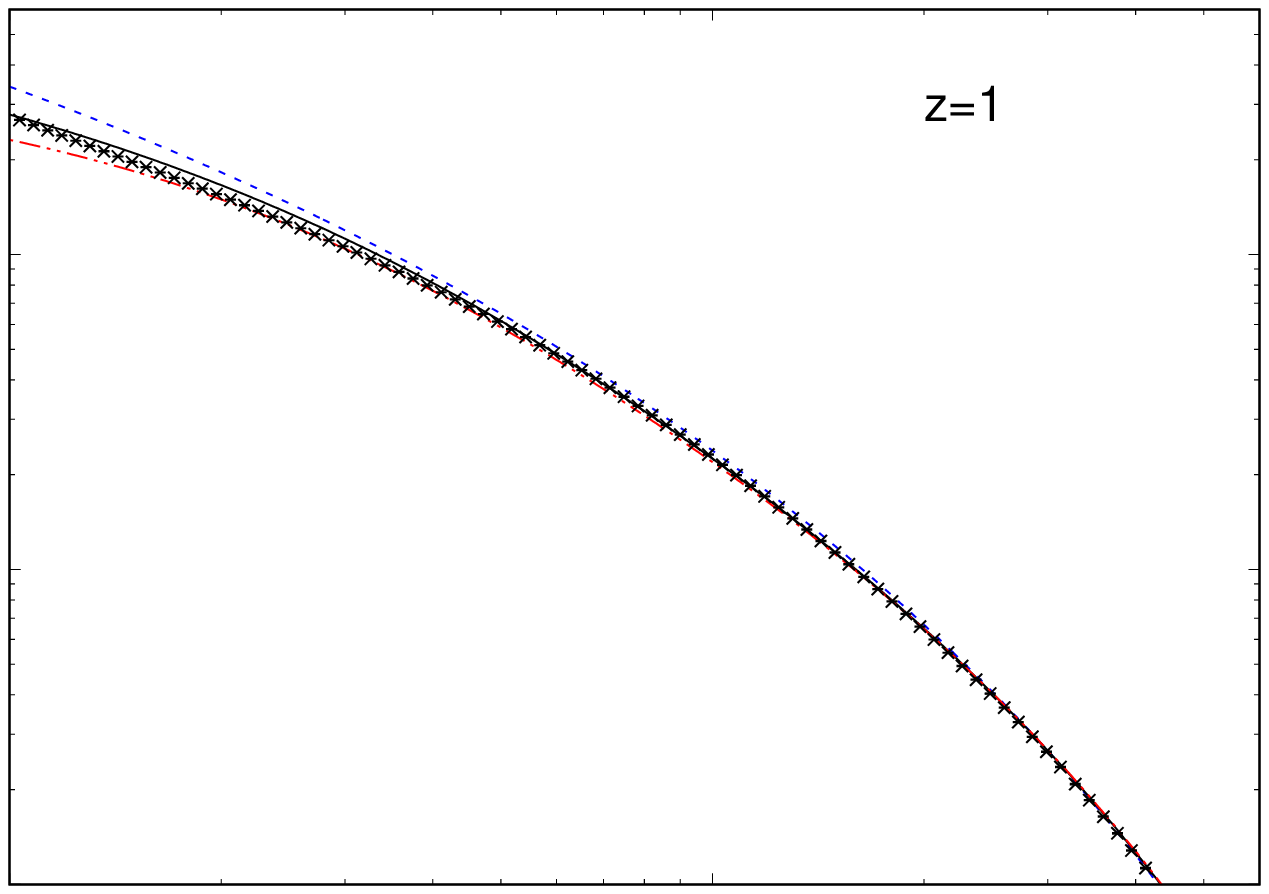}}
\epsfxsize=5.65 cm \epsfysize=4.6 cm {\epsfbox{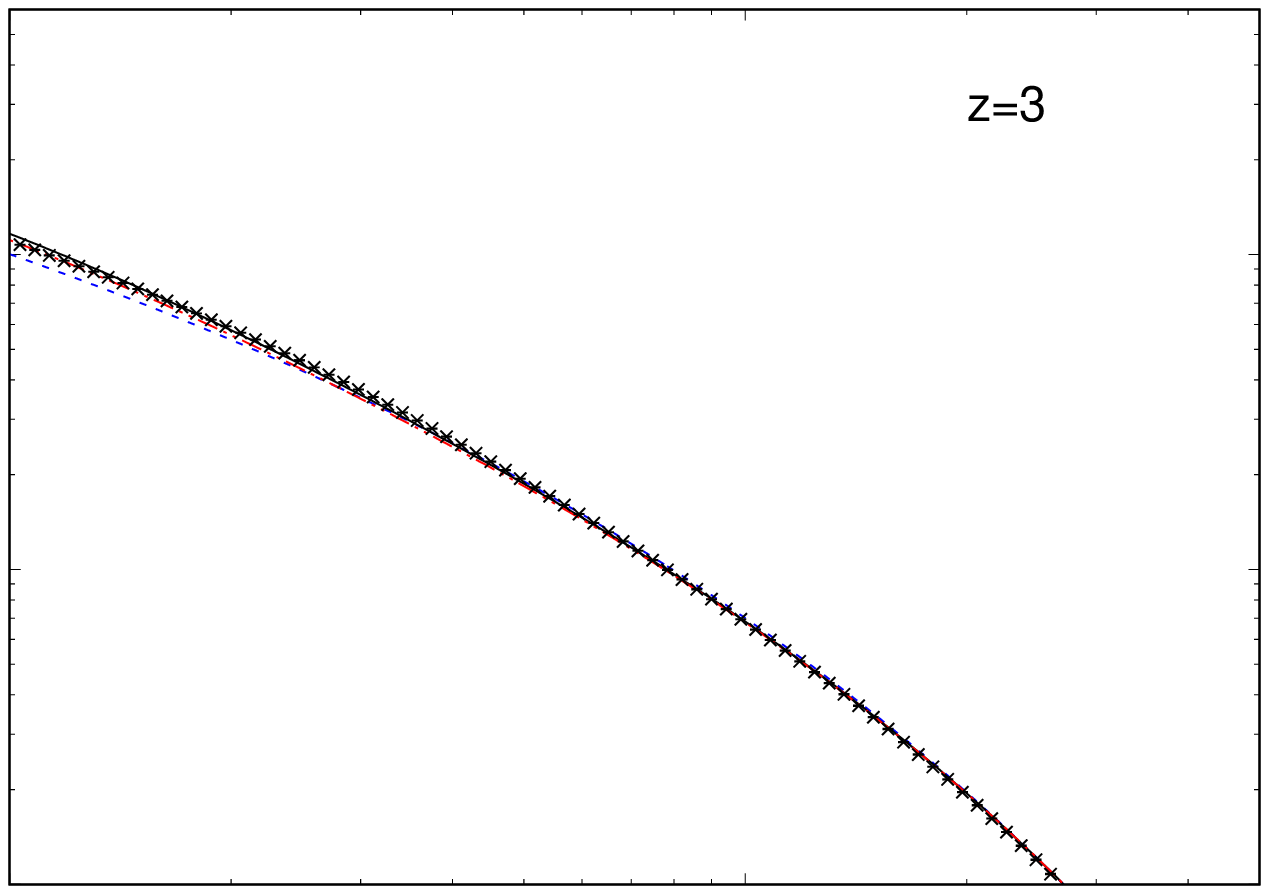}}\\
\epsfxsize=6.4 cm \epsfysize=4.6 cm {\epsfbox{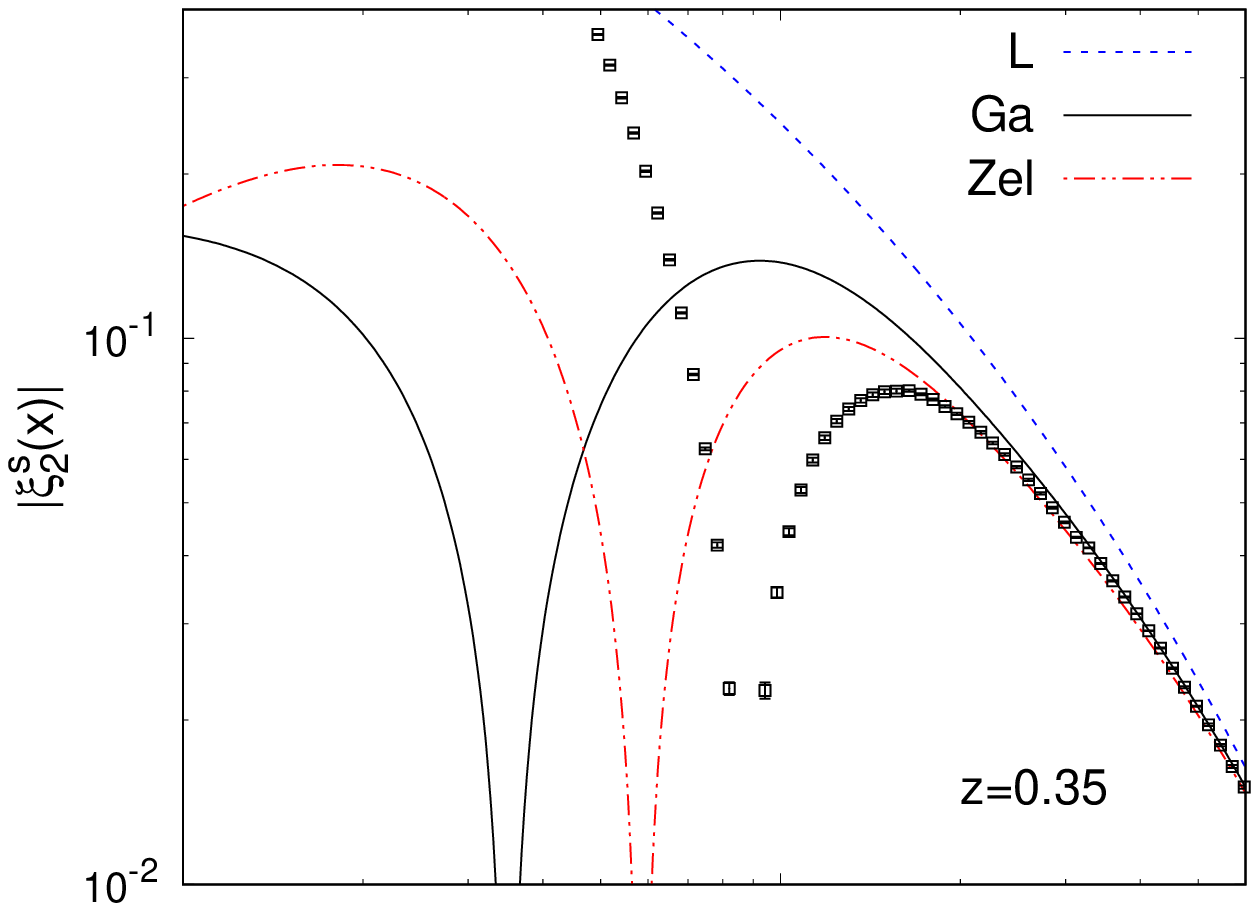}}
\epsfxsize=5.65 cm \epsfysize=4.6 cm {\epsfbox{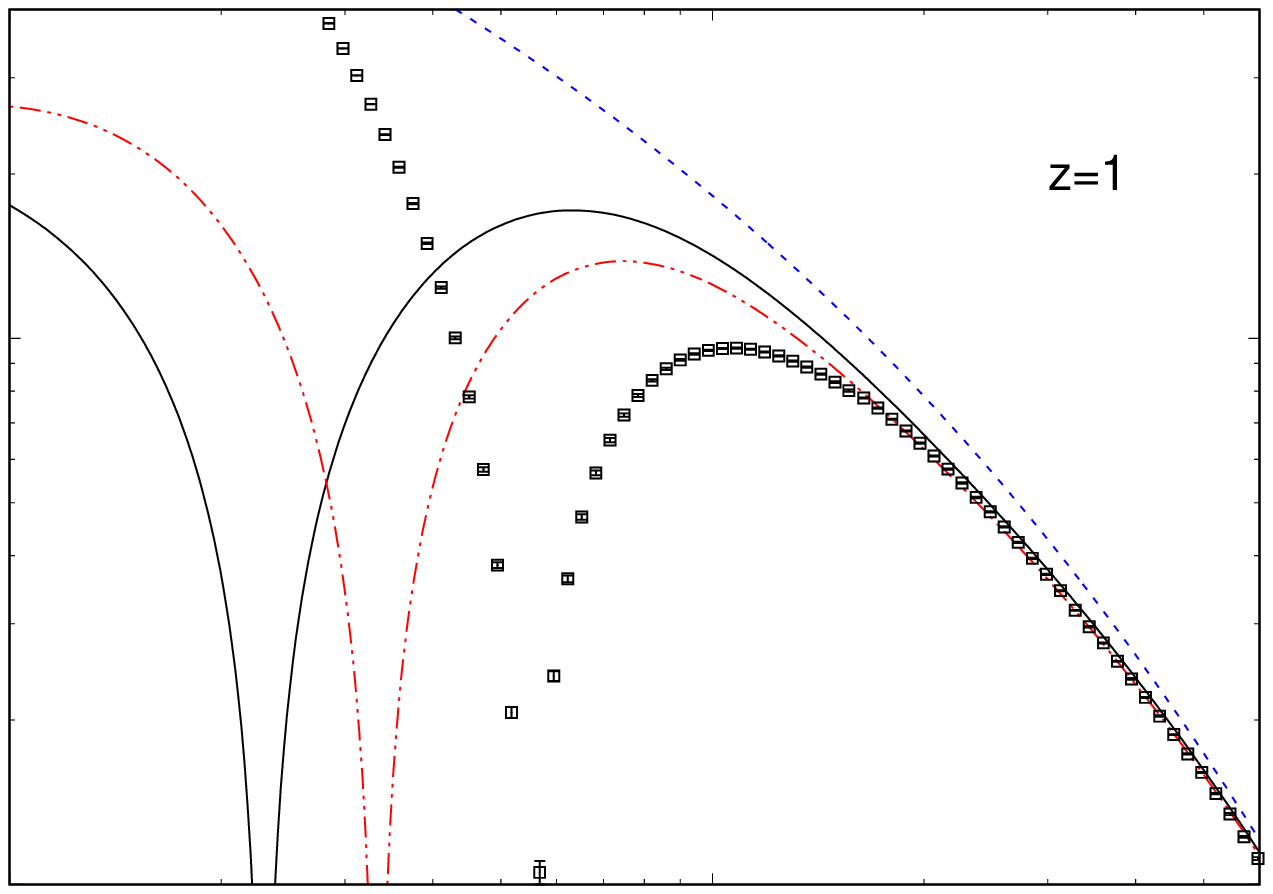}}
\epsfxsize=5.65 cm \epsfysize=4.6 cm {\epsfbox{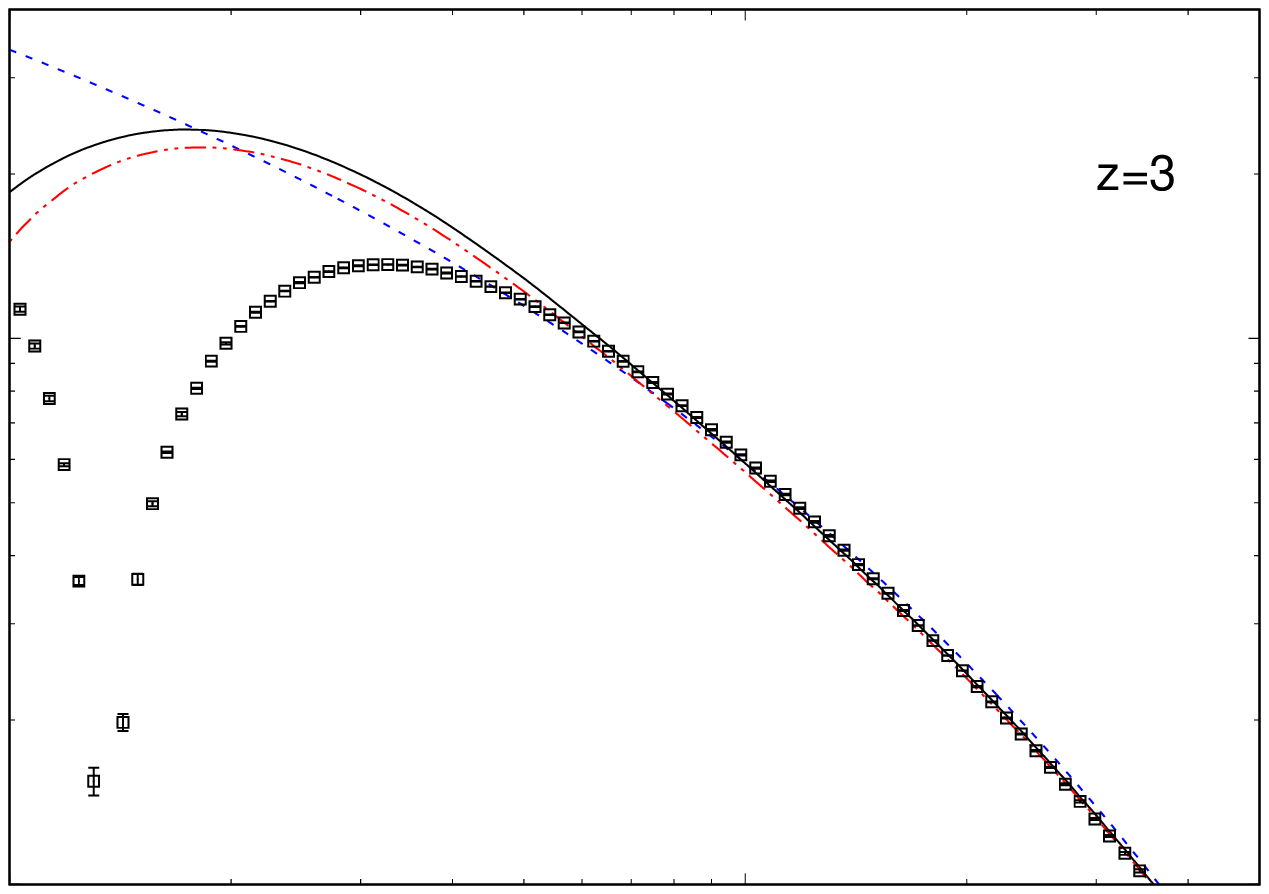}}\\
\epsfxsize=6.4 cm \epsfysize=5.6 cm {\epsfbox{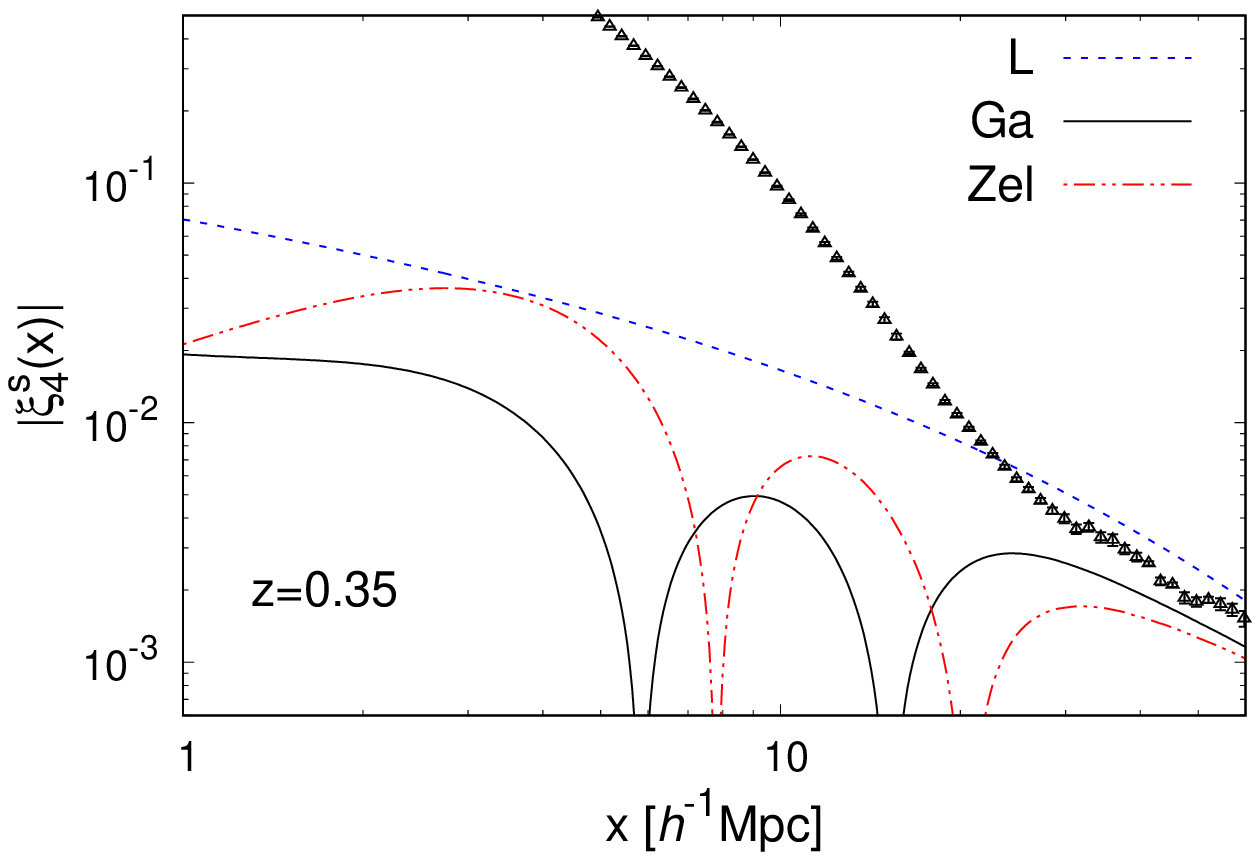}}
\epsfxsize=5.65 cm \epsfysize=5.6 cm {\epsfbox{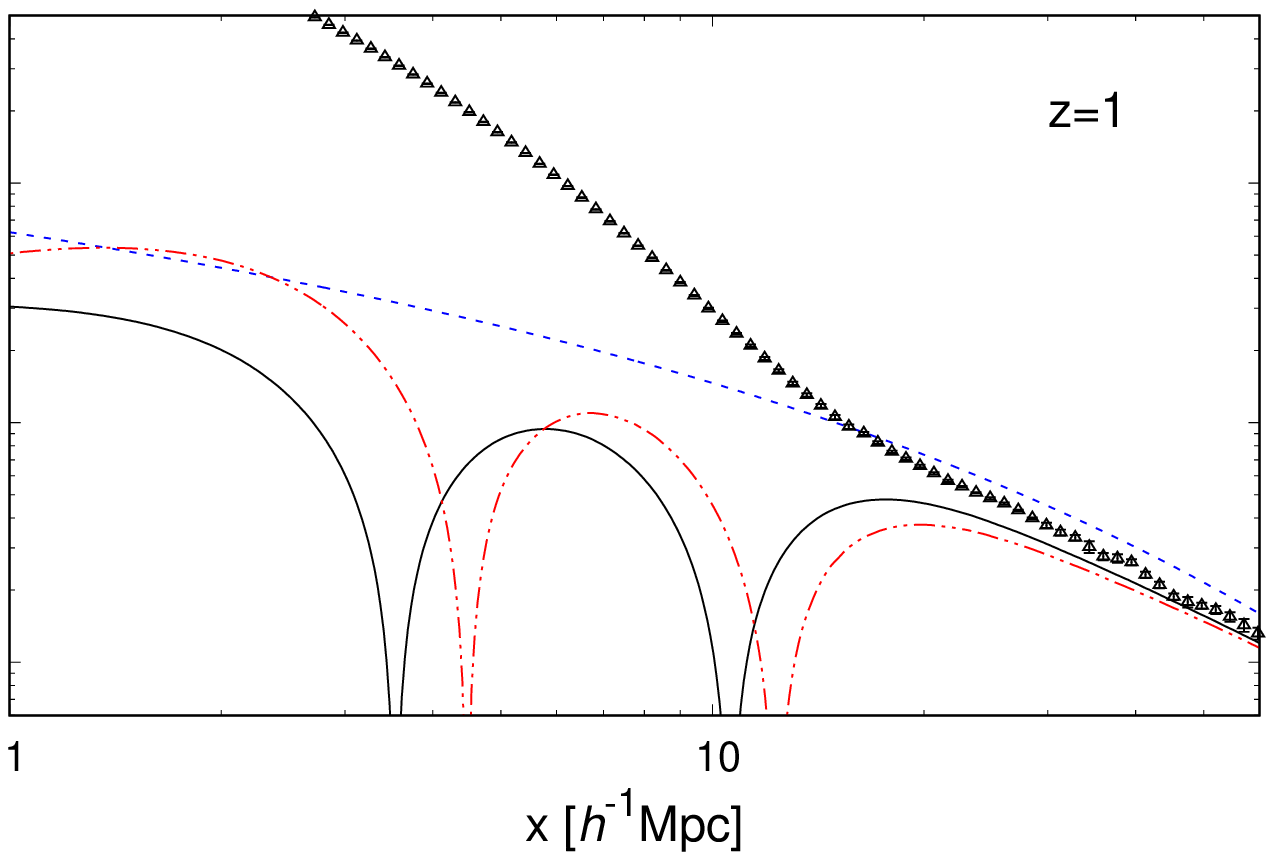}}
\epsfxsize=5.65 cm \epsfysize=5.6 cm {\epsfbox{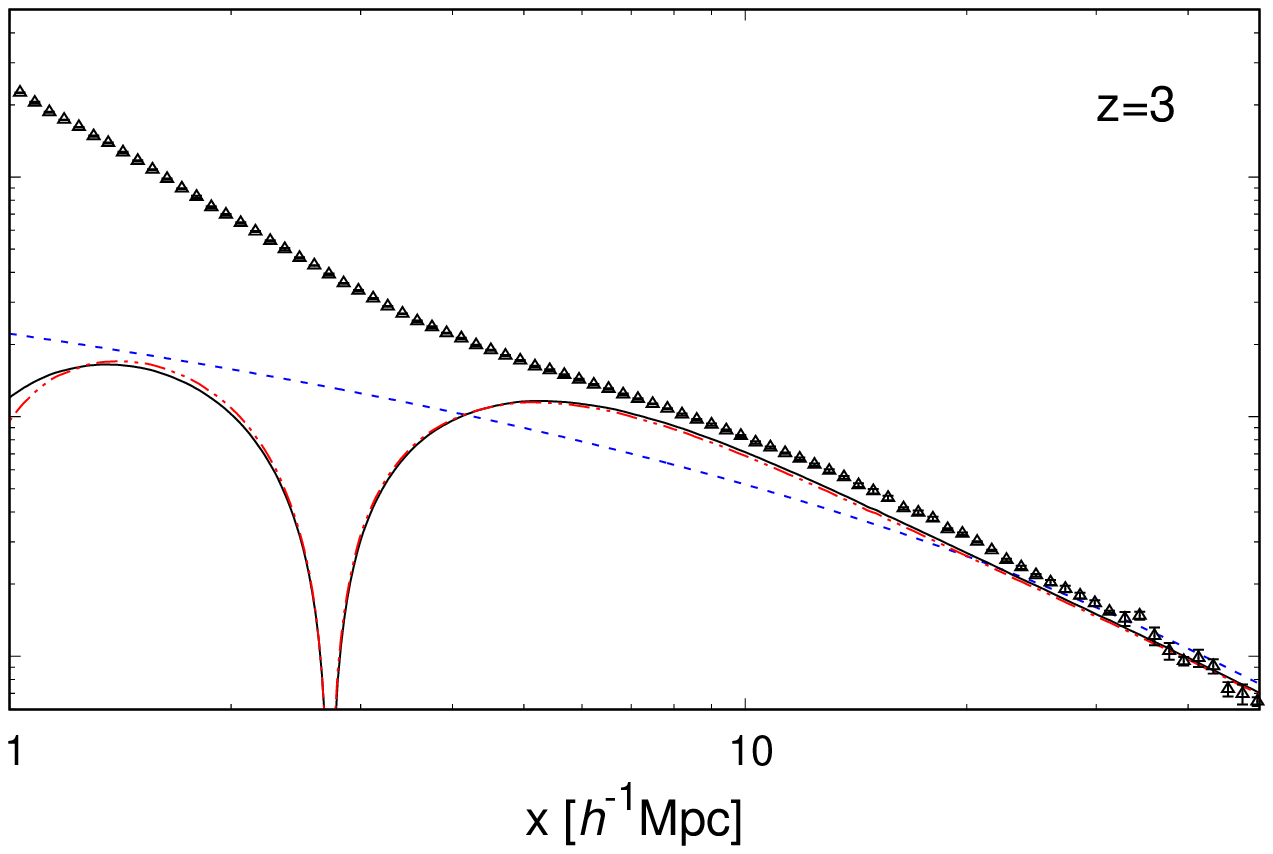}}
\end{center}
\caption{Mulitpoles of the correlation function for the linear prediction ``L'' (blue dashed lines),
our model ``Ga'' (black solid lines) and the Zeldovich approximation ``Zel'' (red dot-dashed lines),
at redshifts $z=0.35$, $1$ and $3$.
We show the multipoles $\ell=0$ (upper row) and the absolute value of the multipoles
$\ell=2$ (middle row) and $\ell=4$ (lower row).}
\label{fig_lxi}
\end{figure*}

We now study the predictions of our Gaussian ansatz for the redshift-space correlation function
$\xi^s({\bf x})$. It is the Fourier transform of the power spectrum,
\be
\xi^s({\bf x}) = \int d{\bf k} \, e^{i{\bf k}\cdot{\bf x}} \, P^s({\bf k}) .
\ee
It also depends on both the distance $x$ and the cosine of the angle with the line of sight,
$\mu=x_z/x$. It can again be expanded over the Legendre polynomials as
\be
\xi^s(x,\mu) = \sum_{\ell=0}^{\infty} \xi^s_{2\ell}(x) {\cal P}_{2\ell}(\mu) .
\ee
We compute the multipoles of the correlation function from the Hankel transforms of the
multipoles of the power spectrum,
\be
\xi^s_{2\ell}(x) = 4\pi (-1)^{\ell} \int_0^{\infty} dk k^2 \, P^s_{2\ell}(k) j_{2\ell}(kx) .
\ee
We obtain in this fashion the redshift-space correlation functions associated with the
linear theory, the Zeldovich approximation and our Gaussian ansatz.

We show in Fig.~\ref{fig_lxi} the redshift-space correlation functions on weakly nonlinear scales,
as compared with numerical simulations. Here, we use simulations newly performed with an improved
measurement of the correlation functions based on a hybrid scheme that combines the Fast Fourier
Transform (FFT) and the direct pair counting (see \cite{Nishimichi:2019}). This is important especially
for high multipole moments because the discreteness and anisotropies of the grids of the FFT-based
method can be problematic on scales close to the inter grid separation. These simulations are performed
in the same WMAP5 cosmology. We employ the ``fixed-and-paired'' technique by \cite{Angulo:2016} to
reduce the sample variance and perform 5 pairs of $1024^3$-body simulations in three different box sizes
($2048$, $1024$ and $512\,h^{-1}$Mpc) to obtain converged results.
As for the real-space correlation function, the Zeldovich approximation gives a redshift-space
correlation function that goes to a constant at small scale (because the power spectrum
decays faster than $k^{-3}$), whereas our Gaussian ansatz shows a logarithmic growth.
However, neither approximations can describe the growth of the correlation function on small
nonlinear scales associated with virialized halos.
As for the power spectra, the agreement with the simulations worsens for higher multipoles
$\ell$.
Again, for the quadrupole, $\xi^s_2$, both the Zeldovich approximation and our Gaussian ansatz
recover the change of sign near the nonlinear scale $x_{\rm NL}$, but they do not
predict its location with a good accuracy.
For the hexadecapole, they also predict two successive changes of sign whereas the numerical
simulations do not show any change of sign.
Whereas we can see a significant improvement over the linear theory for $\ell=0$ and $2$,
for the hexadecapole they only improve over linear theory at high redshift, $z \gtrsim 3$,
over these weakly nonlinear scales.

\begin{figure*}
\begin{center}
\epsfxsize=6.4 cm \epsfysize=4.6 cm {\epsfbox{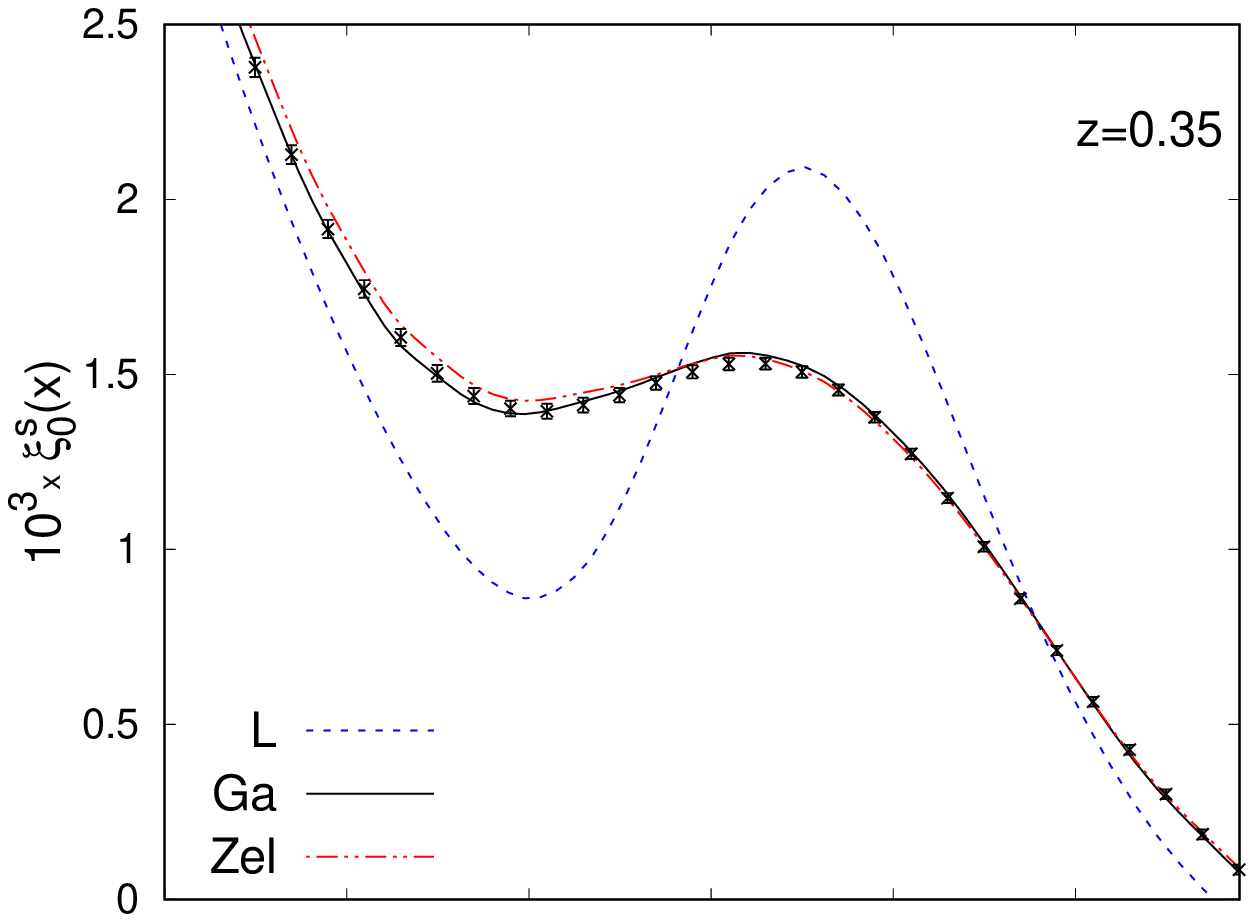}}
\epsfxsize=5.65 cm \epsfysize=4.6 cm {\epsfbox{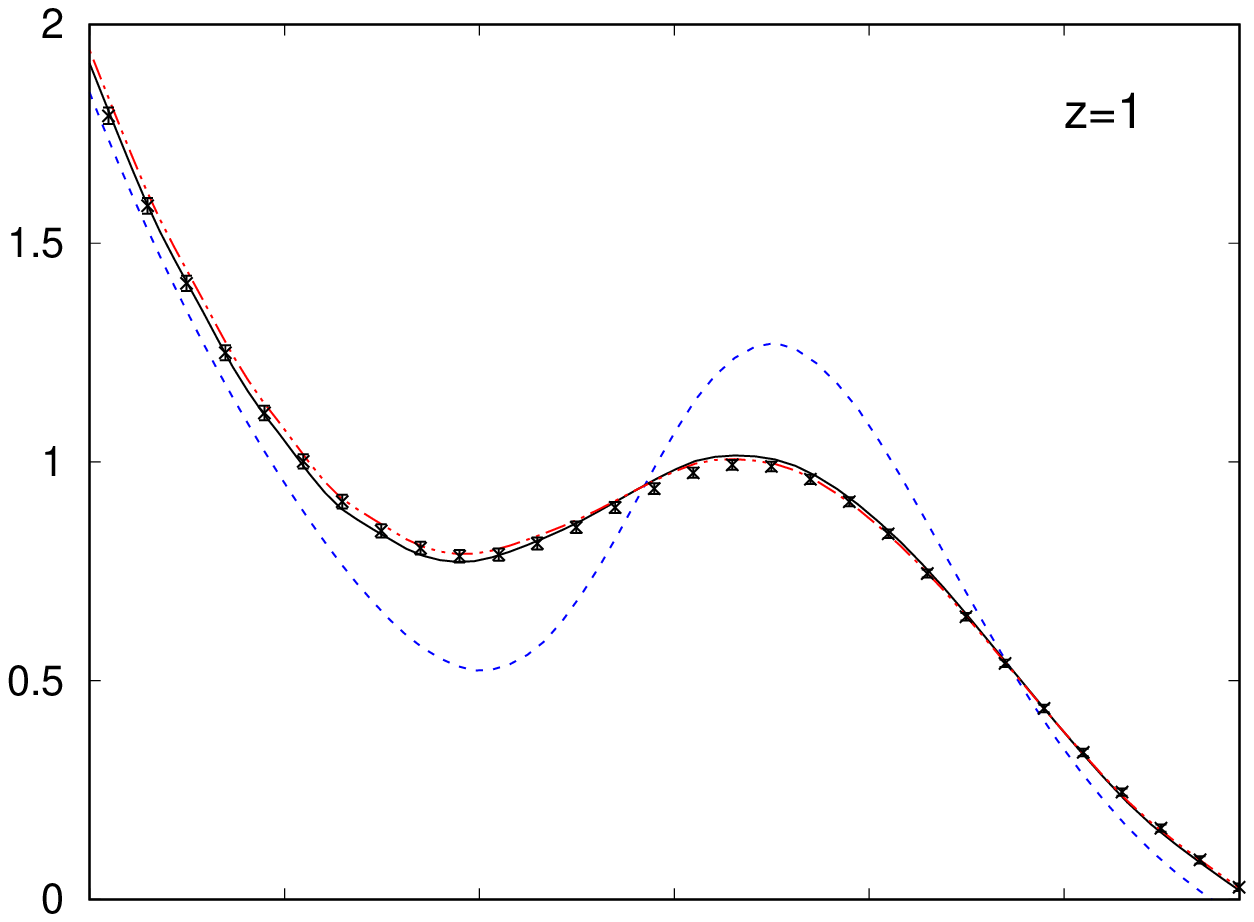}}
\epsfxsize=5.65 cm \epsfysize=4.6 cm {\epsfbox{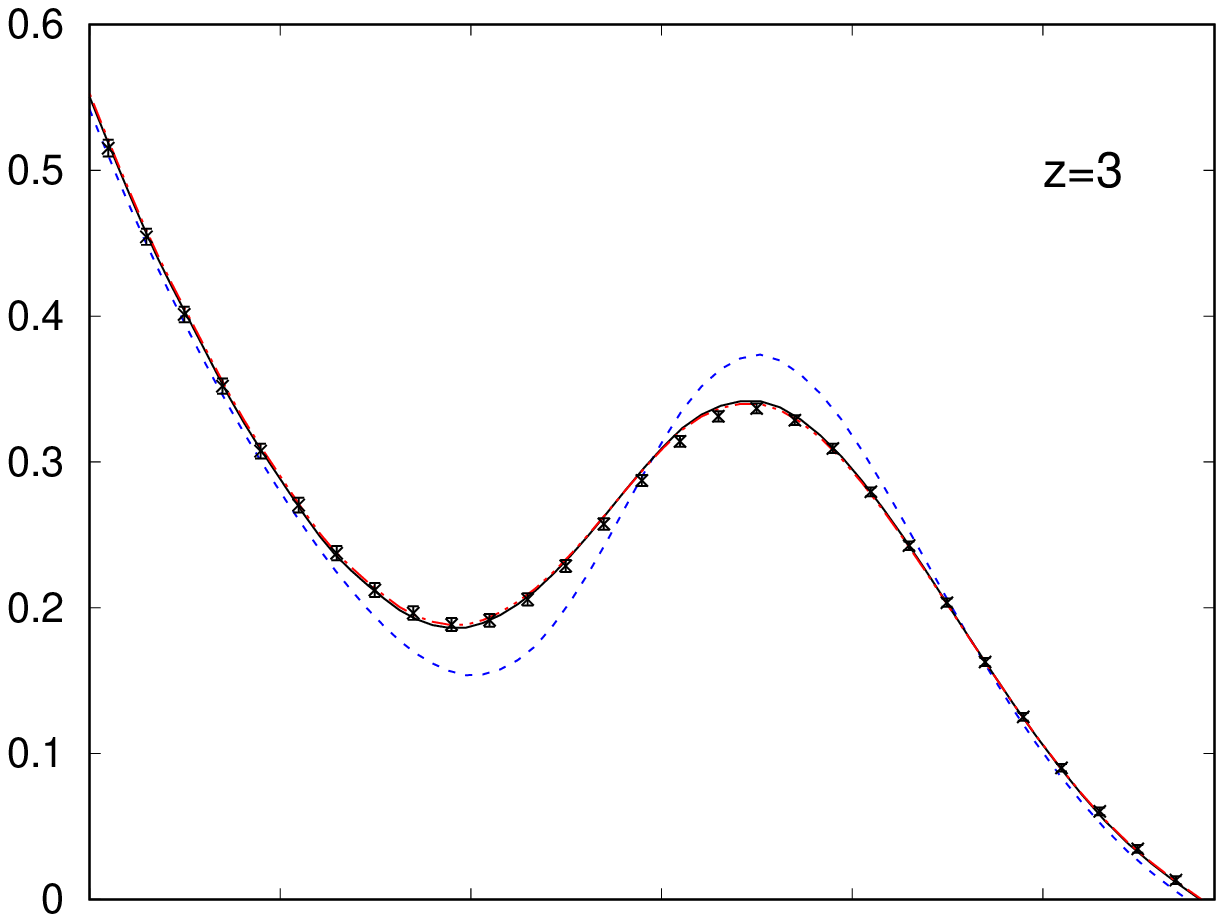}}\\
\epsfxsize=6.4 cm \epsfysize=4.6 cm {\epsfbox{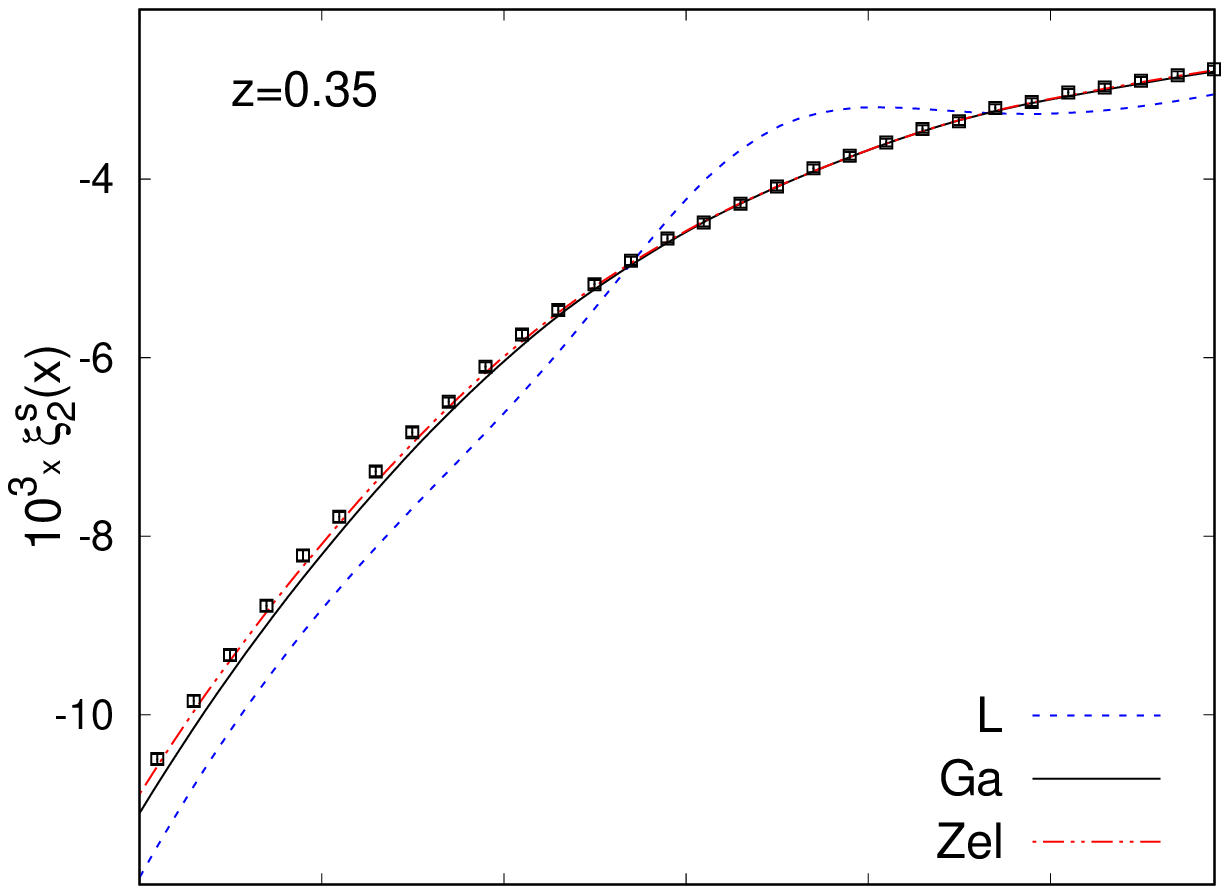}}
\epsfxsize=5.65 cm \epsfysize=4.6 cm {\epsfbox{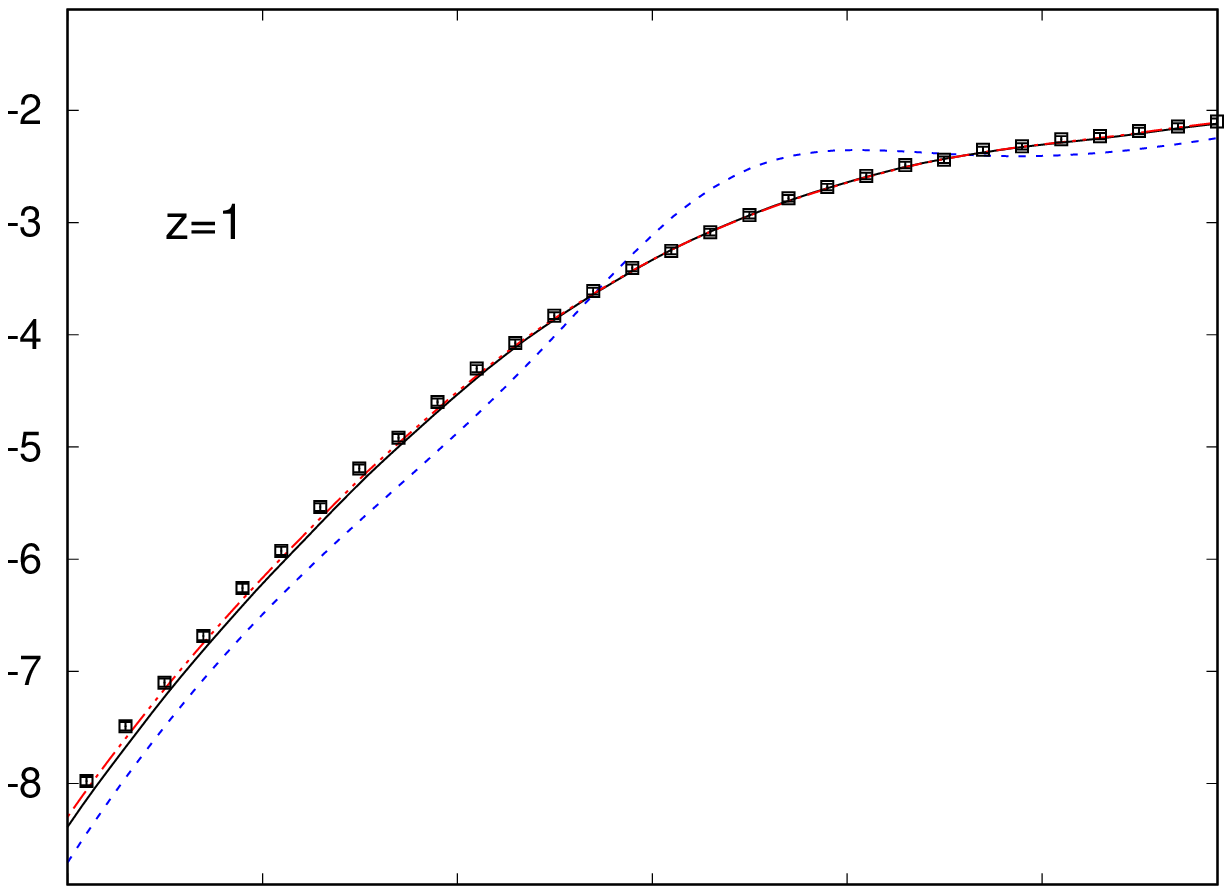}}
\epsfxsize=5.65 cm \epsfysize=4.6 cm {\epsfbox{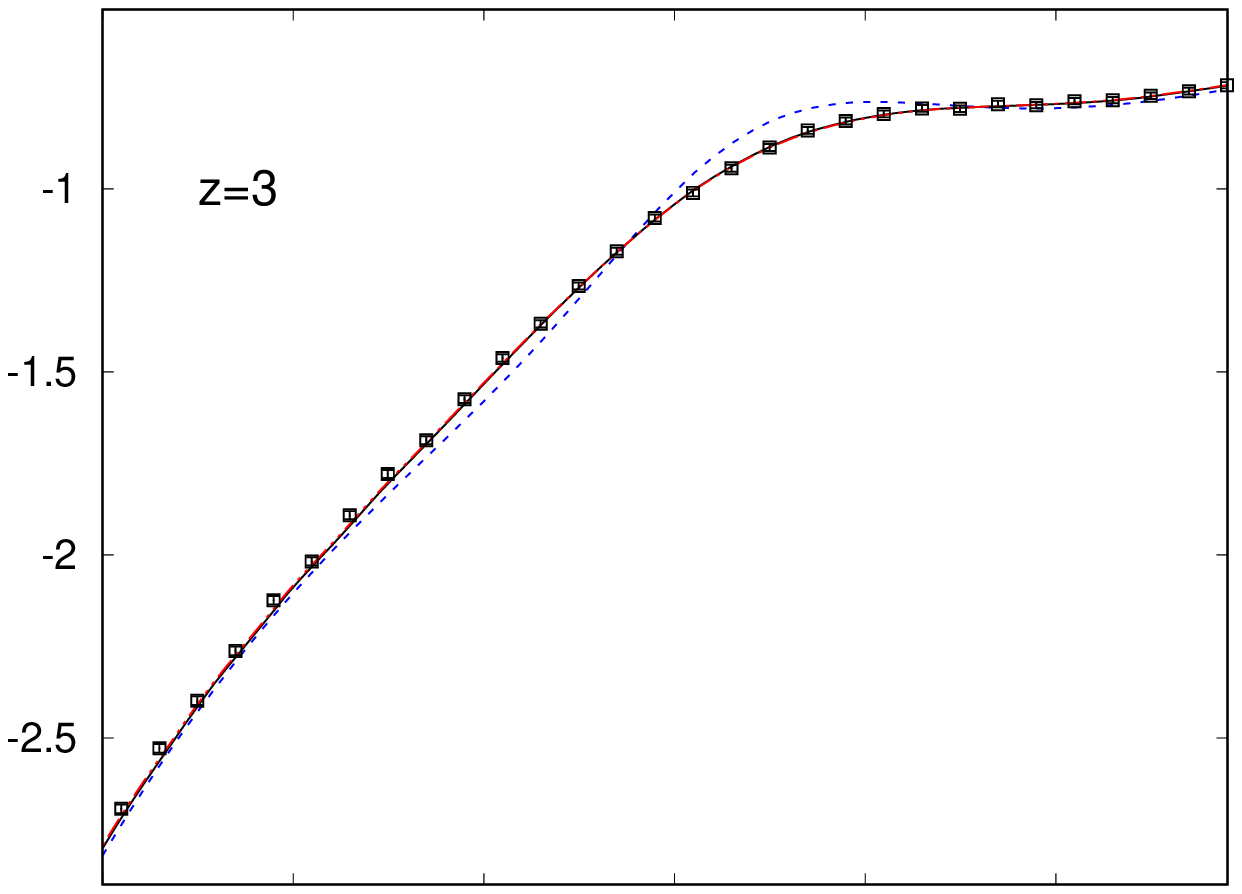}}\\
\epsfxsize=6.4 cm \epsfysize=5.6 cm {\epsfbox{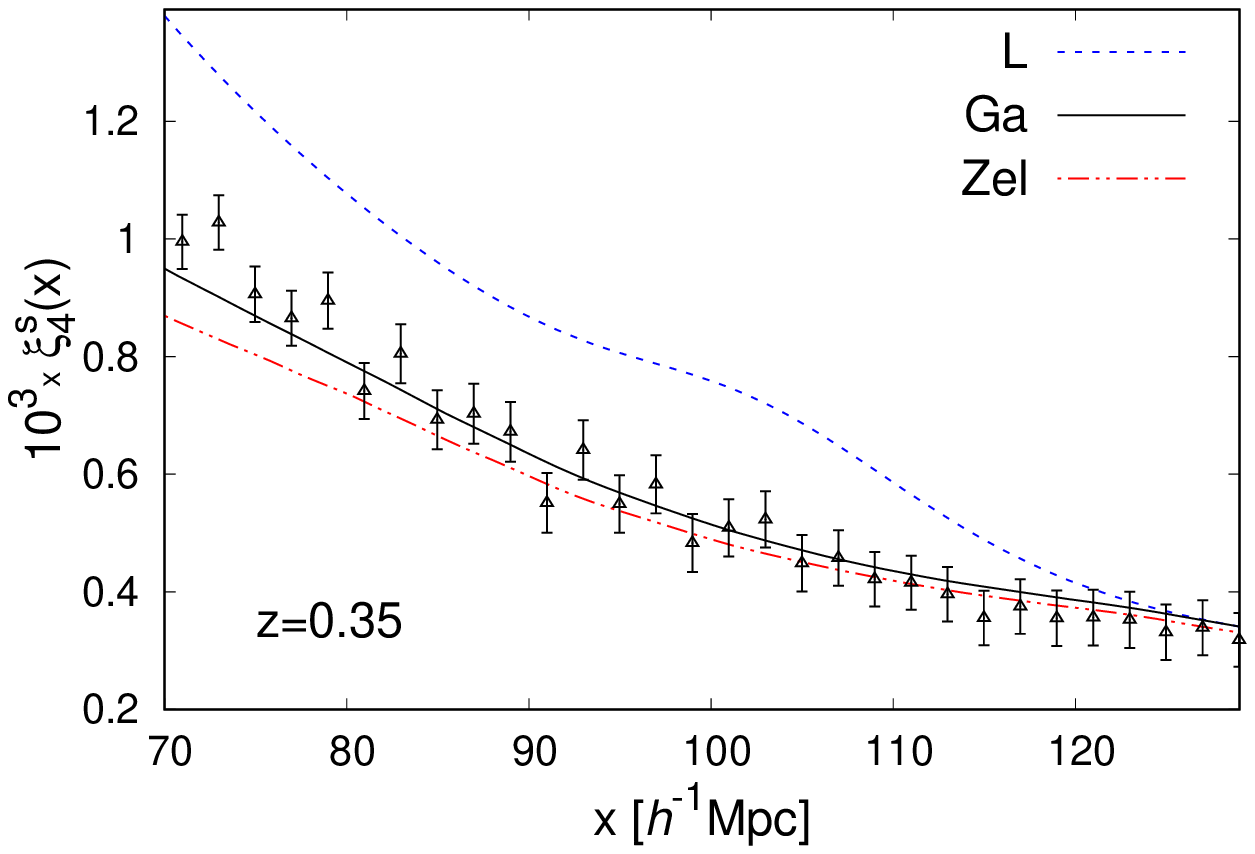}}
\epsfxsize=5.65 cm \epsfysize=5.6 cm {\epsfbox{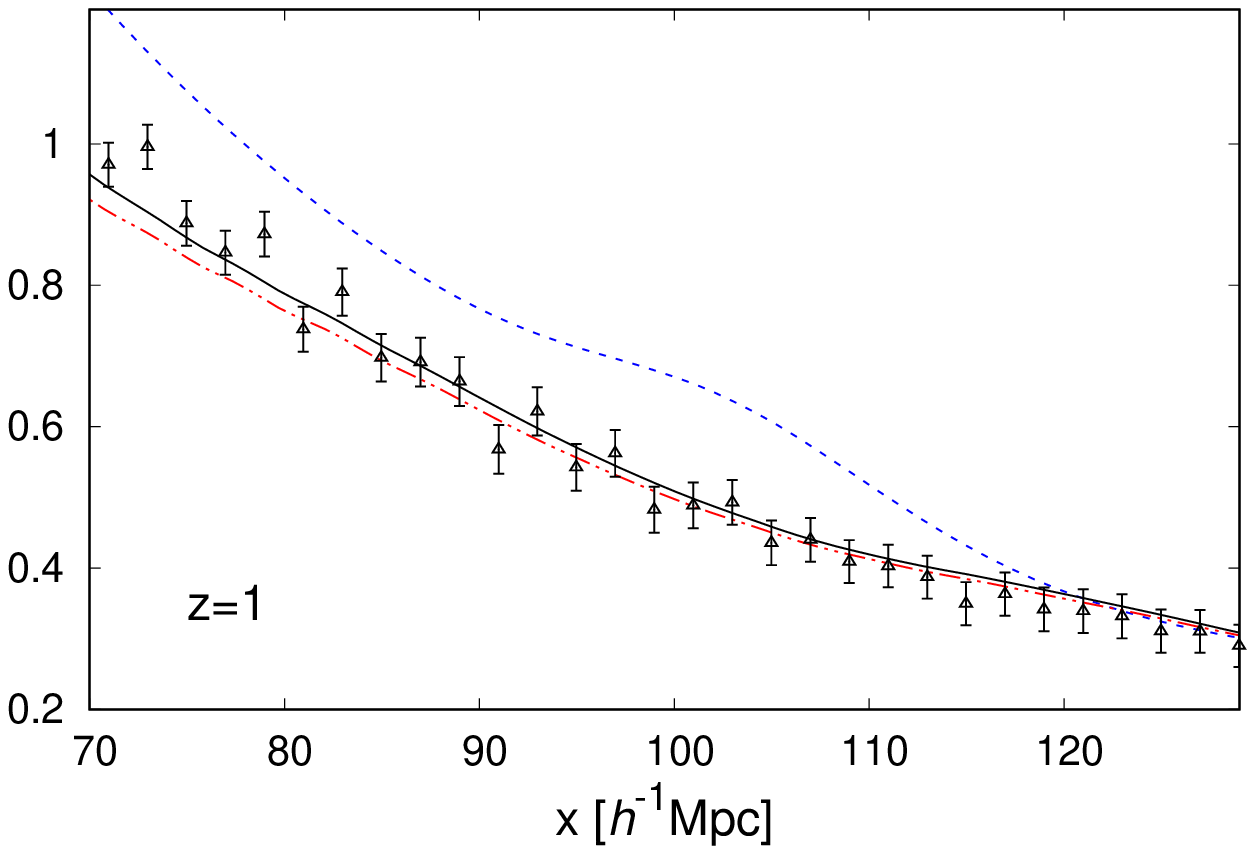}}
\epsfxsize=5.65 cm \epsfysize=5.6 cm {\epsfbox{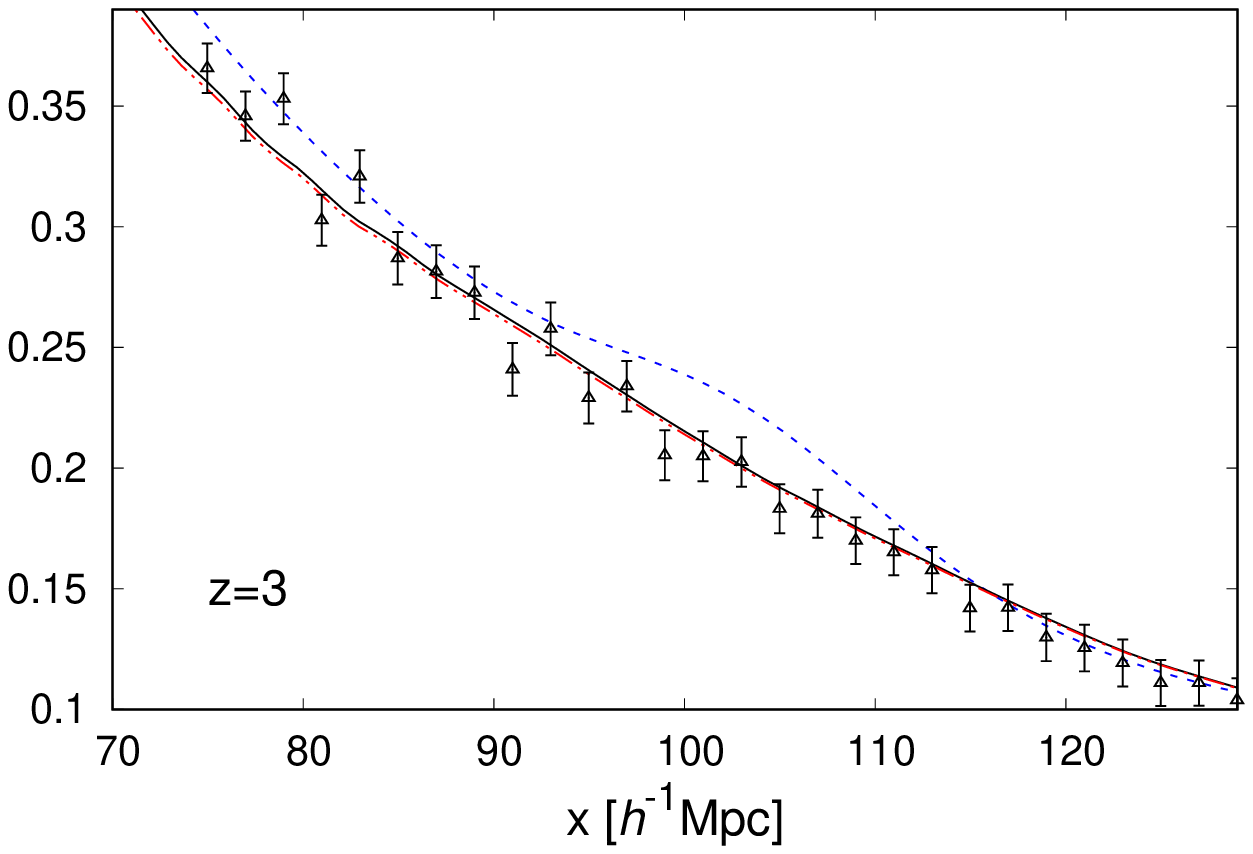}}
\end{center}
\caption{Mulitpoles of the correlation function mutiplied by a factor $10^3$.
We show the multipoles $\ell=0$ (upper row), $\ell=2$ (middle row) and $\ell=4$ (lower row).}
\label{fig_xi}
\end{figure*}

We focus on BAO scales in Fig.~\ref{fig_xi}.
Here we use the 60 realizations of N-body simulations in \cite{Taruya:2012ut},
because we do not see a clear improvement with the new paired-and-fixed simulations.
We recover the fact that the Zeldovich
approximation already gives a great improvement over the linear theory for
the baryonic peak of the monopole correlation function, at $x \sim 105 h^{-1}$ Mpc.
The improvement is also large for the quadrupole and the hexadecapole.
Our Gaussian ansatz further improves over the Zeldovich approximation,
but by a modest amount.
This is again an illustration of the fact that the smoothing of the BAO peak, and more
generally the deviations from linear theory on BAO scales, are governed by
large-scale motions that are well described by Lagrangian approaches and are not sensitive
to displacements on small nonlinear scales,
\cite{Matsubara:2007wj,Valageas:2013hxa,Tassev:2013rta}.

\begin{figure*}
\begin{center}
\epsfxsize=6.4 cm \epsfysize=4.6 cm {\epsfbox{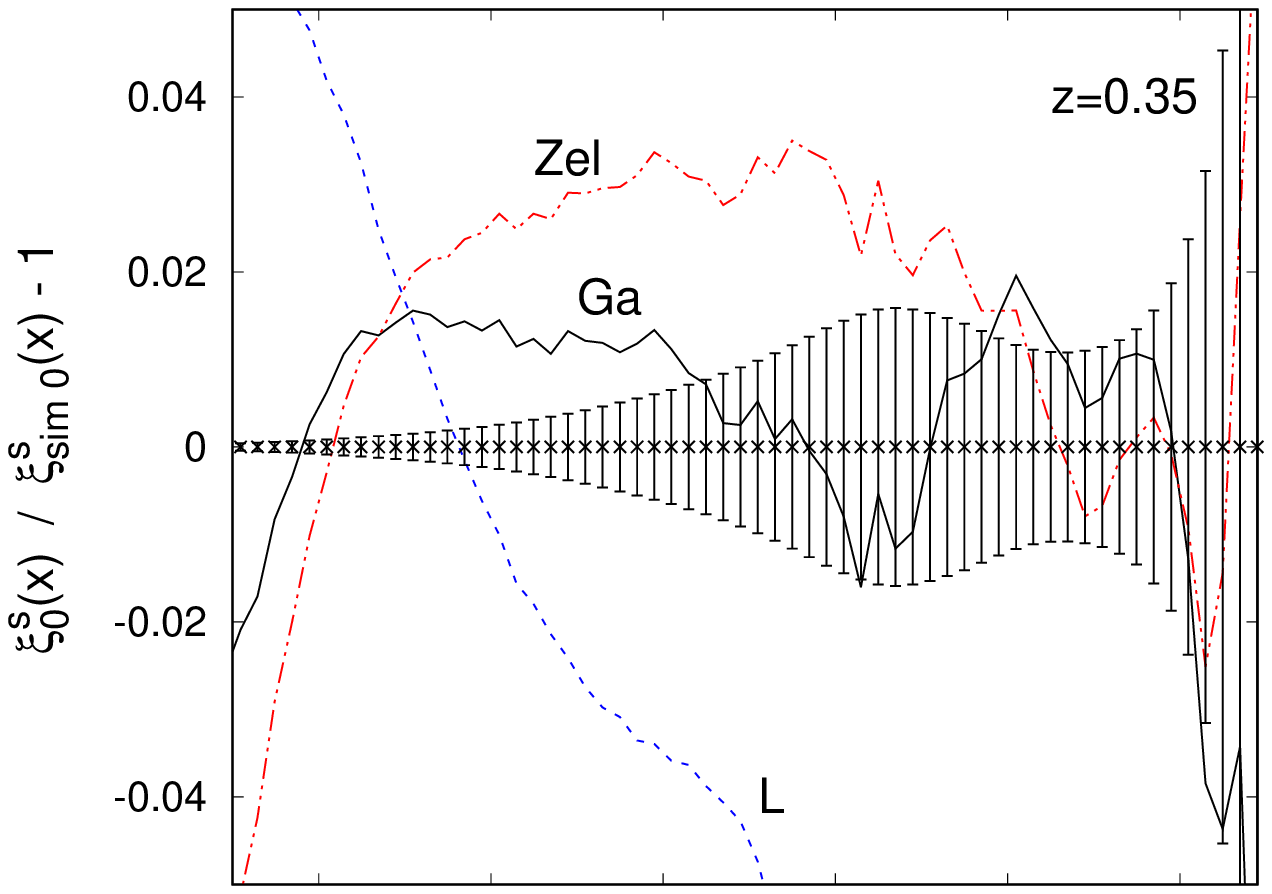}}
\epsfxsize=5.65 cm \epsfysize=4.6 cm {\epsfbox{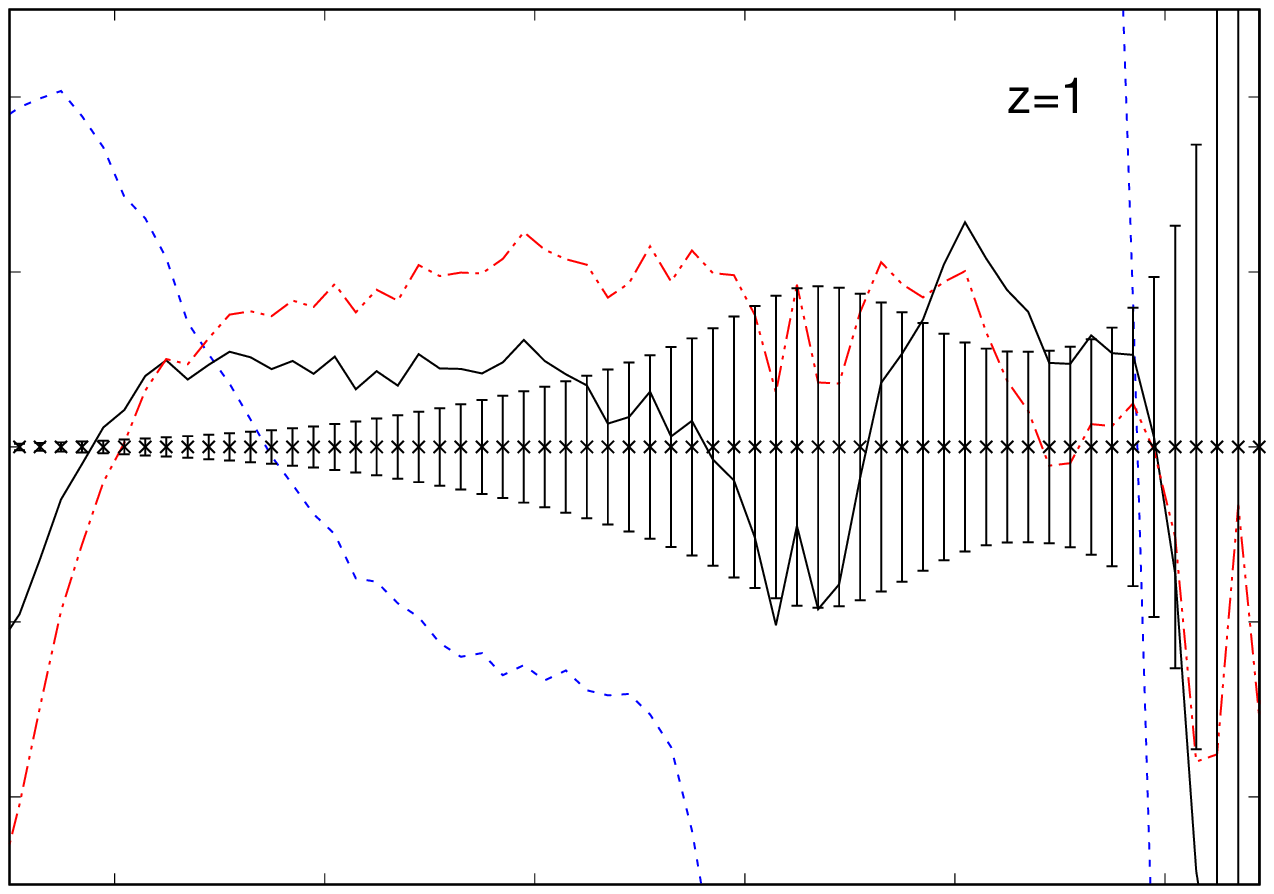}}
\epsfxsize=5.65 cm \epsfysize=4.6 cm {\epsfbox{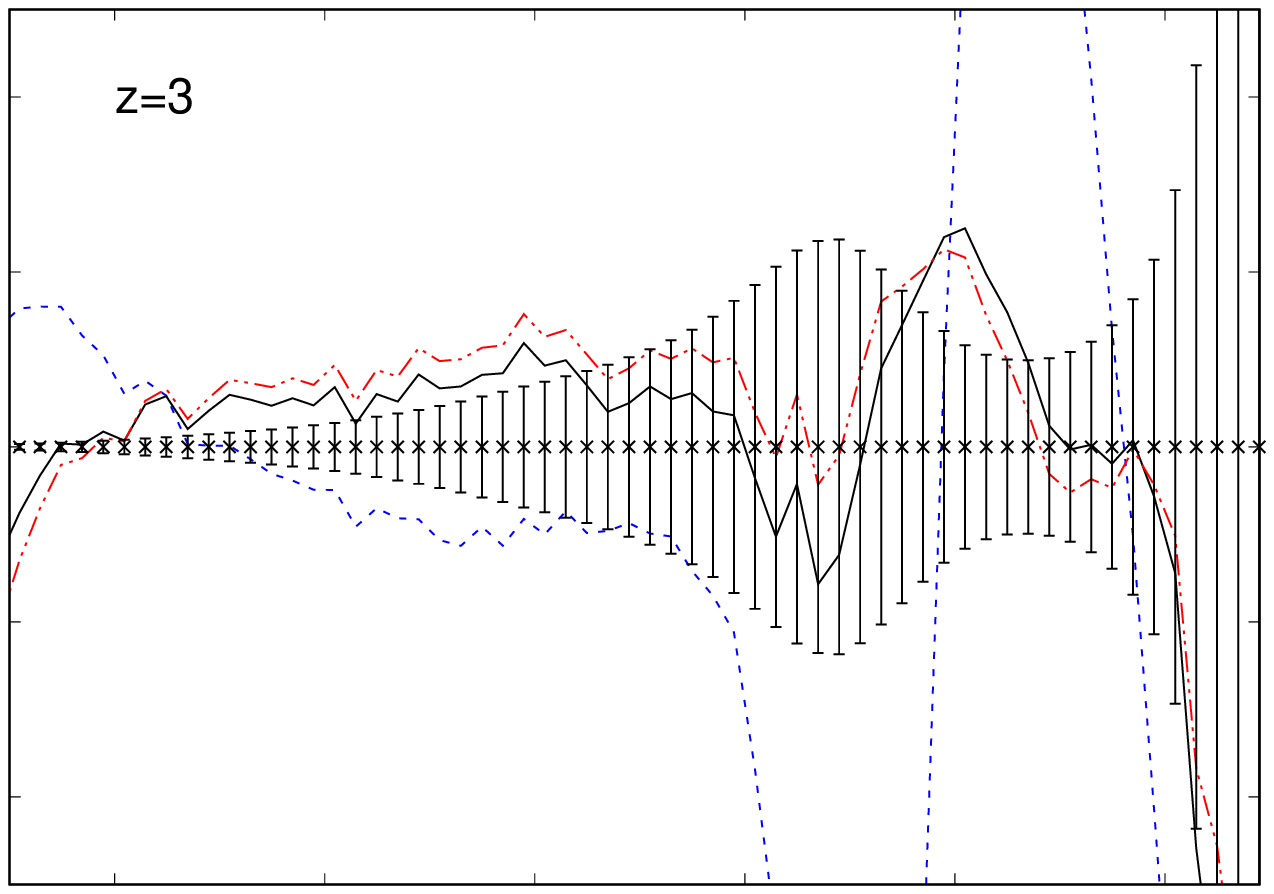}}\\
\epsfxsize=6.4 cm \epsfysize=4.6 cm {\epsfbox{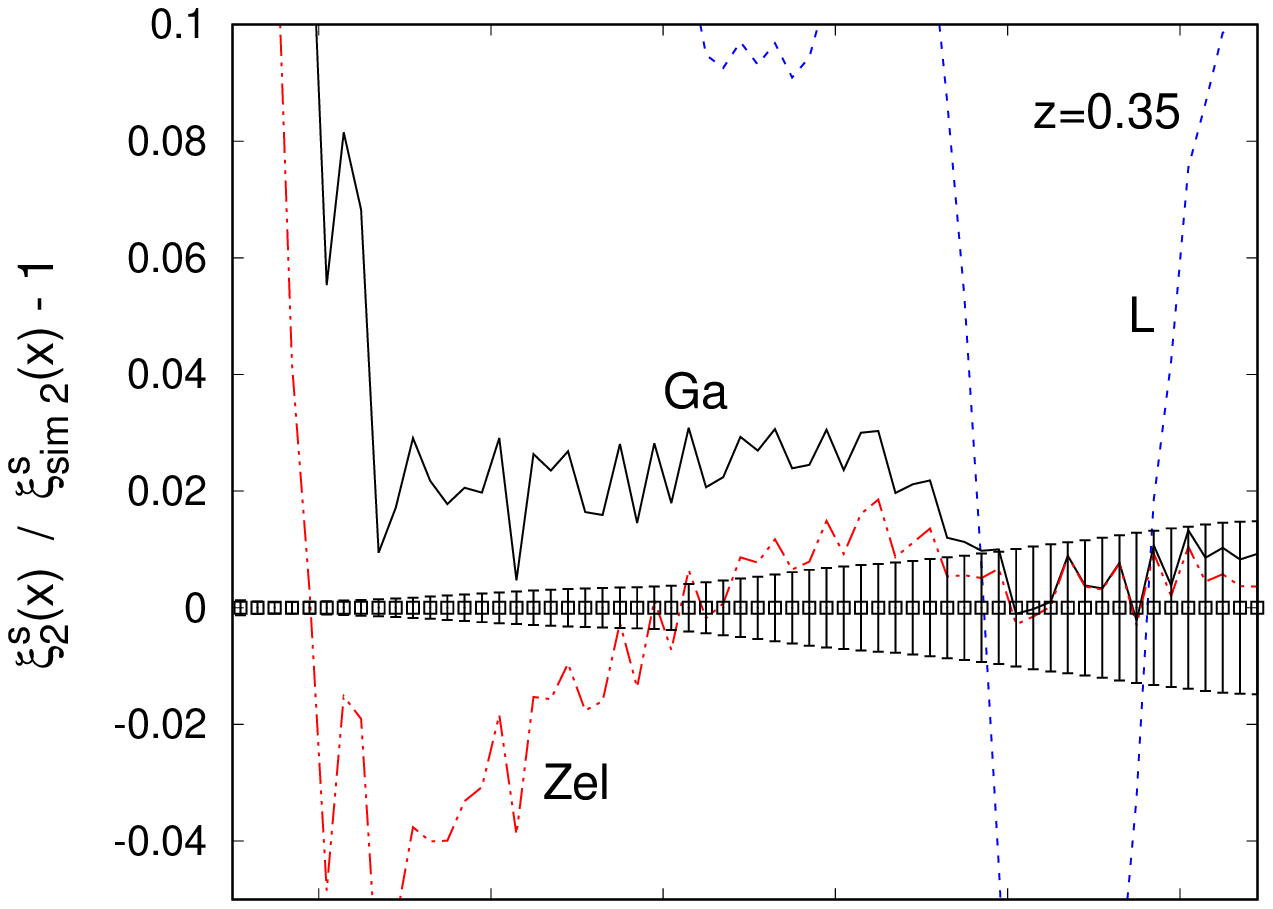}}
\epsfxsize=5.65 cm \epsfysize=4.6 cm {\epsfbox{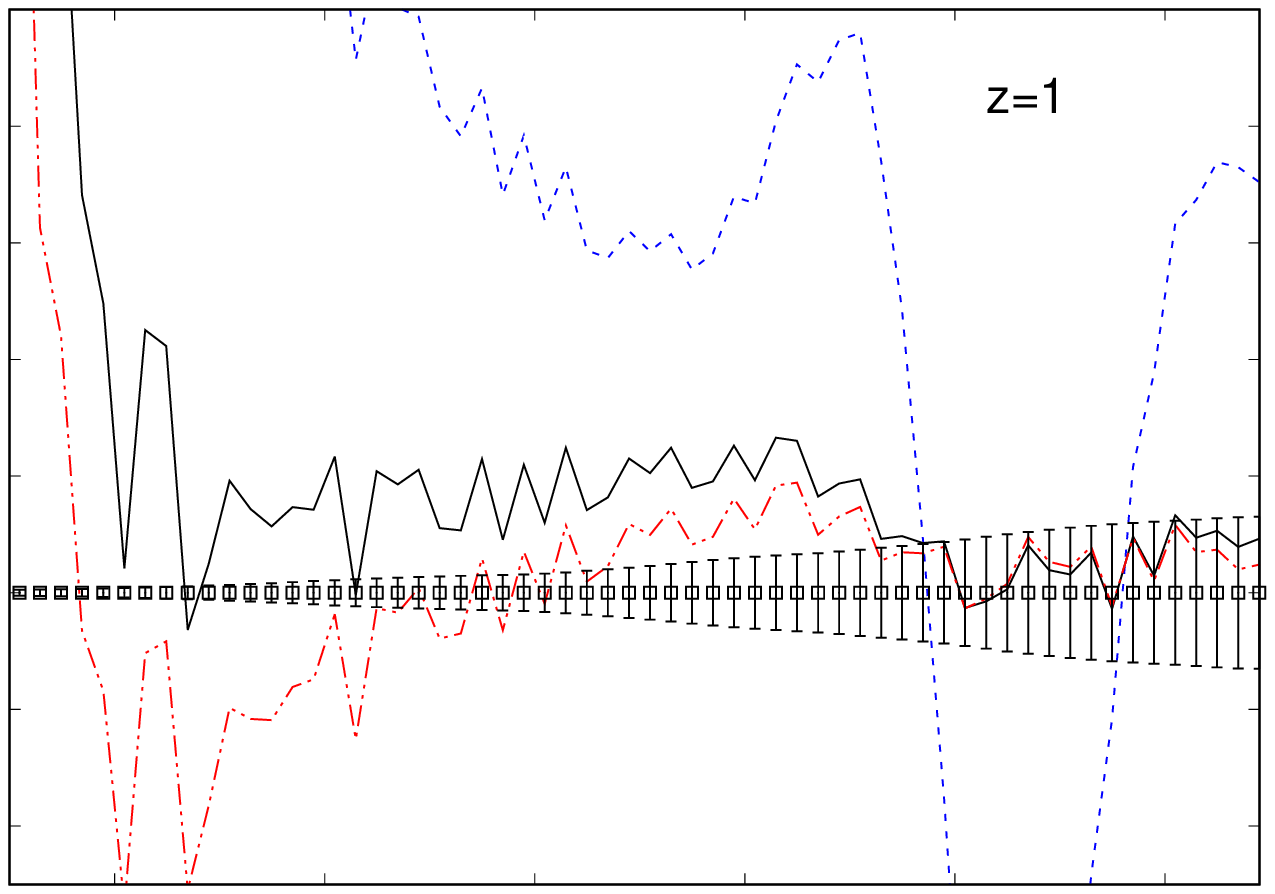}}
\epsfxsize=5.65 cm \epsfysize=4.6 cm {\epsfbox{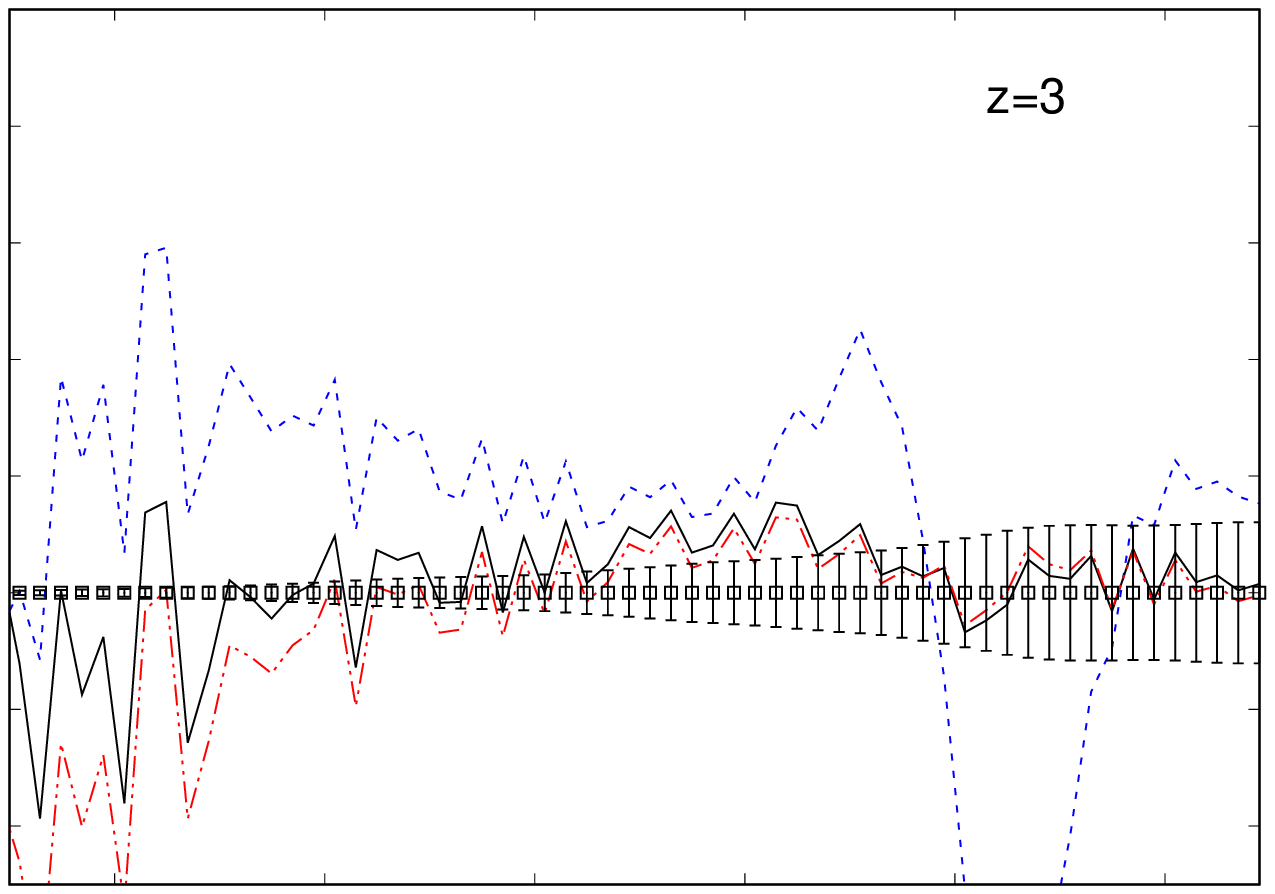}}\\
\epsfxsize=6.4 cm \epsfysize=5.6 cm {\epsfbox{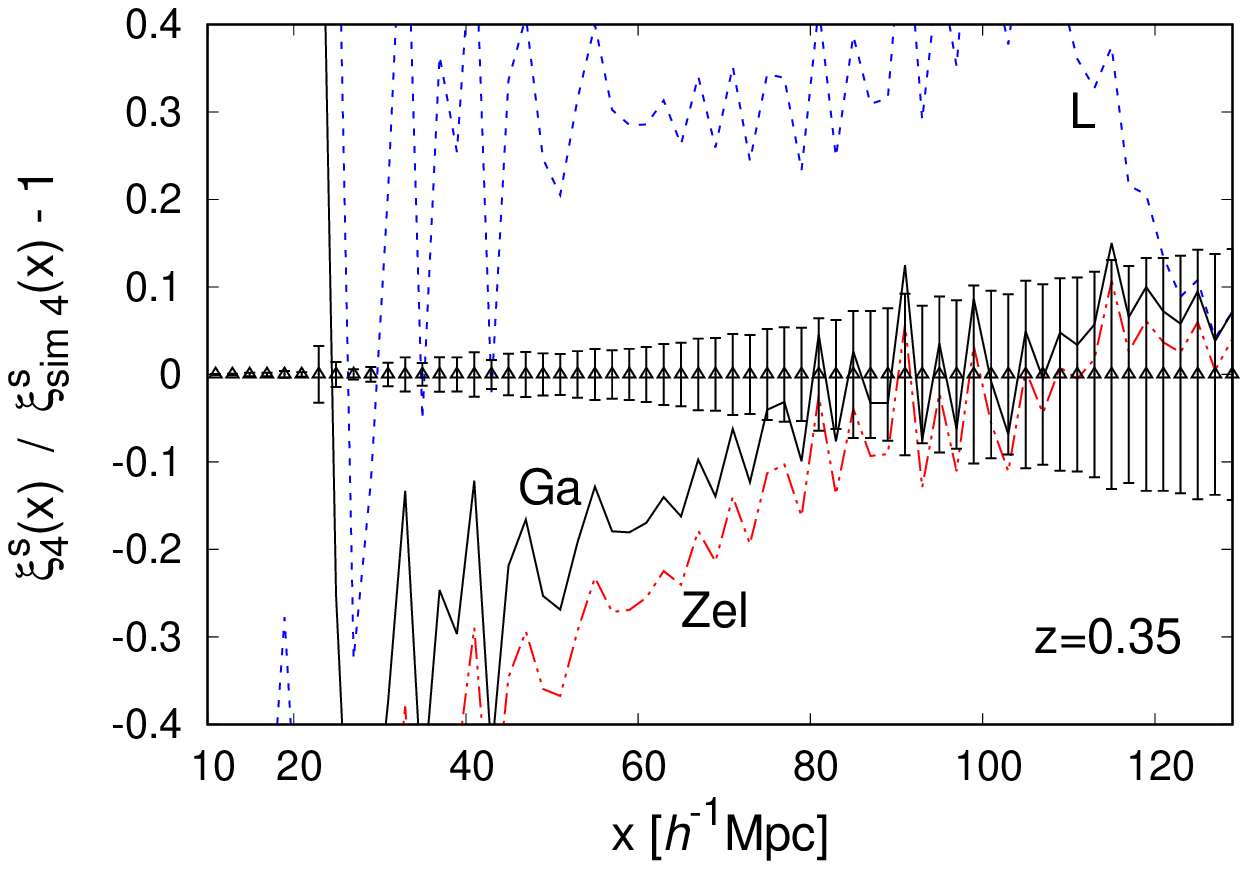}}
\epsfxsize=5.65 cm \epsfysize=5.6 cm {\epsfbox{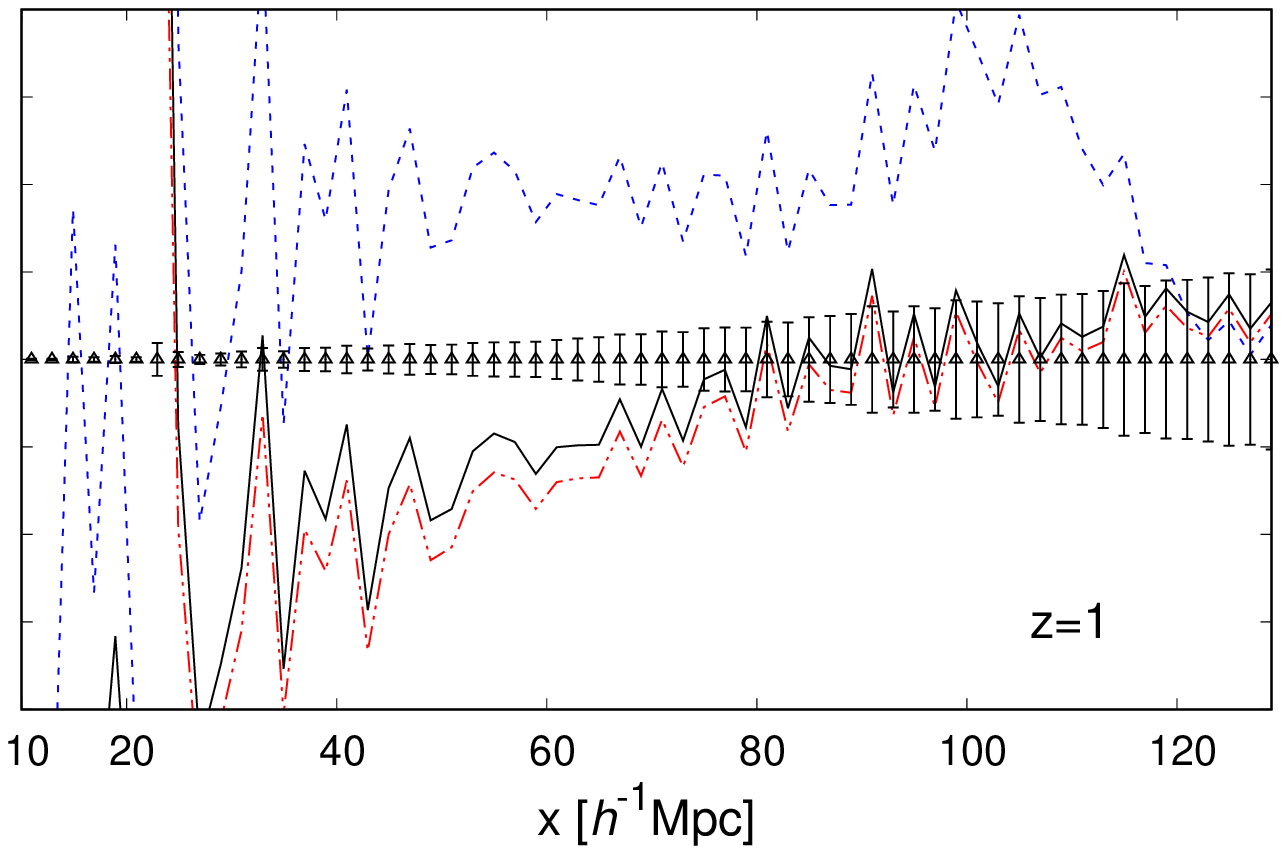}}
\epsfxsize=5.65 cm \epsfysize=5.6 cm {\epsfbox{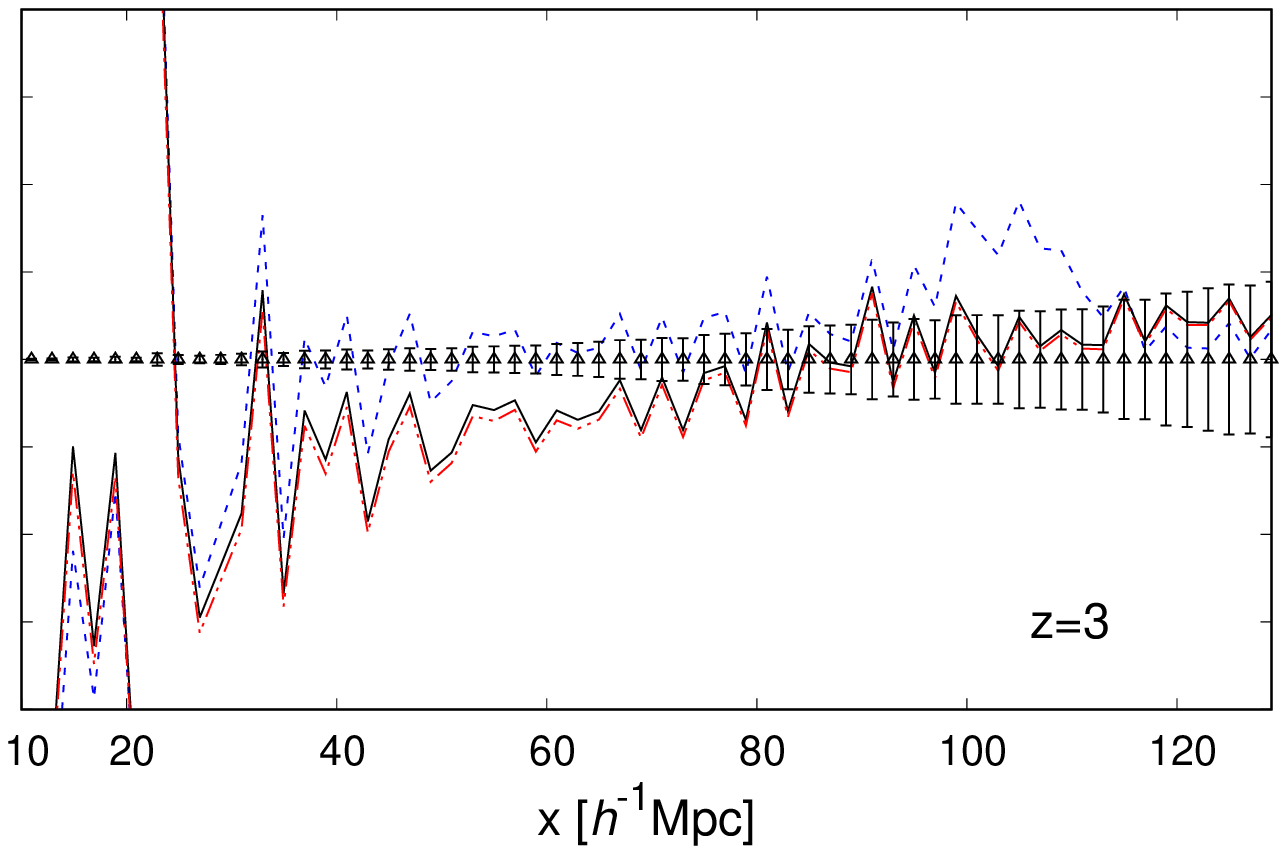}}\\
\end{center}
\caption{Relative deviation of the analytical predictions from the numerical simulations,
for the redshift-space correlation functions.
We show the multipoles $\ell=0$ (upper row), $\ell=2$ (middle row) and $\ell=4$ (lower row).}
\label{fig_dxi}
\end{figure*}

We show in Fig.~\ref{fig_dxi} the relative deviations from numerical simulations
of the multipoles of the correlation functions.
The small wiggles are due to the noise of the numerical simulations and provide
an estimate of their accuracy, beyond the statistical error bars that are shown by
the error bars centered on zero.

In agreement with the previous figures, we find a great improvement over the linear theory
and a modest improvement over the Zeldovich approximation.
This shows that, as expected, making the displacement and velocity power spectra
determined by the equations of motion (\ref{eq:dPchichi-deta})-(\ref{eq:dPthetatheta-deta}),
instead of setting them equal to the linear power spectrum, provides a better description
of the dynamics.
This improvement also agrees with the results of \cite{Kopp:2016crg}, who find
that the halo redshift-space correlation function obtained within a Gaussian streaming
model is improved if one truncates the linear power spectrum, as in the truncated
Zeldovich approximation.
However, the modest level of improvement means that in order
to reach smaller scales, or to obtain a great improvement on large scales, we need
to go beyond the Gaussian ansatz and include higher-order correlations or polyspectra
for the  displacement and velocity fields.

For the monopole, our Gaussian ansatz provides an accuracy of $2\%$ down to
$10 h^{-1}$ Mpc, at $z \geq 0.35$.
For the quadrupole, it gives an accuracy of $3\%$ down to
$26 h^{-1}$ Mpc, and of $10\%$ down to $20 h^{-1}$ Mpc, at $z \geq 0.35$.
For the hexadecapole, it gives an accuracy of $10\%$ down to
$70 h^{-1}$ Mpc, and of $30\%$ down to $44 h^{-1}$ Mpc, at $z \geq 0.35$.

For comparison, we note that \cite{Matsubara:2007wj} obtains similar results on the BAO
scales for the monopole, using a partial resummation of Lagrangian perturbation theory.
The convolution Lagrangian perturbation theory developped in \cite{Carlson:2012bu},
which is an improved resummation, obtains a similar accuracy as our approach.
The Gaussian streaming model used in \cite{Reid:2011ar} gives at $z=0.55$
a percent accuracy down to $10 h^{-1}{\rm Mpc}$ for $\xi^s_0$,
and a $2\%$ accuracy down to $25 h^{-1}{\rm Mpc}$ for $\xi^s_2$.
The TNS model with a partial resummation of Eulerian perturbation theory
and a fitted damping parameter \cite{Taruya:2013my} gives at $z=0.35$
an accuracy of $5\%$ down to $20 h^{-1} {\rm Mpc}$ for $\xi^s_0$, and of $10\%$ down to
$20 h^{-1} {\rm Mpc}$ for $\xi^s_2$.

\begin{figure*}[ht]
\begin{center}
\epsfxsize=5.9 cm{\epsfbox{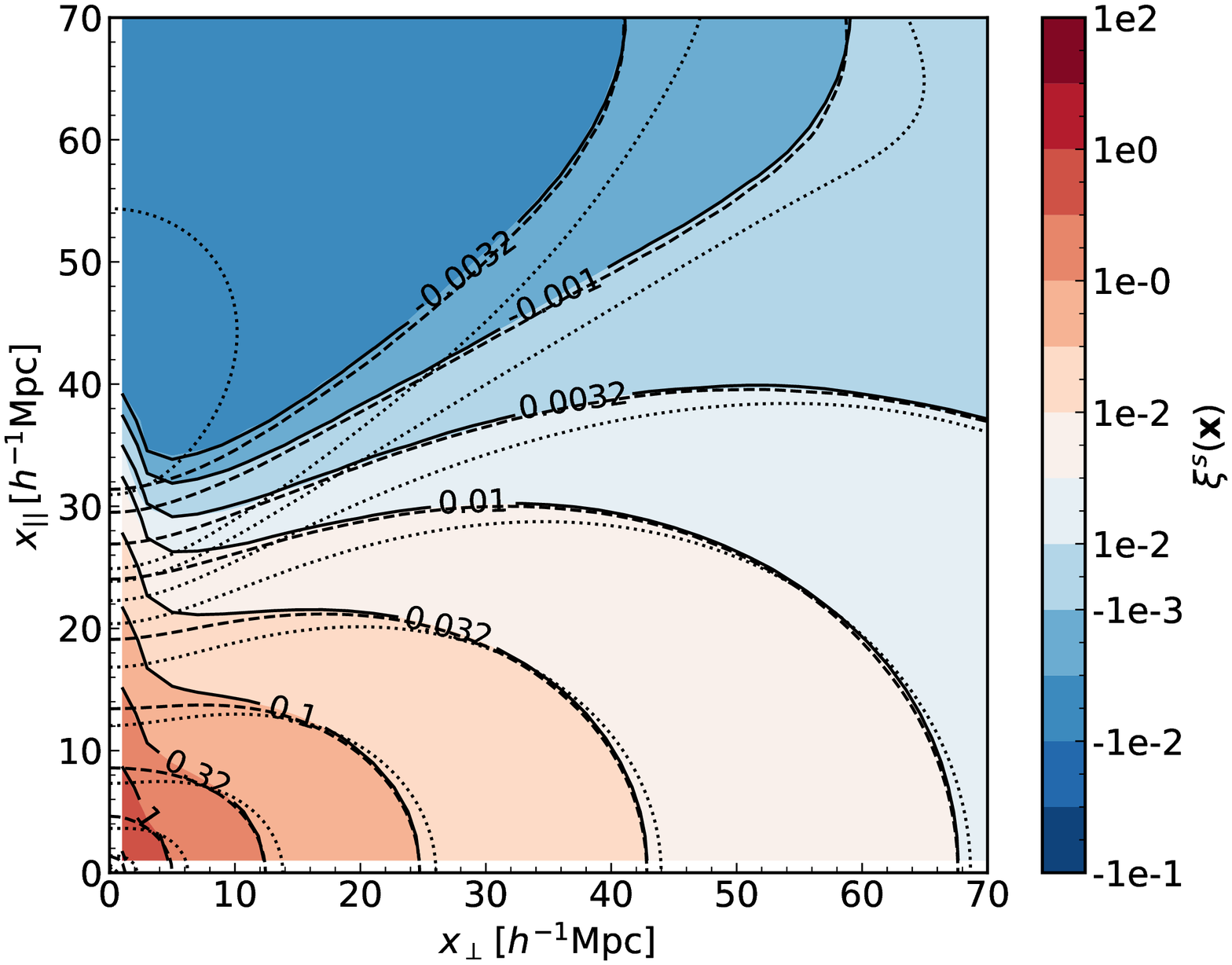}}
\epsfxsize=5.9 cm{\epsfbox{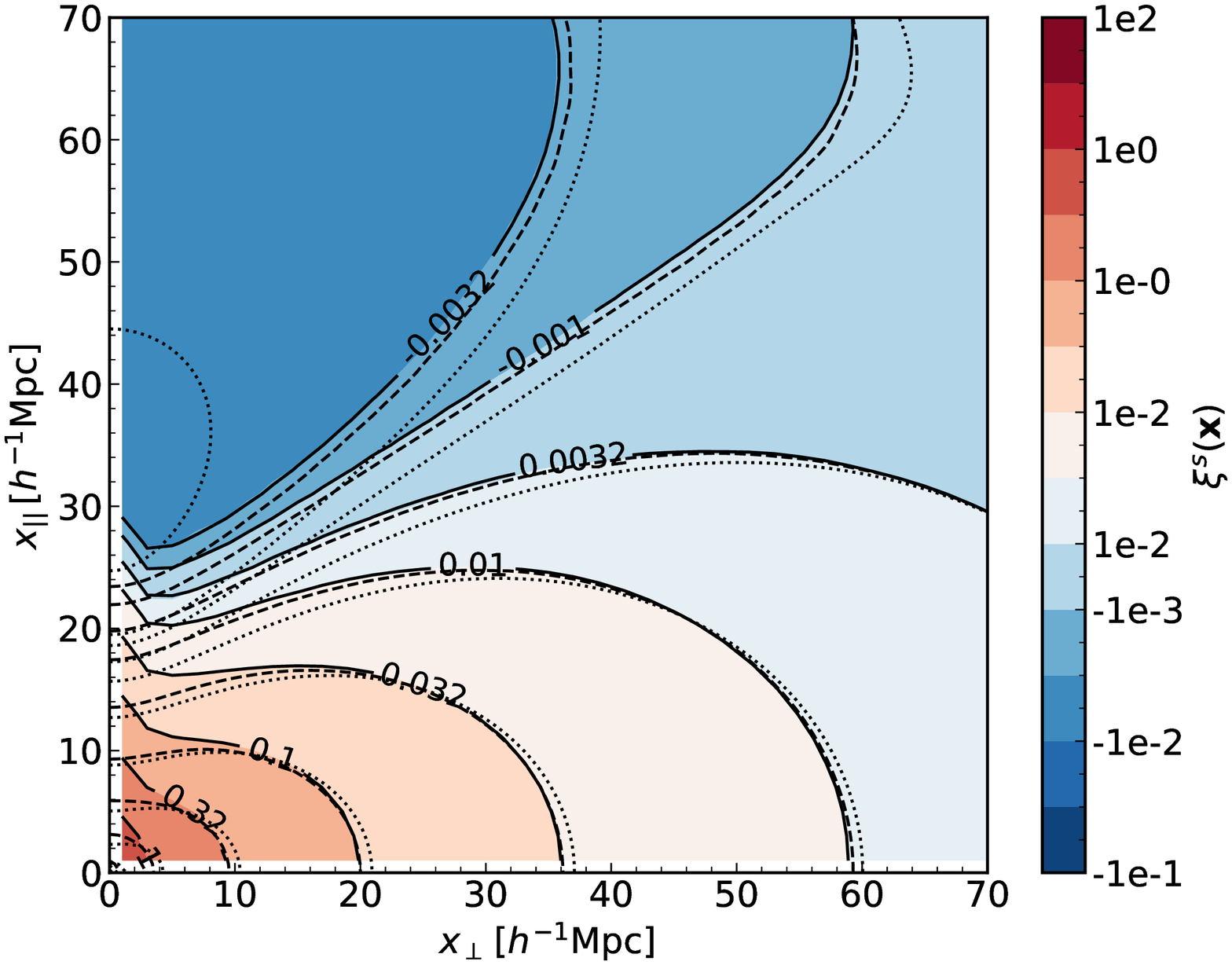}}
\epsfxsize=5.9 cm{\epsfbox{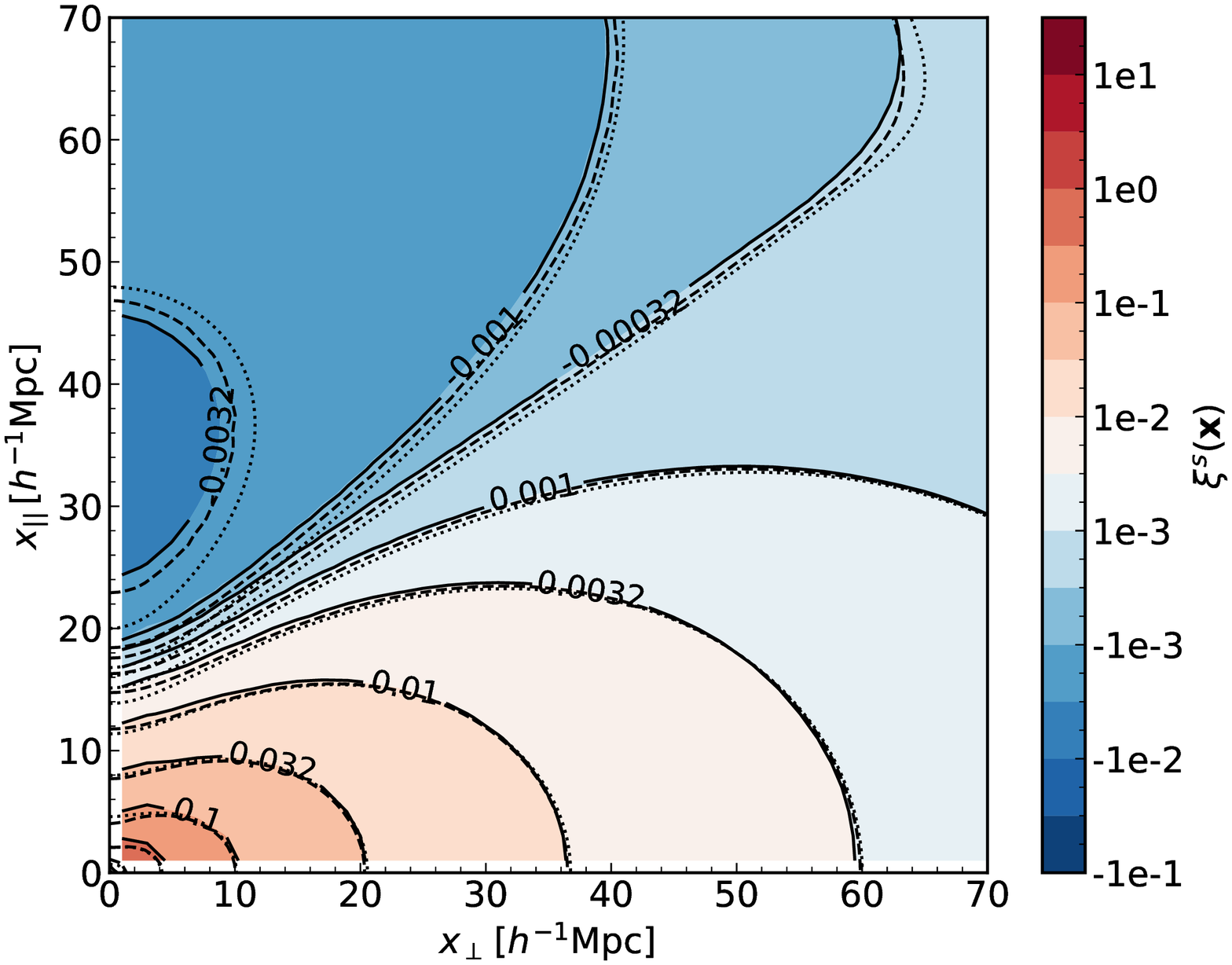}}
\end{center}
\caption{Redshift-space correlation function displayed in two dimensions, $(x_{\perp}, x_{||})$, the pair separation perpendicular and parallel to the line of sight. We show the results at $z=0.35$, $1$ and $3$ from left to right. The solid, dashed and dotted contour lines correspond to the N-body simulations, our model (Ga) and linear theory, respectively. The colorbars correspond to the contour lines for the simulation data. Here the simulation data is from the low resolution runs ($1024^3$ particles, $2048\,h^{-1}$Mpc) with the smallest sample variance error.}
\label{fig_xi2d}
\end{figure*}

\begin{figure*}
\begin{center}
\epsfxsize=5.9 cm{\epsfbox{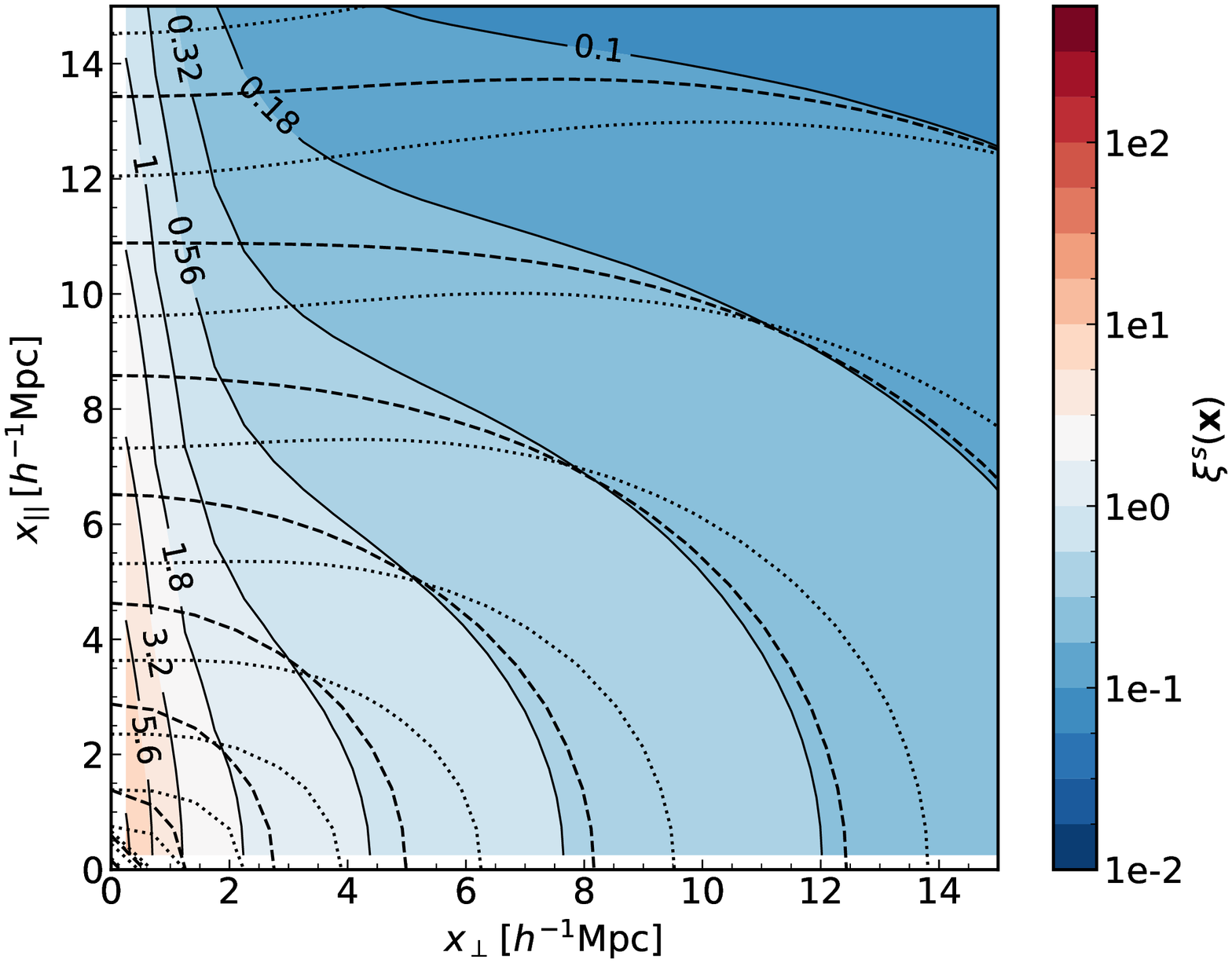}}
\epsfxsize=5.9 cm{\epsfbox{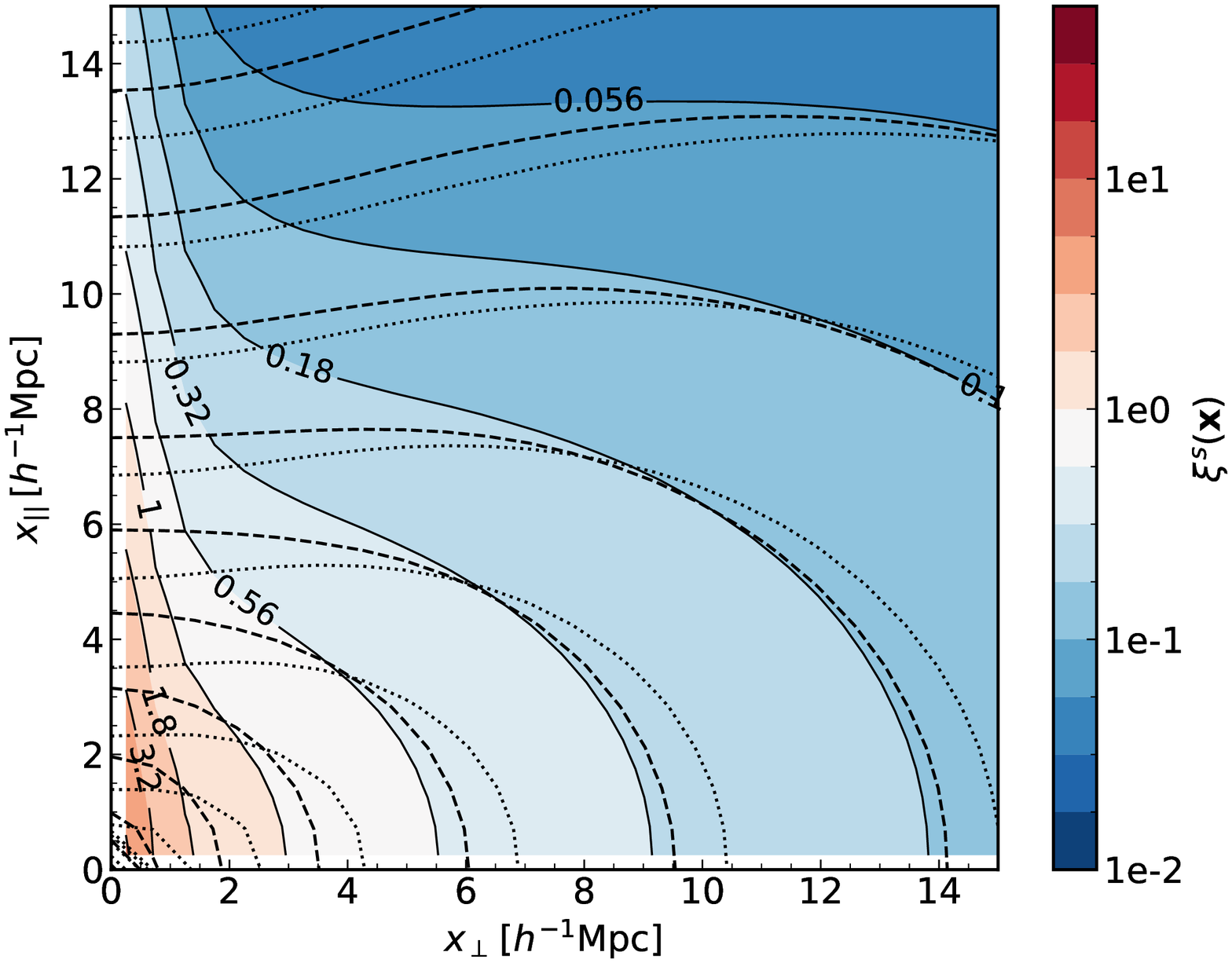}}
\epsfxsize=5.9 cm{\epsfbox{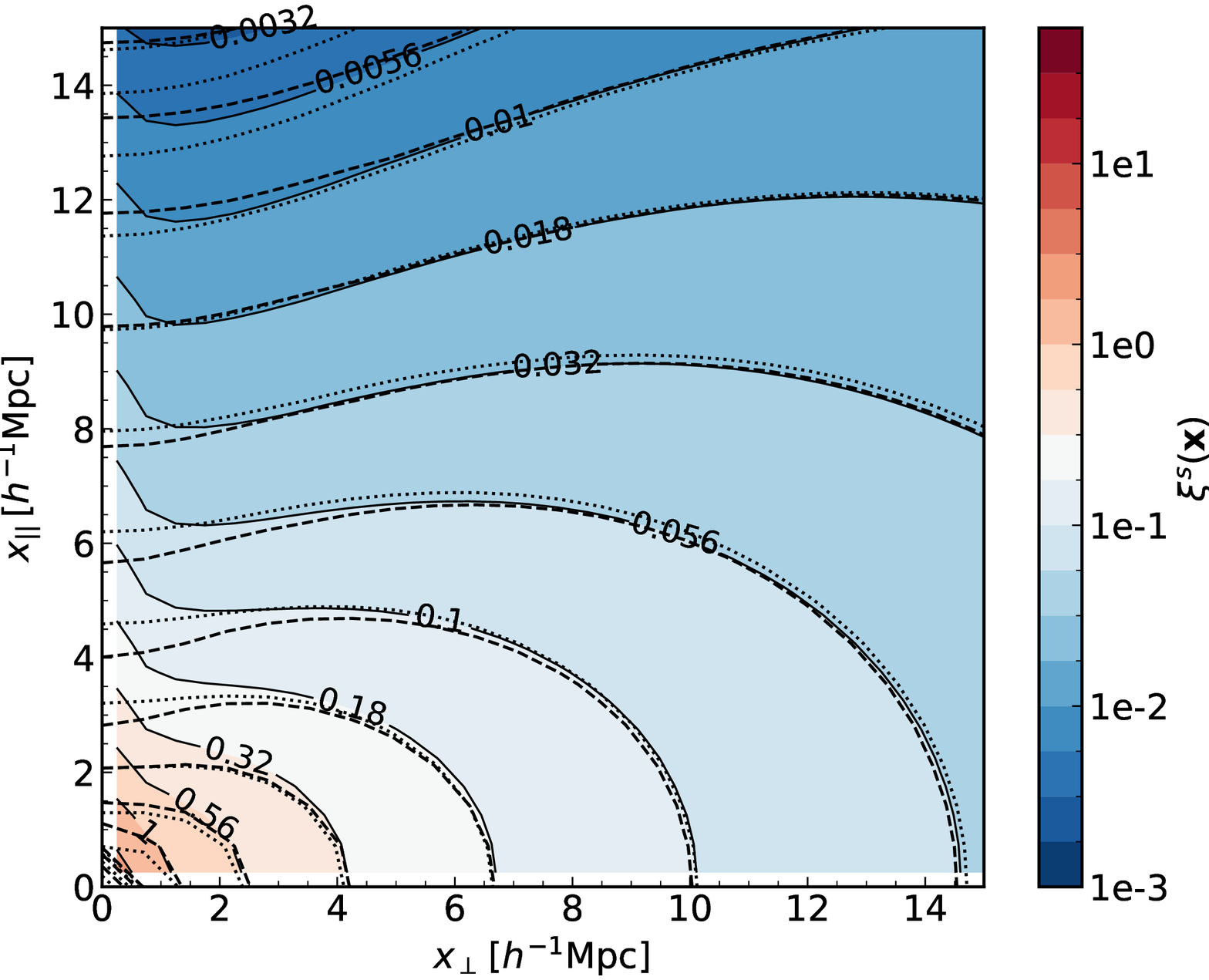}}
\end{center}
\caption{Zoom up of Fig.~\ref{fig_xi2d}, using higer-resolution simulation suite ($1024^3$ particles, $512\,h^{-1}$Mpc).}
\label{fig_xi2d_zoom}
\end{figure*}

Finally, we can see the overall trends of the correlation function from different prescriptions as a function of the separation vector, $(x_\perp, x_{||})$, perpendicular and parallel to the line of sight, in Figs.~\ref{fig_xi2d} and \ref{fig_xi2d_zoom}.
As expected, both our Gaussian ansatz and the N-body simulations approach to the linear theory predictions at large separations and at high redshifts. The distinct feature of the finger-of-god effects is already visible at $z=3$ for the N-body data in the zoom plot near the $x_{||}$ axis ($x_\perp\lesssim 1\,h^{-1}$Mpc), which cannot be recovered neither by linear theory nor the nonlinear model. This feature becomes progressively prominent at lower redshifts. On intermediate scales, the improvement brought by considering our nonlinear ansatz is clear. The deformation of the contour lines from the simplest linear predictions is properly followed by our ansatz, at least to the right direction, except where the finger-of-god effects are severe. A proper description
of the finger-of-god effects on Mpc scales requires taking into account the strong
non-Gaussianities found in virialized objects.

Like the Zeldovich approximation, our Gaussian ansatz does not contain any free
parameter. Therefore, it is competitive with other approaches.
However, in practice, redshift-space statistics are obtained from biased tracers
such as galaxies. This requires adding a bias model to the formalism studied
in this paper, which will degrade the accuracy of the theoretical predictions.
We leave such an investigation for future works.

As already noticed in Sec.~\ref{sec:power-spectrum-numerical}, the agreement with
the numerical simulations is much greater for the configuration-space correlation
function than for the power spectrum.
This is due to our Lagrangian framework and this feature is shared by other Lagrangian
schemes \cite{Matsubara:2007wj,Valageas:2013hxa}.
This also means that configuration-space statistics are much less sensitive to the details
of the dynamics on small nonlinear scales.
Thus, while in the linear theory the power spectrum is superior to the correlation function,
because different Fourier modes are decoupled, the correlation function appears to be
a more robust tool once nonlinear processes come into play
\cite{Valageas:2013hxa,Tassev:2013rta}.
This suggests that the correlation function is a better probe of cosmological models
if we wish to include mildly nonlinear scales in the analysis.

\section{Comparison with some other approaches}
\label{sec:other-approaches}

Ref.~\cite{Matsubara:2007wj} develops a Lagrangian perturbation theory, in a manner similar
to the expansion (\ref{eq:Ps-exp-infty}). It keeps the one-point cumulants,
such as $\alpha_\infty$, in the exponential and expands over the $q$-dependent terms.
It goes beyond the Zeldovich approximation by including higher-order cumulants.
The latter are computed from standard Lagrangian perturbation theory
\cite{Buchert:1993xz,Bouchet:1994xp}, assuming a curl-free velocity field.
In contrast, in the model considered in this paper we do not include higher-order
cumulants beyond the Gaussian, but we do not perform any perturbative expansion
and the Gaussian power spectra themselves are modified from the linear theory
by nonlinear effects, associated with the exact equations of motion
(\ref{eq:dPchichi-deta})-(\ref{eq:dPthetatheta-deta}). This allows us to go beyond shell
crossing. As in \cite{Matsubara:2007wj}, in principle we can go beyond the Gaussian
by taking into account higher-order correlations of the displacement and velocity
fields. However, this may lead to intricate computations and we leave such a study
for future works.

The TNS model \cite{Taruya:2010mx} is based on Eulerian standard perturbation theory.
It goes beyond the linear Kaiser result (\ref{eq:Ps-L}) by going up to one-loop order
and keeping a damping prefactor. This prefactor, which originates from
an exponential term as in Eq.(\ref{eq:Psk-A-def}), is fitted to numerical simulations
to describe the damping due to small-scale motions associated with the ``finger-of-god''
effect.
As for Lagrangian approaches and as for our model shown in Fig.~\ref{fig_rPk},
using the velocity dispersion predicted by linear theory would overestimate the damping
at high redshifts \cite{Taruya:2010mx}.
Nevertheless, with the appropriate damping factor it gives a good match to numerical
simulations and improves over standard perturbation theory
\cite{Taruya:2010mx,GilMarin:2012nb}.

EFT approaches to the redshift-space power spectrum have been presented in
\cite{Lewandowski:2015ziq,delaBella:2017qjy}.
As compared with the real-space power spectrum, this requires a few additional
counterterms factors, because of the composite operators brought by the nonlinear
mapping to redshift space. These new counterterms may also be associated with
the successive terms of the expansion of $P^s(k,\mu)$ in powers of $\mu^2$
\cite{delaBella:2017qjy}.
In our approach we do not need such counterterms (unless we consider
biased tracers or baryonic effects), as we use the exact equation of motion
(\ref{eq:Psi-eom}), which is valid beyond shell crossing.
In fact, the equivalent of the EFT counterterms is provided by the self-truncation
at high $k$ of the displacement and velocity power spectra. This damping
arises from the equations of motion (\ref{eq:dPchichi-deta})-(\ref{eq:dPthetatheta-deta})
in a nonperturbative manner. This can be seen from the effective damping factor
$\lambda_{\infty} = - e^{-1/(12\alpha_0)}/(6\sqrt{3\pi}\alpha_0^{3/2})$
that arises from the dynamics in the Gaussian ansatz, see \cite{Valageas:2020} for details.

These EFT Eulerian-space methods also need to perform a partial resummation
to take into account the damping of the baryon acoustic oscillations by large-scale
motions. This uses the Lagrangian picture, and its exponential damping as in
(\ref{eq:Psk-A-def}), as a starting point to infer an effective damping kernel
that is inserted in the Eulerian power spectrum \cite{Lewandowski:2015ziq}.
This may be done through a semi-phenomenological split of the linear power spectrum
into a smooth ``no-wiggle'' component and the oscillatory ``wiggly'' component
\cite{Vlah:2015zda,delaBella:2017qjy}.
Then, whereas the no-wiggle component is expanded as in SPT, the wiggly component
keeps a non-expanded Gaussian prefactor, which corresponds to part of the exponent
in Eq.(\ref{eq:Psk-A-def}). This provides a damping of only the wiggly part.
This can be related to the behavior found in Fig.~\ref{fig_rPk}.
As we use a Lagrangian approach, the damping due to large-scale motions
is automatically included and ``resummed''. In fact, it is never expanded, as we keep
the exponential (\ref{eq:Psk-A-def}). However, because we do not treat in different manners
the smooth and wiggly components, the damping applies to the full power spectrum.
This explains why we find in Fig.~\ref{fig_rPk} an excessive damping of the smooth
component, as compared with the numerical simulations and such Eulerian schemes
with semi-phenomenological splitted damping.
This excessive damping is a typical feature of Lagrangian approaches
\cite{Matsubara:2007wj,Valageas:2013hxa,Vlah:2014nta}.
Because our goal is to investigate the Lagrangian-space Gaussian ansatz
introduced in \cite{Valageas:2020}, we do not try to cure this problem by an ad-hoc procedure.
Indeed, the spirit of the general method presented in \cite{Valageas:2020}, beyond the Gaussian ansatz
computed in this paper, is to keep as much as possible exact expressions,
such as the equations of motion (\ref{eq:dPchichi-deta})-(\ref{eq:dPthetatheta-deta})
and the power spectrum (\ref{eq:Ps-sq}). This allows us to interprete the excessive
damping of the smooth component on BAO scales as due to the neglect of higher-order
correlations of the displacement field and to the failure to describe highly nonlinear
overdensities such as virialized halos. This is also suggested by the good agreement
with the configuration-space correlation function, except on small scales
below $10 h^{-1}$Mpc.

Ref.~\cite{Ivanov:2018gjr} uses the ``time-sliced perturbation theory'' introduced
in \cite{Blas:2015qsi}. As in our approach, instead of considering the dynamical
fields this method directly works at the level of their probability distribution.
However, whereas we use a nonperturbative scheme on the probability distribution
of the Lagrangian-space displacement and velocity fields,
the method of \cite{Blas:2015qsi,Ivanov:2018gjr} uses a perturbative expansion
on the probability distribution of the Eulerian-space density and velocity fields.
As for Eulerian-based EFT, they perform a partial resummation to take care
of infrared effects associated with large-scale motions.

Streaming models \cite{Scoccimarro:2004tg} relate the redshift-space correlation function
to a convolution of the real-space correlation function by the probability distribution
of the pairwise line-of-sight velocity.
In the popular Gaussian streaming model \cite{Reid:2011ar}, the velocity distribution
is Gaussian, as predicted by linear theory, but it is possible to include the skewness
\cite{Uhlemann:2015hqa} or exponential tails \cite{Bianchi:2016qen,Kuruvilla:2017kev}.
However,
this requires measurements of the velocity distribution or low-order moments from simulations.
These approaches are related to our Gaussian ansatz as they recover the Zeldovich
approximation at lowest order if Gaussian terms are kept exponentiated \cite{Carlson:2012bu}.
The difference is that in our method the Gaussian term itself is modified in a nonperturbative
manner by the requirement to fulfil the equations of motion
(\ref{eq:dPchichi-deta})-(\ref{eq:dPthetatheta-deta}).

\section{Conclusion}
\label{sec:conclusion}

In this paper we have investigated the redshift-space matter density power spectrum
and correlation function predicted by a new Lagrangian Gaussian ansatz.
We have also derived the redshift-space power spectrum for arbitrary
Gaussian displacement and velocity fields, and provided some explicit expressions
for numerical computations.

As for the real-space statistics, we find that the damping of the BAO oscillations
in the power spectrum is well recovered but the amplitude is off by a smooth drift,
so that this approach is not competitive as compared with other methods.
However, if one can extract the oscillatory pattern from the data, or if one adds
a few free parameters to describe the smooth drift, this scheme may become efficient.
We leave an investigation of this point for future work.

The accuracy is much greater for the configuration-space correlation function.
This is generally expected for Lagrangian-space schemes. It also suggests that
nonlinear processes are easier to separate in configuration space.
As usual, the accuracy degrades for higher orders, as one goes from the monopole
to the quadrupole and the hexadecapole, but in all cases we obtain a significant improvement
over the linear theory and a modest improvement over the Zeldovich approximation.
In particular, for the monopole, we obtain an accuracy of $2\%$ down to
$10 h^{-1}$ Mpc, at $z \geq 0.35$.
For the quadrupole, we find an accuracy of $3\%$ down to
$26 h^{-1}$ Mpc, and of $10\%$ down to $20 h^{-1}$ Mpc, at $z \geq 0.35$.

This work suggests several points for further investigations.
The practical analysis of galaxy surveys will require a biasing scheme in order
to describe biased tracers.
To improve the accuracy for the power spectrum or to reach smaller scales,
it will be necessary to go beyond the Gaussian ansatz and to include the higher-order
correlations of the displacement and velocity fields.
Indeed, it is well known that the pairwise velocity distribution is not Gaussian but
asymmetric with exponential tails, even on large scales, which has an impact on
redshift-space statistics \cite{Scoccimarro:2004tg,Kuruvilla:2017kev}.

\acknowledgments
This work is supported in part by World Premier
International Research Center Initiative (WPI Initiative),
MEXT, Japan, and by MEXT/JSPS KAKENHI
Grant Numbers JP17K14273, and JP19H00677.
This work was also supported by JST AIP Acceleration Research Grant Number JP20317829, Japan.
Numerical computations were carried out on Cray XC50 at Center for Computational
Astrophysics, National Astronomical Observatory of Japan.

\appendix

\section{Numerical computation of the redshift-space power spectrum}
\label{app:power-spectrum}

We present here the expressions of the redshift-space power spectrum that we
use for our numerical computations. Similar and alternative methods for the particular
case of the Zeldovich power spectrum are described in
\cite{Taylor:1996ne,Valageas:2010rx,Vlah:2018ygt}.
Our method gives an expression that keeps the same form as the
expansions of the real-space power spectra for the Zeldovich approximation
\cite{1995MNRAS.273..475S} and the Gaussian ansatz \cite{Valageas:2020}.
It only involves spherical Bessel functions and polynomials (the series associated
with the hypergeometric function in Eq.(\ref {eq:Ps-kmu-2F1}) below terminates at
a finite number of terms).

Choosing the coordinate axis so that ${\bf k} = (0,0,k)$, ${\bf e}_z=(\sqrt{1-\mu^2},0,\mu)$
and ${\bf q} = (q \sqrt{1-\nu^2} \cos\varphi, q \sqrt{1-\nu^2} \sin\varphi, q \nu)$,
Eq.(\ref{eq:Ps-k-alpha-beta}) reads
\ba
&& P^s(k,\mu) = \int_0^{\infty} \frac{d q}{(2\pi)^3} q^2 \;
e^{-A} \int_{-1}^{1} d\nu \; e^{ikq\nu-B \nu^2} \nonumber \\
&& \times \int_0^{2\pi} d\varphi \;
e^{- C \nu \sqrt{1-\nu^2} \cos\varphi - D (1-\nu^2) \cos^2\varphi} ,
\ea
with
\ba
&& A= k^2 [ \alpha_{\chi\chi} + f \mu^2 (2 \alpha_{\chi\theta}+f \alpha_{\theta\theta}) ] ,
\label{eq:Aq-def} \\
&& B = k^2 [ \beta_{\chi\chi} + f \mu^2 (2\beta_{\chi\theta} + f \mu^2 \beta_{\theta\theta} ) ] , \\
&& C = k^2 ( \beta_{\chi\theta} + f\mu^2 \beta_{\theta\theta} ) 2 f \mu \sqrt{1-\mu^2} , \\
&& D = k^2 \beta_{\theta\theta} f^2\mu^2 (1-\mu^2) .
\ea
Expanding the exponentials of the $\cos\varphi$ and $\cos^2\varphi$ terms and using
\be
\int_0^{2\pi} d\varphi \, (\cos\varphi)^{2n} = 2\pi \frac{(2n)!}{2^{2n} (n!)^2} ,
\ee
the integration over $\varphi$ gives
\ba
&& P^s(k,\mu) = \int_0^{\infty} \frac{d q}{2\pi^2} q^2 \; e^{-A}
\int_0^{1} d\nu \; \cos(kq\nu) e^{-B \nu^2} \nonumber \\
&& \times \sum_{\ell,m=0}^{\infty}
\frac{(2\ell+2m)!}{(2\ell)! \, m! \, [(\ell+m)!]^2} \frac{C^{2\ell} (-D)^{m}}{2^{2\ell+2m}}
\nu^{2\ell} (1-\nu^2)^{\ell+m} . \nonumber \\
&&
\ea
To recover the series associated with the real-space power spectrum,
we expand the exponential over $(1-\nu^2)$ instead of $\nu^2$
\cite{1995MNRAS.273..475S} \cite{Valageas:2020}.
Reorganizing the series, we obtain
\ba
&& P^s(k,\mu) = \int_0^{\infty} \frac{d q}{2\pi^2} q^2 \; e^{-(A+B)}
\sum_{n=0}^{\infty} B^n \sum_{\ell=0}^n \left( \frac{-C^2}{4B} \right)^{\! \ell} \nonumber \\
&& \times \sum_{m=0}^{n-\ell} \left( \frac{-D}{4B} \right)^{\! m}
\frac{(2\ell+2m)!  \; I_{n,2\ell}(kq)}{(n-\ell-m)! \,(2\ell)! \, m! \, [(\ell+m)!]^2} , \hspace{0.8cm}
\label{eq:Ps-kmu-In2l}
\ea
where we introduced
\ba
&& I_{n,2\ell+1}(z) = (-1)^{\ell+1} \int_0^{1} dx \, \sin(z x) \, (1-x^2)^{n} \, x^{2\ell+1} , \nonumber \\
&& I_{n,2\ell}(z) = (-1)^{\ell} \int_0^{1} dx \, \cos(z x) \, (1-x^2)^{n} \, x^{2\ell}  .
\ea
We recover the real-space power spectrum by setting $f=0$ or $\mu=0$.
In this case, $C=D=0$ and only the terms $\ell=m=0$ contribute.
We derive explicit expressions for the integrals $I_{n.2\ell}$ by recursion.
At order $\ell=0$, we have \cite{Gradshteyn1965,1995MNRAS.273..475S}
\be
I_{n,0}(z) = n! \; 2^n z^{-n} j_n(z) ,
\label{eq:In-0}
\ee
while higher orders obey the recursion
\be
I_{n,\ell+1}(z) = \frac{d}{dz} I_{n,\ell}(z) .
\label{eq:In+1-In}
\ee
Using the properties of spherical Bessel functions,
\be
\frac{d}{dz} \left( \frac{j_n}{z^n} \right) = - \frac{j_{n+1}}{z^n} , \;\;\;
j_{n-1}+j_{n+1} = \frac{2n+1}{z} j_n ,
\ee
we can show by recursion from Eqs.(\ref{eq:In-0})-(\ref{eq:In+1-In}) that the
functions $I_{n,\ell}(z)$ take the form
\ba
I_{n,\ell}(z) & = & n! \; 2^n z^{-n-\ell+1} [ P_{n,\ell}(z) j_{n+\ell-1}(z) \nonumber \\
&& + Q_{n,\ell}(z) j_{n+\ell}(z) ] ,
\label{eq:In-PQ}
\ea
where the functions $P_{n,\ell}(z)$ and $Q_{n,\ell}(z)$ are polynomials of order
$\ell-2$ and $\ell-1$, except for $Q_{n,0}$, and satisfy the recursion
\ba
&& P_{n,\ell+1} = (2n+2\ell+1) P'_{n,\ell} - z P_{n,\ell} + Q_{n,\ell} + z Q_{n,\ell}' ,
\hspace{0.3cm} \nonumber \\
&& Q_{n,\ell+1} = - z P'_{n,\ell} - z Q_{n,\ell} ,
\ea
where the prime denotes the derivative with respect to $z$.
The lowest orders are
\ba
&& P_{n,0} = P_{n,1} = 0, \;\; P_{n,2} = -1 , \nonumber \\
&& Q_{n,0} = \frac{1}{z} , \;\; Q_{n,1} = -1 , \;\; Q_{n,2} = z .
\ea
Substituting Eq.(\ref{eq:In-PQ}) into Eq.(\ref{eq:Ps-kmu-In2l}) gives
\ba
&& P^s(k,\mu) = \int_0^{\infty} \frac{d q}{2\pi^2} q^2 \; e^{-(A+B)} kq
\sum_{n=0}^{\infty} \left( \frac{2B}{kq} \right)^{\! n} \nonumber \\
&& \times \sum_{\ell=0}^n \left( \frac{-C^2}{4B k^2q^2} \right)^{\! \ell}
[ P_{n,2\ell} j_{n+2\ell-1}(kq) + Q_{n,2\ell} j_{n+2\ell}(kq) ] \nonumber \\
&& \times \sum_{m=0}^{n-\ell} \left( \frac{-D}{4B} \right)^{\! m}
\frac{n! \; (2\ell+2m)!}{(n-\ell-m)! \,(2\ell)! \, m! \, [(\ell+m)!]^2} . \hspace{1.cm}
\label{eq:Ps-kmu-PQl}
\ea
Using the summation in terms of the hypergeometric function,
\ba
\sum_{m=0}^{n-\ell} && \!\!\!\!\! \frac{(2\ell+2m)!}{(n-\ell-m)! \, m! \, [(\ell+m)!]^2} (-x)^m =
\frac{4^{\ell} \Gamma[\ell+1/2]}{\sqrt{\pi} (n-\ell)! \, \ell !} \nonumber \\
&& \times \; _2 F_1(\ell+1/2,-n+\ell;\ell+1;4x) ,
\ea
we obtain
\ba
&& P^s(k,\mu) = \int_0^{\infty} \frac{d q}{2\pi^2} q^2 \; e^{-(A+B)} kq
\sum_{n=0}^{\infty} \left( \frac{2B}{kq} \right)^{\! n} \nonumber \\
&& \times \sum_{\ell=0}^n \left( \frac{-C^2}{4B k^2q^2} \right)^{\! \ell}
[ P_{n,2\ell} j_{n+2\ell-1}(kq) + Q_{n,2\ell} j_{n+2\ell}(kq) ] \nonumber \\
&& \times \frac{ n! }{(n-\ell)! \, (\ell !)^2}  \; _2 F_1(\ell+1/2,-n+\ell;\ell+1;D/B) . \hspace{1.cm}
\label{eq:Ps-kmu-2F1}
\ea
For $f=0$ or $\mu=0$, which give $C=D=0$, we recover the expression of the real-space
power spectrum \cite{Valageas:2020}.

As for the real-space power spectrum \cite{Valageas:2010rx} \cite{Valageas:2020},
for numerical computations it is convenient
to improve the convergence of the integral over $q$ by separating the linear part.
Thus, defining the one-point variance
\be
\alpha_{\infty **} = \frac{4\pi}{3} \int_0^\infty dk \, P_{**}(k) ,
\label{eq:alpha-infty}
\ee
which is also the limit of the variance (\ref{eq:alpha-def}) at large separations $q$,
the redshift-space power spectrum (\ref{eq:Psk-A-def}) also reads as \cite{Crocce:2005xz}
\ba
&& P^s({\bf k}) = e^{-k^2 [ \alpha_{\infty\chi\chi}+f\mu^2(2\alpha_{\infty\chi\theta}
+ f\alpha_{\infty\theta\theta})]} \int \frac{d{\bf q}}{(2\pi)^3} e^{i {\bf k}\cdot{\bf q}} \nonumber \\
&& \times e^{\int d{\bf k}' e^{i{\bf k}'\cdot{\bf q}} [ ({\bf k}\cdot{\bf k}')^2 P_{\chi\chi}
+ 2 f ({\bf k}\cdot{\bf k}') k_z k'_z P_{\chi\theta} + f^2 (k_z k'_z)^2 P_{\theta\theta}]/k'^4 } .
\nonumber \\
&& \label{eq:Ps-exp-infty}
\ea
Defining $A_{\infty}$ as the infinite-separation limit of $A(q)$ introduced in
Eq.(\ref{eq:Aq-def}), obtained from the infinite-separation variance (\ref{eq:alpha-infty}),
and expanding the exponential, gives the alternative expansion
\be
P^s({\bf k}) = e^{-A_\infty} \sum_{n=0}^{\infty} \frac{1}{n!} P^{s(n)}({\bf k}) ,
\label{eq:Ps-Ainfty}
\ee
where $P^{s(n)}$ is of order $n$ in the displacement and velocity power spectra.
The integration over ${\bf q}$ gives a Dirac factor in each term $P^{s(n)}$.
As usual, the term $n=0$ vanishes for $k>0$ while the linear term reads
\be
P^{s(1)}(k,\mu) = P_{\chi\chi}(k) + 2f\mu^2 P_{\chi\theta}(k) + f^2 \mu^4 P_{\theta\theta}(k) .
\ee
Then, we subtract the first term of the expansion (\ref{eq:Ps-Ainfty}) from
the expression (\ref{eq:Ps-kmu-2F1}). This gives
\ba
&& P^s(k,\mu) = e^{-A_\infty} P^{s(1)}(k,\mu) +  e^{-A_\infty}
\int_0^{\infty} \frac{d q}{2\pi^2} q^2 \nonumber \\
&& \times \Biggl \lbrace j_0(kq)
\left[ e^{A_{\infty}-A-B} - 1 - (A_{\infty}-A-B) \right] \nonumber \\
&& + j_1(kq) B (1-D/2) \left[ e^{A_{\infty}-A-B} - 1 \right]
+ e^{A_{\infty}-A-B} kq \nonumber \\
&& \times BC [4 j_2(kq) - kq j_1(kq) ]
+ e^{A_{\infty}-A-B} kq \sum_{n=2}^{\infty} \left( \frac{2B}{kq} \right)^{\! n} \nonumber \\
&& \times \sum_{\ell=0}^n \left( \frac{-C^2}{4B k^2q^2} \right)^{\! \ell}
[ P_{n,2\ell} j_{n+2\ell-1}(kq) + Q_{n,2\ell} j_{n+2\ell}(kq) ] \nonumber \\
&& \times \frac{ n! }{(n \! - \! \ell)! \, (\ell !)^2}  \; _2 F_1(\ell+1/2,-n+\ell;\ell+1;D/B)
\Biggl \rbrace . \hspace{1.cm}
\label{eq:Ps-kmu-regular}
\ea
This improves the convergence of the integral at large $q$ and makes the numerical
computation easier. The one-point variance $A_\infty$ is only an auxiliary quantity
for the numerical scheme. The power spectrum (\ref{eq:Ps-k-alpha-beta}) does not
depend on its value and remains well defined even if $A_\infty$ is infinite,
see \cite{Valageas:2020} for an explicit example on the case of the real-space power spectrum
with a power-law initial condition $P_L(k) \propto k^{-2}$.
Thus, because our approach is based on a Lagrangian-space framework,
it does not suffer from the infrared divergences or artificially large contributions
that affect Eulerian approaches and require specific care
\cite{Senatore:2014via,Vlah:2015zda,Blas:2016sfa,Ivanov:2018gjr}.

\bibliography{ref1}


\end{document}